\documentclass{siamltex}

\usepackage[dvips]{graphics}
\usepackage{amssymb,amsmath}
\newcommand{\mathtext}[1]{\ \hbox{\rm #1} \ }
\newcommand{\de}[1]{\textit{#1}}
\def\cS{{\mathcal S}}
\def\cC{{\mathcal C}}
\def\bX{{\bf X}}
\def\torus{{\mathbb T}}
\def\complexes{{\mathbb C}}
\def\reals{{\mathbb R}}
\def\integers{{\mathbb Z}}
\def\dz{{\dot z}}
\def\raw{\rightarrow}
\def\diag{\hbox{\rm diag}}
\def\relper{relative periodic orbit}
\def\relpers{relative periodic orbits}
\def\relsoln{relative periodic solution}
\def\relsolns{relative periodic solutions}
\def\releq{relative equilibrium}
\def\releqs{relative equilibria}
\def\pos{periodic orbits}
\def\po{periodic orbit}
\def\phit{{\phi_t}}
\def\relmon{relative monodromy}
\def\tM{{\tilde M}}
\def\hG{{\Gamma}}
\def\AA{{\tilde A}}
\newcommand{\mmbf}[1]{\mbox{\boldmath ${#1}$}}
\newcommand{\refEqn}[1]{(\ref{#1})}
\newcommand{\refEqns}[2]{\mbox{(\ref{#1})--(\ref{#2})}}
\newcommand{\kernel}[1]{ker({#1})}
\newcommand{\LM}{\mbox{Levenberg-Marquardt}\ }
\newcommand{\newtons}{\mbox{Newton's}\ }
\newcommand{\F}{\mathbf{F}}
\newcommand{\FL}{\F_{\mathrm{L}}}
\newcommand{\FNL}{\F_{\mathrm{NL}}}
\newcommand{\Feqzero}{\mbox{$\F = \mmbf{0}$}}
\newcommand{\nleargs}{\mmbf{\hat{a}}, \varphi, S, T}
\newcommand{\Fargs}{\mbox{$\F(\nleargs)$}}
\newcommand{\FLargs}{\mbox{$\FL(\nleargs)$}}
\newcommand{\FNLargs}{\mbox{$\FNL(\mmbf{\hat{a}})$}}
\newcommand{\ii}{\mathrm{i}}
\newcommand{\e}{\mathrm{e}}
\newcommand{\dd}{\mathrm{d}}
\newcommand{\Fix}{\mathrm{Fix}}
\newcommand{\sss}{\scriptscriptstyle}
\newcommand{\tblrelsolns}{A.1}
\newcommand{\tblsf}{A.2}

\newcommand{\prava}[1]{\small\it
\begin{flushleft}
Copyright \copyright \  {#1}
\end{flushleft}}
\title{Relative Periodic Solutions of the \\ Complex Ginzburg-Landau Equation
       \thanks{This work was partially supported by NSF Grant CTS 97-29189 and by the endowment for
       the Fulton Watson Copp Chair in Computer Science at the University of Illinois.}}

\author{Vanessa L\'{o}pez\footnotemark[2]\ \footnotemark[5]
       \and Philip Boyland\footnotemark[3]
       \and Michael T. Heath\footnotemark[2]
       \and Robert D. Moser\footnotemark[4]}

\begin{document}

\maketitle

\renewcommand{\thefootnote}{\fnsymbol{footnote}}
\footnotetext[2]{Department of Computer Science, University of Illinois, Urbana, IL, USA
    ({\tt vlopez@cse.uiuc.edu, heath@cs.uiuc.edu}).}
\footnotetext[3]{Mathematics Department, University of Florida, Gainesville, FL, USA
    ({\tt boyland@math.ufl.edu}).}
\footnotetext[4]{Department of Theoretical and Applied Mechanics, University of Illinois, 
    Urbana, IL, USA.  Current address: Mechanical 
    Engineering Department,  The University of Texas at Austin,
    1 University Station C2200, Austin, TX, USA ({\tt rmoser@mail.utexas.edu}).}
\footnotetext[5]{Corresponding author. 
    Current address: Lawrence Berkeley National Lab,
    1 Cyclotron Road, Mail Stop 50F1650,
    Berkeley, CA 94720, USA.     
%%    Tel: 1-510-486-5627.
%%    Fax: 1-510-486-5812.
    E-mail: {\tt vlopez@hpcrd.lbl.gov}.}

\renewcommand{\footnoterule}{}
{\renewcommand{\thefootnote}{}
 \footnote{\prava{2005 Society for Industrial and Applied Mathematics}}} 

\renewcommand{\thefootnote}{\arabic{footnote}}

\begin{abstract}
A method of finding \relpers\ for differential equations with continuous symmetries is
described and its utility demonstrated by computing \relsolns\ for the one-dimensional complex 
Ginzburg-Landau equation (CGLE) with periodic boundary conditions.  A \relsoln\ is a solution that is 
periodic in time, up to a transformation by an element of the equation's symmetry group.  With the 
method used, \relsolns\ are represented by a space-time Fourier series modified to include the symmetry 
group element and are sought as solutions to a system of nonlinear algebraic equations for the Fourier 
coefficients, group element, and time period. The 77 \relsolns\ found for the CGLE exhibit a wide variety 
of temporal dynamics, with the sum of their positive Lyapunov exponents varying from 5.19 to 60.35 and 
their unstable dimensions from 3 to 8.  Preliminary work indicates that weighted averages over the 
collection of \relsolns\ accurately approximate the value of several functionals on typical trajectories.
\end{abstract}



\pagestyle{myheadings}
\thispagestyle{plain}
\markboth{L\'{O}PEZ, BOYLAND, HEATH, MOSER}{RELATIVE PERIODIC SOLUTIONS OF THE CGLE}

%%% new section %%%
\section{Introduction}

The complex Ginzburg-Landau equation (CGLE) is a widely studied partial differential equation with
applications in many areas of science. (See~\cite{aranson02,encyc} and references therein.)  It has
become a model problem for the study of nonlinear evolution equations with chaotic spatio-temporal
dynamics.  In this paper we work with the CGLE with cubic nonlinearity in one spatial dimension,
%\end{eqnarray}
\begin{equation}   \label{eq:cgle_pde}
    \frac {\partial A}{\partial t}  = 
    R A + (1 + \ii \nu) \frac{\partial^2 A}{\partial x^2} - (1 + \ii \mu) A|A|^2,
\end{equation}
with periodic boundary conditions
\begin{equation}   \label{eq:cgle_bcs}
    A(x,t)  =  A(x+2\pi,t),
\end{equation}
and, as a convention, when we refer to \de{the CGLE} we mean equation~\refEqn{eq:cgle_pde} with the
boundary conditions~\refEqn{eq:cgle_bcs}, unless otherwise noted.  Equation~\refEqn{eq:cgle_pde}
describes the evolution of a complex-valued field $A(x,t)$.  The parameters $R$, $\nu$, and $\mu$
are real.  When $R<0$ all solutions converge to $A = 0$ in forward time, but when $R>0$ there is,
in general, nontrivial spatio-temporal behavior.  The parameters $\nu$ and $\mu$ are measures of the
linear and nonlinear dissipation, respectively.  (See, for example, \cite{aranson02,doering88,
levermore96,vanSaarloos94} for further details.)  It is known that in the form~\refEqn{eq:cgle_pde}
the CGLE generates a continuous semiflow on a variety of spaces~\cite{Bart,Doelman,levermore96,
temambook}.

The CGLE has a three-parameter group $\torus^2 \times \mathbb{R}$ of continuous symmetries
generated by space-time translations and a rotation of the complex field $A$.  That is, if
$A(x, t)$ is a solution of the CGLE, then so is $\e^{\ii\eta_1} A(x+\eta_2,  t+\tau)$ for any
element $(\eta_1, \eta_2)$ of the two-torus $\torus^2$ and $\tau \in \reals$.  Thus, we focus our
study on time-periodic solutions of the CGLE relative to the $\torus^2$-symmetry, namely,
solutions of the CGLE that satisfy
\begin{equation}  \label{eq:Atpsol}
    A(x,t)  =  \e^{\ii\varphi}A(x+S, t+T)
\end{equation} 
for some $(\varphi, S) \in \torus^2$ and $T > 0$.   Note that after a period of time $T$, such a
solution returns not to itself, but rather to an element in its $\torus^2$-orbit.  Such
\emph{relative periodic solutions} represent invariant three-tori in the CGLE flow.  Previous
studies of time-periodic solutions of the CGLE~\refEqn{eq:cgle_pde} have centered on two types of
solutions.  The first are single-frequency solutions of the form $A(x,t) = B(x) \e^{\ii\omega t}$
(see, for example, \cite{dangvu00,doelman89,kapitula96}); these are referred to as \emph{stationary
solutions}. The second type are generalized traveling waves, also called \emph{coherent structures},
for which $A(x,t) = \rho(x-vt) \e^{\ii\phi(x-vt)} \e^{\ii\omega t}$, where $\rho$ and $\phi$ are
real-valued functions and $\omega$ is some frequency (see, for example, \cite{aranson02,brusch01,
encyc,vanHecke02}).  These coherent structures reduce to single-frequency solutions by a change to
a moving frame, that is, by a change of variables $x \rightarrow x+vt$, $t \rightarrow t$.
The relative periodic solutions of interest in this study are different in that they exhibit more
complicated temporal behavior than the single-frequency solutions just described. 
Studies of bifurcations to (stable) invariant two and three-tori for the CGLE are described in
\cite{takac98} and references therein.  
The \relsolns\ we compute, which represent invariant three-tori, are obtained by working with fixed
parameter values for the CGLE, as described below (and in more detail in \S~\ref{sec:dynatparam}
and \S~\ref{sec:compute_relsolns}), not via a bifurcation study.

We employ a spectral-Galerkin procedure to compute \relsolns\ of the CGLE.  A solution to
equation~\refEqn{eq:Atpsol} is constructed and is then used to represent the \relsolns.  This
ansatz consists of a Fourier series in both space and time, modified to include the $\torus^2$
element $(\varphi,S)$ as an unknown.  Substitution of the ansatz into the differential equation
yields a system of nonlinear algebraic equations, the solutions of which give the desired \relsolns.
Using this procedure we were able to identify 77 distinct \relsolns\ of the CGLE with parameters
$R=16$, $\nu=-7$, and $\mu=5$.  These parameter values were chosen because for them the system
exhibits temporal chaos, yet the value of $R$ is such that a relatively small number of spatial
modes can capture the dynamics of the system.

The task of computing \relsolns\ was undertaken as a first step in understanding the dynamics of
the CGLE with the chosen parameters.  The 77 \relsolns\ are neither stationary nor coherent
structures (as defined earlier), but two stationary solutions were also found. Aside from these two
single-frequency solutions, we found no (truly) time-periodic solutions.  For each \relsoln,
the largest Lyapunov exponent ranges from 1.88 to 17.20, the sum of the positive Lyapunov exponents
from 5.19 to 60.35, and the unstable dimension from 3 to 8. In particular, the system displays
unstable dimension variability, a phenomenon that a number of recent papers have indicated is
central to the understanding of high-dimensional dynamical systems; see, for example,
\cite{udvpaper2,udvpaper1}.

In a general context, the study of periodic orbits has been a basic tool in the theory of dynamical
systems for at least 100 years.  Typically, there are infinitely many different periodic solutions
embedded in a chaotic attractor.  Using tools from periodic orbit theory, statistical averages that
provide a description of the asymptotic behavior of a chaotic dynamical system can be approximated
from the short-term dynamics of the (unstable) periodic solutions on an attractor of the chaotic
system~\cite{bk:chaosbook,cvitanovic00}. Periodic solutions of chaotic dynamical systems are also
used in techniques for controlling chaos.  Given a set of unstable periodic solutions of a chaotic
dynamical system, the chaos could be mitigated by applying small, time-dependent perturbations to
stabilize trajectories onto chosen periodic solutions.  This is the basis of the
\mbox{Ott-Grebogi-Yorke} approach for controlling chaos; see, for example,~\cite{bk:ott}. 

(True) periodic solutions of dynamical systems have been computed with a variety of methods.  The
use of Fourier series along with the numerical solution of systems of nonlinear algebraic equations
is common (as in~\cite{lau82,ling87,urabe66}), but the use of multiple shooting algorithms for boundary
value problems, which combine numerical integration and the solution of systems of nonlinear
algebraic equations, is also popular (see, for example, \cite{choe99,bk:chaosbook,guckenheimer00}).
Furthermore, expansions other than Fourier series, for example, Taylor series
expansions~\cite{guckenheimer00} and piecewise polynomials~\cite{auto}, are also utilized.  More
recently, a variational method for finding (true) periodic solutions was proposed~\cite{lan03}.  

Often, the suitability of a numerical method for computing periodic solutions is illustrated by
applying the method to low-dimensional dynamical systems (2--4 equations).  Higher dimensional
systems have been considered in \cite{christiansen97}, \cite{lan03}, and \cite{zoldi98} (with 16,
32, and 99 equations, respectively), where (true) periodic solutions of the Kuramoto-Sivashinsky
equation were computed.  The system studied in~\cite{christiansen97} has a low-dimensional chaotic 
attractor; those in~\cite{zoldi98} and~\cite{lan03} have higher intrinsic dimension.  In particular,
the system considered in~\cite{zoldi98} has a Lyapunov dimension of 8.8~\cite{manneville85}, with
typical trajectories having four positive and one zero Lyapunov exponent.  We also work with a
high-dimensional system (62 real ordinary differential equations) when computing \relsolns\ of the CGLE.
The system we study has a Lyapunov dimension of approximately 11.52; typical trajectories in our system
have five positive and three zero Lyapunov exponents.

%%% new section %%%
\section{Symmetries and Differential Equations}
\label{sec:symmetries_overview}

In this section we review some standard definitions and results concerning differential equations
with continuous symmetries. For more information see, for example, \cite{bk:golubitsky,Marsden,
bk:miller,bk:olver}.  Properties of relative equilibria and \relpers\ are given in \cite{Field1,
Field2,Krupa,Wulff3,Wulff1,Wulff2}.

\subsection{Basic Definitions}
\label{sec:symmetries_basic}
Under suitable hypotheses,  a vector field $\bX$, a differential equation 
\begin{equation}   \label{genode}
    \dz = \bX(z)
\end{equation}
on a space $B$, and an $\reals$-action or flow $\phi: \reals\times B \raw B$ are connected by
\begin{equation*}
    \bX(\phit(z)) = \frac{\partial\phit(z)}{\partial t}, 
\end{equation*}
where $\phit(z) $ is commonly used for $\phi(t, z)$.  The \de{trajectory} of a flow with initial
condition $z_0$ is $\{ \phit(z_0) : t \in \reals \}$ and is also called a \de{flow orbit} or a
\de{solution}.  A curve $\gamma(t)$ is said to be a solution of \refEqn{genode} if
$\gamma(t) = \phit(z_0)$ for some $z_0\in B$.

Continuous symmetries are expressed by the smooth left action of a Lie group $G$ on $B$, where we
write $g \cdot z$ for the action of the element $g\in G$ on $z\in B$. For simplicity of exposition
we assume that $G$ is compact and connected and acts linearly on the linear space $B$.  The
differential equation \refEqn{genode} is said to be \de{equivariant under the group action} if
$g \cdot \bX(z) = \bX( g \cdot z)$ for all $g\in G$ and $z \in B$. If $\phi_t$ is the flow of $\bX$,
then $G$-equivariance of $\bX$ is equivalent to
\begin{equation} \label{equivar}
    g \cdot \phi_t(z) = \phit(g \cdot z)
\end{equation}
for all $g, z$, and $t$, and one also speaks of the flow being $G$-equivariant. Since the vector
field $\bX$ is assumed to be time-independent, the equation \refEqn{genode} is invariant
under the product action of $G \times \reals$ given by $(g, \tau) \cdot z(t) = g\cdot z(t + \tau)$.

The \de{group orbit} of $z\in B$ under the $G$-action is $G \cdot z = \{ g \cdot z: g \in G\}$. If
the group acting is clear from the context, this orbit will also be denoted by $\mathcal{O}_z$.
The \de{isotropy subgroup or stabilizer} of $z$ is $G_z = \{ g\in G : g \cdot z = z\}$, and  
the group orbit $G\cdot z$ is diffeomorphic to the quotient $G/G_z$.  As a consequence of
\refEqn{equivar}, the stabilizer is constant on a trajectory, that is, $G_{\phi_t (z)} = G_z$ for
all $t$, so one may speak unambiguously of the $G$-stabilizer of a flow trajectory or of the
topological type of its $G$-orbit. In addition, if $H\subset G$ is a subgroup, then its fixed
subspace $\Fix(H) = \{ z \in B : h \cdot z = z \mathtext{for all}  h\in H\}$ is flow invariant.

\subsection{Relative Equilibria and Relative Periodic Orbits}
\label{releq_relper}

There are two equivalent points of view from which one can discuss the influence of symmetries on
the dynamics of~\refEqn{genode}.  The first is to consider the family of trajectories that arise by
applying symmetry operations to a given trajectory. The second is to note that~\refEqn{equivar}
implies that the flow on $B$ descends to a flow on the quotient space $B/G$, and then study the
dynamics of this \de{reduced flow}.

To pursue the first point of view, note that~\refEqn{equivar} implies that given the trajectory with
initial condition $z_0$, the trajectory with initial condition $g\cdot z_0$ can be obtained by
applying the symmetry $g$ to the trajectory of $z_0$. Thus, the equivariance of the flow implies
that trajectories come in families related by the symmetries.  For a given $z_0$, the union of all
the members of this family is equal to the group orbit of $z_0$ under the joint action of
$\hG :=(G \times \reals)$ on $B$ given by $(g, \tau) \cdot z = \phi_\tau(g \cdot z) = g \cdot
\phi_\tau(z)$.  Thus, each initial condition generates a flow invariant set $\hG \cdot z_0$, and we
may consider these $\hG$-orbits as the principal dynamical objects associated with the differential
equation with symmetry.

The simplest case of a $\hG$-orbit occurs when $z_0$ projects to a rest point of the reduced flow.
This occurs exactly when the $G$-orbit of $z_0$ is flow invariant, and so the $\hG$-orbit of $z_0$
is equal to its $G$-orbit. In this case the set $\hG \cdot z_0$ is called a \de{relative equilibrium},
and the point $z_0$ is said to be an element of the relative equilibrium.  When $z_0$ is an element of
a relative equilibrium, its flow trajectory is always equal to the action on $z_0$ of a one-parameter
subgroup $\sigma(t)$ of $G$, that is, 
\begin{equation}\label{releq}
    \phit(z_0) = \sigma(t) \cdot z_0.
\end{equation}

The case of primary interest here are $\hG$-orbits that project to periodic orbits of the reduced
flow. The sets are called \de{\relpers}. Thus, a point  $z_0$ is an element of a \relper\ if and only if
\begin{equation}\label{relper}
    z_0 = g_0 \cdot \phi_T(z_0)
\end{equation}
for some $g_0 \in G$ and $T>0$. Since a \releq\ in the form \refEqn{releq} will satisfy \refEqn{relper}
for $g_0 = \sigma(t)$ and $T = t$, for any $t$, we explicitly include in the definition the requirement
that a \relper\ is {\em not} a relative equilibrium. The group element $g_0$ is called the \de{drift},
and $T$ is the \de{period}.  If  $z_0$ is an element of a \relper, then its $\hG$-orbit is diffeomorphic
to $(G \cdot z_0) \times S^1$.  We will refer to the flow orbit of a point $z_0$ satisfying
\refEqn{relper} as a \de{$(g_0,T)$-\relper\ or solution}, and when it is clear from the context (in a
slight abuse of terminology), simply as a relative periodic orbit or solution.

Note that if \refEqn{relper} holds, then $\phi_t(z_0) = g_0 \cdot \phi_{t + T}(z_0)$ for all $t$.  Also
note that \refEqn{relper} does not uniquely determine $g_0$ and $T$:  a $(g_0, T)$-\relper\ is also a
$(g_0^k,kT)$-\relper\ for any integer $k$, as well as a $(g_0 g',T)$-\relper\ for any $g' \in G_{z_0}$.
However, if $z_0$ is an element of a \relper, then there is a least positive $T$ for which $\phi_t(z_0)$
is a $(g_0,T)$-\relper\ for some $g_0\in G$.  Thus, we adopt the convention that when a \relper\ is said
to have drift and period $(g_0, T)$ the period is always this least positive $T$. Even with this
convention, the drift $g_0$ is still defined only up to the stabilizer $G_{z_0}$.

Finally, we remark that there is a variety of terminology used in the literature for \relpers\ and
\releqs. In particular, the meaning of the term ``drift'' is not completely uniform, and in Hamiltonian
dynamics what is called the drift here is called, depending on the context, a reconstruction, geometric,
or other variety of phase.  Also note that the definitions and results of this section extend easily to
semiflows with continuous symmetries.

%%% new section %%%
\section{CGLE: Symmetries and Temporal Dynamics}

The CGLE in the form \refEqn{eq:cgle_pde} with periodic boundary conditions generates a continuous
semiflow on $\cS := L^2(S^1)$ as well as on other spaces requiring greater regularity~\cite{Bart,
Doelman,levermore96,temambook}.  The semiflow on $\cS$ is known to have a finite dimensional attracting
set or universal attractor as well as a Lipschitz inertial manifold~\cite{Bart,constart,doering88,Ghir,
promislow2}.  Further, on the universal attractor the CGLE generates a flow (i.e., solutions that are
bounded and defined for all time), and elements of the attractor are in fact $C^\omega$-functions
\cite{Doelman}.

\subsection{Symmetries and Associated ODEs}

The CGLE has a number of well known symmetries that are central to its behavior and our study (see, 
for example, \cite{aston99, aston00}).  If $A(x, t)$ is a solution of  (\ref{eq:cgle_pde}), then so are
\begin{align}   
    \e^{\ii\eta_1} A(x, t)&,    \label{cglesym1} \\
    A(x + \eta_2, t)&,          \label{cglesym2} \\
    A(x, t + \tau)&,            \label{cglesym3} \\
    A(-x, t)&,                  \label{cglesym4}
\end{align}
where $\eta_1, \eta_2$, and $\tau$ are any real numbers. Since the $\eta_i$ can be treated mod $2 \pi$, 
\refEqn{cglesym1} and \refEqn{cglesym2} say that the CGLE semiflow is equivariant under the action of the
two-torus group $\torus^2$ on $\cS$ with the action on  $\alpha \in \cS$ given by $(\eta_1, \eta_2)
\cdot \alpha(x) = \e^{\ii\eta_1} \alpha(x + \eta_2)$.  The symmetry \refEqn{cglesym3} corresponds to
translation along the semiflow.  The $\mathbb{Z}_2$-action expressed by \refEqn{cglesym4} will be
important in describing properties of specific solutions in \S~\ref{sec:specific_solns}, but in
most of this paper the continuous symmetries will be our primary concern.
 
Since the chosen boundary conditions for the CGLE~\refEqn{eq:cgle_pde} are periodic in $x$, we may use
spatial Fourier series to transform into $\cC :=\ell^2_b(\complexes)$, the space of all square-summable
bi-infinite, complex sequences.  Thus, we write
\begin{equation}  \label{eq:xFseries}
    A(x,t)   =  \sum_{m \in \mathbb{Z}} a_m(t) \e^{\ii mx},
\end{equation}
substitute into \refEqn{eq:cgle_pde}, and obtain an infinite system of ODEs,
\begin{equation}   \label{eq:odes}
    \frac {\dd a_m}{\dd t}  =  R a_m - m^2 (1 + \ii\nu) a_m - (1 + \ii\mu) \sum_{m_1+m_2-m_3=m}
	a_{m_1} a_{m_2} a_{m_3}^{*},
\end{equation}
for the complex-valued functions $a_m(t)$.

Under this transformation the symmetries of \refEqn{eq:cgle_pde} become symmetries of~\refEqn{eq:odes}.
Thus, if $(a_m(t))$ is a solution of \refEqn{eq:odes}, then so are 
\begin{align}   
    (\e^{\ii\eta_1} a_m(t))&,    \label{wavesym1} \\
    (\e^{\ii m\eta_2 } a_m(t))&, \label{wavesym2} \\
    (a_m( t + \tau))&,           \label{wavesym3} \\
    (a_{-m}(t))&.                \label{wavesym4}
\end{align}
In particular, \refEqn{wavesym1} and \refEqn{wavesym2} say that the ODEs~\refEqn{eq:odes} are invariant
under an action of $\torus^2$ on $\cC$, with the action on $\mmbf{a} = (a_m)\in \cC$ given by 
\begin{equation}  \label{torusact}
    (\eta_1, \eta_2)\cdot (a_m) = (\e^{\ii\eta_1} \e^{\ii m \eta_2} a_m).
\end{equation}

In all our computations with the CGLE we work with the spectral Galerkin projection obtained by fixing
an even number $N_x$, truncating the expansion \refEqn{eq:xFseries} to include the indices $m$ with 
$-N_x/2+1 \leq m \leq N_x/2-1$, and considering only the corresponding ODEs from \refEqn{eq:odes}. 
Much accumulated theory and computation demonstrates that for sufficiently large $N_x$ the behavior of
this truncation captures the essential features of the dynamics of~\refEqn{eq:cgle_pde} \cite{Doelman,
Jolly}.

The system of $N_x \! - \! 1$ complex ODEs resulting from the Galerkin projection has the same
symmetries as the infinite system.  The $\torus^2$-action is linear on the finite-dimensional space
$\complexes^{N_x-1}$, with a group element $(\eta_1,  \eta_2) \in \torus^2$ acting via multiplication by
the unitary matrix $\diag(\e^{\ii\eta_1} \e^{\ii m \eta_2})$, where it is understood that the indexing is by
increasing value of $m$, that is, $m = -N_x/2+1, \ldots, N_x/2-1$.  Henceforth this ordering is assumed
whenever applicable.  For convenience, we denote the truncated system of ODEs as
\begin{equation}  \label{eq:trunc_odes}
    \frac{\dd \mmbf{a}}{\dd t}  =  \mathbf{X}(\mmbf{a}).  
\end{equation}

\subsection{Stabilizers, Group Orbits, and Invariant Subspaces}
\label{sec:cglestabs}

For our discussion of the temporal dynamics  and  \relsolns\ of the CGLE it will be useful to have a
catalog of the various stabilizers, group orbits and fixed subspaces.  These facts are well known and
elementary.  It is simplest to work with the $\torus^2$-action on $\cC$ given by \refEqn{torusact} and
then transfer the conclusions to $\cS$. 

The only point in $\cC$ fixed by all of $\torus^2$ is $\mmbf{a} = \mmbf{0}$, so henceforth we only
consider nonzero elements and proper subgroups of $\torus^2$.  The nontrivial isotropy subgroups are of
two classes. The first class consists of finite cyclic subgroups of $\torus^2$ generated by elements of
the form $(2\pi k/q, 2\pi/q)$, for integers $q$ and  $k$ with $0 < q$ and $0 \leq k < q$.  Note that
$k$ and $q$ may have common factors.  These groups will be denoted $C(k, q)$. The fixed subspaces
corresponding to these cyclic subgroups are $\Fix(C(k,q)) = \{ (a_m) : a_m = 0\ \mbox{if}\ m \not =
-k \mod q \}$, and the stabilizer of $(a_m)$ is $C(k,q)$ exactly when $(a_m) \in \Fix(C(k,q))$ and the
least $m > 0$ with $a_m \not = 0$ is $m = q - k$. The points with this stabilizer have group orbits 
diffeomorphic to a two-torus and correspond to $\alpha \in \cS$ which satisfy $\alpha(x) = \e^{\ii 2\pi k/q}
\alpha(x + 2 \pi/q)$.

The second class of stabilizers consists of one-dimensional subgroups of $\torus^2$ of the form
$\{ (p u, u) : u \in S^1 \}$ for integers $p$.  These groups will be denoted $D(p)$. The fixed
subspaces corresponding to these one-dimensional subgroups are $\Fix(D(p)) =  \{ (a_m) : a_m = 0\
\mbox{if}\  m \not = p \}$, and the stabilizer of $(a_m)$ is $D(p)$ exactly when $(a_m) \in \Fix(D(p))$.
The points with this stabilizer have group orbits diffeomorphic to the circle and correspond to
$\alpha \in \cS$ of the form $\alpha(x) = a_p \e^{\ii p x}$ for a fixed nonzero $a_p \in \complexes$.

Now recall that each subspace fixed by a subgroup of $\torus^2$ is invariant under the CGLE semiflow.
Using the ODEs~\refEqn{eq:odes}, the semiflow restricted to each $\Fix(D(p))$ is easily understood.
If $p^2 \geq R$, then all trajectories in $\Fix(D(p))$ converge to zero in forward time. If $p^2 < R$,  
then all trajectories in $\Fix(D(p))$ are attracted to what is usually called the \de{plane wave},
\begin{equation} \label{planewave}
    A(x, t) = c \e^{\ii\omega t} \e^{\ii px},
\end{equation} 
for $ R = p^2 + c^2$ and $\omega = -\nu p^2 - \mu c^2$.

The dynamics on the subspaces $\Fix(C(k,q))$ are in general quite complicated.  But it is worth noting
that the transformation $A(x, t)  = q^2 B( qx, q^2 t)$ is a bijection between solutions $A(x, t)$ of
\refEqn{eq:cgle_pde} with isotropy subgroup $C(0,q)$ and solutions $B$ of \refEqn{eq:cgle_pde} with the
same $\nu$ and $\mu$ but with the $R$ coefficient equal to $R/q^2$.  Thus, all the \relsolns\ for $R=16$ 
described in \S~\ref{sec:results} are present, after a change of coordinates, in invariant
subspaces for $R = 16 n^2$, for all integers $n$.

Finally, since the focus here is on \relsolns\ with respect to the $\torus^2$-symmetry, we mention only
briefly the $\integers_2$-symmetry acting as in \refEqn{cglesym4} or \refEqn{wavesym4}. The new class of
fixed subspaces consist of elements of $\cS$ that satisfy
\begin{equation} \label{Z2sym}
    \alpha(-x + \eta_2) = \alpha(x)\ \  \mbox{or} \ \
    -\alpha(-x + \eta_2) = \alpha(x),
\end{equation}
that is, they are even or odd with respect to reflection about $\eta_2/2$.

\subsection{Relative Equilibria and Relative Periodic Solutions}
\label{sec:cglerelper}

One-parameter subgroups of $\torus^2$ can be written as $(\omega_1 t, \omega_2 t)$, so the relative
equilibria of the CGLE \refEqn{eq:cgle_pde} contain solutions of the form
\begin{equation*}
    A(x, t) = \e^{\ii \omega_1 t} \alpha(x + \omega_2 t)
\end{equation*}
for $\alpha\in\cS$.  When $\omega_2 = 0$, these solutions have a single temporal frequency and are
called \de{steady solutions}.  When $\omega_1 = 0$, the solution is a traveling wave, and when both
group elements are nonzero the solution is what is called a \de{generalized traveling wave} or
\de{coherent structure}.  These have been extensively studied in the literature (see, for example,
\cite{aranson02,encyc,vanHecke02}).

A \relper\ of the CGLE with drift $(\varphi, S)$ and period $T$ contains solutions that satisfy
$A(x,t) = \e^{\ii\varphi} A(x+ S, t+T)$ for all $t$. The corresponding solution of the system of ODEs
\refEqn{eq:odes} thus satisfies 
\begin{equation}   \label{eq:am_rpo}
    a_m(t)  =  \e^{\ii\varphi} \e^{\ii m S} a_m(t +T)
\end{equation}
for all $m$ and $t$.  Note that the only recurrent trajectories with stabilizers $D(p)$ are the plane
waves \refEqn{planewave}, which are relative equilibria. Since by definition a \relper\ is not a \releq,
any \relper\ of the CGLE has a $\torus^2$-stabilizer that is either trivial or finite, and thus its
group orbit is a two-torus.  Thus, a \relper\ of the CGLE is diffeomorphic to a three-torus, so each
solution described in \S~\ref{sec:results} yields an invariant three-torus in the CGLE semiflow.

By virtue of \refEqn{cglesym4}, if $A(x,t)$ is a solution of the CGLE, then so is $A(-x,t)$. This
implies that a \relsoln\  typically yields two distinct invariant three-tori in the CGLE semiflow.
However, if the solution possesses the symmetry \refEqn{Z2sym}, these tori are identified.

We finally remark that the task of finding a \releq\ for the CGLE can be reduced to solving a
three-dimensional system of ordinary differential equations with parameters (see, for example,
\cite{aranson02}). No such simplification appears to be possible for finding \relsolns\ of the CGLE.

\subsection{Dynamics for Selected Parameter Values} 
\label{sec:dynatparam}

The parameter values $R=16$, $\nu=-7$, and $\mu=5$ were selected for our study.  For these parameter
values the numerical computation of the largest Lyapunov exponent $\lambda_1$ of 50 typical trajectories
evolved from time $t=0$ to $t=600$ gave a mean value of $\lambda_1 \approx 5.36$, with a standard
deviation of 0.11.  We obtained a value of approximately 11.52 for the Lyapunov (or Kaplan-Yorke)
dimension, with a standard deviation of 0.04.  As a consequence of the $\torus^2$-symmetry, each 
trajectory has two Lyapunov exponents equal to zero~\cite{zerolyap}, in addition to the zero exponent
in the flow direction. These three exponents were included in calculation of the Lyapunov dimension.
A typical trajectory has five positive Lyapunov exponents, and thus an unstable manifold with real
dimension 5. The sum of the positive Lyapunov exponents (which will henceforth be called the
\de{total instability}) was typically equal to  approximately 15.71, with a standard deviation of 0.40.

The fairly small deviations from the mean for these dynamical quantities justify referring to them as
``typical values'', but we use this term without any claim of the existence of any particular class of
ergodic invariant measure.  Similarly, we will speak of the typical trajectories as being contained in
a main attractor without claiming that it is indecomposable in any sense.  If we accept the
Kaplan-Yorke and related conjectures, then the computed Lyapunov dimension indicates that the main
attractor has fractal dimension equal to 11.52.  The terminology ``main attractor'' refers to the
closure of the typical trajectories and is distinguished from the ``universal attractor'' whose existence
is proved in \cite{Bart,constart,doering88,Ghir}.  The latter, which would be called a global attracting
set in topological dynamics, refers to the intersection of the forward images of a forward invariant
set that is globally attracting. In particular, the universal attractor can contain trajectories that
are not typical in the sense just described.  For example, the Lyapunov spectra of the rest point
$\mmbf{0}$ and the plane waves, as reported in Appendix~A, Table~\tblsf, are very different
from those of typical trajectories and from those of the \relsolns\ reported in Table~\tblrelsolns.

When comparing the results here with those in the literature it is important to note that in addition
to \refEqn{eq:cgle_pde} there are other forms of the CGLE that appear in the literature. The most
common alternative is to fix the coefficient of $A$ as 1, 
\begin{equation}   \label{eq:altpde}
    \frac {\partial \AA}{\partial t}  =  \AA + (1 + \ii \nu) \frac{\partial^2 \AA}{\partial x^2} 
     - (1 + \ii \mu) \AA|\AA|^2,
\end{equation}
and have periodic boundary conditions, $\AA(x + L,t ) = \AA(x,t)$. The transformation
\begin{equation} \label{coordchange}
    A(x, t) = \sqrt{R}\; \AA( \sqrt{R}\; x, R\; t)
\end{equation}
takes a solution $A$ of \refEqn{eq:cgle_pde} to a solution $\AA$ of \refEqn{eq:altpde} with period
$L = 2 \pi \sqrt{R}$ in $x$. In particular, the parameter value of $R = 16$ used here corresponds to
solving \refEqn{eq:altpde} with $L = 8 \pi \approx 25.13$ and the commonly used value of $L = 512$ in
\refEqn{eq:altpde} corresponds to $R \approx 6640$ in \refEqn{eq:cgle_pde}.  The range of periods
$0.02 < T <  0.46$  described in \S~\ref{sec:results} corresponds to periods $0.32 < T < 7.36$ for
\refEqn{eq:altpde}. Note also that the rescaling of time in \refEqn{coordchange} rescales Lyapunov
exponents and so will alter the total instability of a trajectory but not its Lyapunov dimension or its
unstable dimension.

The value $R=16$ used in this paper for the CGLE in the form \refEqn{eq:cgle_pde} is comparable to that
of other studies of the CGLE temporal dynamics (for example, \cite{keefe85,Moon}), but is much less
than the corresponding $R$ value used in most studies of the spatiotemporal behavior of the CGLE. The
parameters as chosen have the advantage that the temporal dynamics are chaotic and indeed, have a much
higher intrinsic dimension than is usual in dynamical studies, but $R$ is sufficiently small that a
relatively small number of spatial Fourier modes suffices to capture the dynamics.  As has been noted
(for example in \cite{lowdimisdiff}), the dynamics of the CGLE at small $R$ (or $L$) is quite different
from that at larger values. As for the commonly used characterizations in terms of phase and defect
turbulence, we note that for a typical trajectory of the CGLE at the chosen parameter values the 
modulus $|A(x,t)|$ frequently vanishes (defects) and the average phase gradient (i.e., the winding 
number about the origin of the image of $x \mapsto A(x, t)$ in the complex plane as shown, for example,
in Figure~\ref{fig:sol14_realvsimag}) varies somewhat irregularly from $-2$ to $2$. The chosen values
of $\nu=-7$ and $\mu=5$ are in the Benjamin-Feir unstable region $1 + \mu\nu < 0$, and the plane waves
are unstable to sideband perturbations and have unstable dimension of 8 (see Appendix~A,
Table~\tblsf).

%%% new section  %%%
\section{Computing Relative Periodic Solutions}
\label{sec:compute_relsolns}

We now describe the procedure used to compute relative periodic solutions of the system of
ODEs~\refEqn{eq:trunc_odes} and, thus, of the CGLE.  Although the discussion is specifically for the
problem of computing relative periodic solutions for the CGLE, the procedure is applicable to the
general problem of computing relative periodic solutions of evolution equations with continuous
symmetries.

\subsection{From ODEs to Nonlinear Algebraic Equations}

In seeking \relsolns\ $\mmbf{a}(t)$ of the system of ODEs~\refEqn{eq:trunc_odes}, we use the ansatz 
\begin{equation}   \label{eq:am_ansatz}
    a_m(t)  =  \e^{-\ii\frac{\varphi}{T}t} \e^{-\ii m\frac{S}{T}t}
            \sum_{n} \hat{a}_{m,n} \e^{\ii n\frac{2 \pi}{T}t}
\end{equation}
and work with the Galerkin projection obtained by fixing an even number $N_t$, so that the summation
index in~\refEqn{eq:am_ansatz} runs over the range $-N_t/2+1 \le n \le N_t/2-1$.  The $a_m(t)$ given
by~\refEqn{eq:am_ansatz} solve the equations~\refEqn{eq:am_rpo} and hence are an appropriate
representation for a \relsoln\ of the system of ODEs.
Note that a $(g,T)$-\relsoln\ of the system of ODEs is in fact being represented as
\begin{equation}  \label{eq:avec_ansatz}
    \mmbf{a}(t)  =  \e^{-t L_g/T} \mmbf{b}(t/T),
\end{equation}
where $g = (\varphi,S)$, $L_g$ is the matrix $\diag(\ii\varphi + \ii mS)$, and $\mmbf{b}$ is periodic with
period one.  Since the action of $\torus^2$ is linear on the space $\complexes^{N_x-1}$,
substituting~\refEqn{eq:avec_ansatz} into the system of ODEs~\refEqn{eq:trunc_odes} and using the
$\torus^2$-equivariance of $\mathbf{X}$ yields a system of equations for $\mmbf{b}$ and $(g,T)$,
\begin{equation*} 
    \frac{1}{T} \left( \frac{\dd \mmbf{b}}{\dd t} - L_g \mmbf{b} \right)  =  \mathbf{X}(\mmbf{b}),
\end{equation*}
from which one computes relative periodic solutions.  We point out that~\refEqn{eq:avec_ansatz} can be
viewed as the passage into a moving frame in which the \relsoln\ is in fact periodic.  The moving frame
is constructed using the appropriate one-parameter subgroup as in~\cite{Wulff1}.

Indeed, in the general case of finding \relpers\ of a $G$-equivariant ODE, ${\dot z} = \mathbf{X}(z)$,
one adopts the ansatz  $z(t) = \e^{-\xi t/T}\cdot \beta(t/T)$, where $\xi$ is in the Lie algebra of $G$
and $\beta(t+1) = \beta(t)$.  Substitution into the ODE and some manipulation yields that $z(t)$ is an
$(\e^{\xi}, T)$-\relper\ if it satisfies
\begin{equation} \label{eq:genfundeq}
    \frac{1}{T} \left( \frac{\dd \beta}{\dd t} -\mathbf{Y}_\xi\circ \beta \right)  =  \mathbf{X}\circ \beta,
\end{equation}
where $\mathbf{Y}_\xi$ is the infinitesimal generator corresponding to $\xi$.  The terms in 
\refEqn{eq:genfundeq} represent vector fields along the closed curve $\beta$ with $\mathbf{Y}_\xi
\circ \beta$ corresponding to the velocity vector induced by passage into the moving frame.

To be concrete, substituting~\refEqn{eq:am_ansatz} into the truncated system of ODEs~\refEqn{eq:odes}
results in a system of nonlinear algebraic equations,
\begin{multline}   \label{eq:cgle_nleqns}
    \ii \left (\frac{2\pi n}{T} - \frac{\varphi}{T} - m\frac{S}{T} \right )
	\hat{a}_{m,n}  \ = \ 
        R \hat{a}_{m,n} - m^2 (1 + \ii\nu) \hat{a}_{m,n}    \\
        - (1 + \ii\mu) \sum_{m_1+m_2-m_3=m}
    \left ( \sum_{n_1+n_2-n_3=n}
    \hat{a}_{m_1,n_1} \hat{a}_{m_2,n_2} \hat{a}_{m_3,n_3}^{*} \right ),
\end{multline}
for the complex Fourier coefficients $\{\hat{a}_{m,n}\}$, the drift $(\varphi,S)$, and the time period
$T$.  Therefore, \refEqn{eq:cgle_nleqns} is an underdetermined system of $(N_x-1)(N_t-1)$ complex
equations in $(N_x-1)(N_t-1)$ complex unknowns plus three real unknowns or, after splitting the
equations into their real and imaginary parts, $2(N_x-1)(N_t-1)$ real equations in $2(N_x-1)(N_t-1) + 3$
real unknowns.  Solutions to this system of equations will give the desired relative periodic solutions
of the truncated system of ODEs via the expansion \refEqn{eq:am_ansatz}.

It is convenient to separate \refEqn{eq:cgle_nleqns} into its linear and nonlinear parts and write it as
\begin{equation}  \label{eq:Feq0}
    \Fargs  \: = \:  \FLargs + \FNLargs  \: = \:  \mmbf{0},
\end{equation}
where $\mmbf{\hat{a}}$ is a vector with components given by the coefficients $\{\hat{a}_{m,n}\}$,
\FLargs\ is a vector whose components are given by 
\[ \left (\ii \left (\frac{2\pi n}{T} - \frac{\varphi}{T} - m\frac{S}{T}
   \right ) - R + m^2 (1 + \ii\nu) \right ) \hat{a}_{m,n}, \]
and \FNLargs\ is a vector with components given by 
\[ (1 + \ii\mu) \sum_{m_1+m_2-m_3=m} \left ( \sum_{n_1+n_2-n_3=n}
    \hat{a}_{m_1,n_1} \hat{a}_{m_2,n_2} \hat{a}_{m_3,n_3}^{*} \right ). \]
Note that the components of \FNLargs\ are just the coefficients in the truncated Fourier series expansion
(in both space and time) of the function $(1 + \ii\mu) A|A|^2$.  We remark that in defining the vector
$\mmbf{\hat{a}}$ (and similarly for $\FL$ and $\FNL$) we are implicitly assigning
an ordering on the coefficients $\{\hat{a}_{m,n}\}$ that uniquely determines an index for the components
of $\mmbf{\hat{a}}$.  Henceforth, such a convention should be understood whenever applicable.  In
addition, we will use the notation in~\refEqn{eq:Feq0} to denote both the system of complex
equations~\refEqn{eq:cgle_nleqns} and the system of equations \refEqn{eq:cgle_nleqns} split into its
real and imaginary parts, as it should be clear from the context which case applies.

\subsection{Symmetries of \Feqzero}
\label{sec:symmetriesF}

The symmetries~\refEqns{wavesym1}{wavesym4} of the system of ODEs \refEqn{eq:odes} induce symmetries of
the system of nonlinear algebraic equations~\refEqn{eq:cgle_nleqns}.  Note that if $(\{\hat{a}_{m,n}\},
\varphi,S,T)$ is a solution of \Feqzero, then for any $(\eta_1, \eta_2, \tau) \in \torus^3$
\begin{align}
    (\{ \e^{\ii\eta_1} \hat{a}_{m,n} \}, \varphi, S, T)&,      \label{Fsym1}    \\
    (\{ \e^{\ii m\eta_2} \hat{a}_{m,n} \}, \varphi, S, T)&,    \label{Fsym2}    \\
    (\{ \e^{\ii n \tau} \hat{a}_{m,n} \}, \varphi, S, T)&,     \label{Fsym3}    \\
    (\{ \hat{a}_{-m,n} \}, \varphi, -S, T)&,                   \label{Fsym4}
\end{align}
are also solutions.  From the continuous symmetries~\refEqns{Fsym1}{Fsym3}, it follows that the set of
solutions to \Feqzero\ splits into orbits $\mathcal{O}_{(\nleargs)}$ of the symmetry group $\torus^3$,
\begin{equation*}
    \mathcal{O}_{(\nleargs)}  :=  \left \{ (\eta_1,\eta_2,\tau) \cdot 
    (\nleargs) \; : \: (\eta_1,\eta_2,\tau) \in \torus^3 \right \},
\end{equation*}
where the action of $\torus^3$ on a point $(\nleargs)$ is defined by
\begin{equation}  \label{eq:torus3act}
     (\eta_1,\eta_2,\tau) \cdot (\nleargs)  \ = \ (\{ \e^{\ii\eta_1} \e^{\ii m \eta_2}
     \e^{\ii n \tau}  \hat{a}_{m,n} \}, \varphi, S, T). 
\end{equation}
That is, $\torus^3$ acts on $\mmbf{\hat{a}}$ via multiplication by the matrix $\diag(\e^{\ii\eta_1}
\e^{\ii m \eta_2} \e^{\ii n \tau})$, and it acts trivially on $(\varphi, S, T)$.  Since the solutions of
\Feqzero\ come in continuous families, when computing solutions to \Feqzero\ numerically we will augment
the system with additional equations (and hence solve a system with equal numbers of equations and
unknowns), in effect fixing a particular solution from each orbit $\mathcal{O}_{(\nleargs)}$.

\subsection{Kernel of Jacobian Matrix at Solutions of \Feqzero}
\label{sec:jacobian}

Recall that, split into its real and imaginary parts, the system \Feqzero\ is an underdetermined system
of $2(N_x-1)(N_t-1)$ real equations in $2(N_x-1)(N_t-1) + 3$ real unknowns.  Thus, the kernel,
\kernel{$J$}, of the Jacobian matrix $J$ of $\F$ is at least three-dimensional.  From the 
$\torus^3$-equivariance of $\F$ (implied by the symmetries \refEqns{Fsym1}{Fsym3}), one obtains
three linearly independent vectors in \kernel{$J$} at a solution $(\nleargs)$ of \Feqzero.  Such vectors
result from a basis for the space of infinitesimal generators of the action~\refEqn{eq:torus3act}
of $\torus^3$ on the point $(\nleargs)$, which can be obtained as
\begin{eqnarray}  
    \mmbf{\hat{v}}_1  \ := \
        \left. \frac{\dd}{\dd \eta_1} \, \left (\{\e^{\ii \eta_1}
        \hat{a}_{m,n}\}, \varphi, S, T  \right ) \right |_{\eta_1=0}
        & =  & \left ( \{\ii \: \hat{a}_{m,n}\}, 0, 0, 0 \right ),   
        \label{eq:nullvec1}  \\
    \mmbf{\hat{v}}_2  \ := \
        \left. \frac{\dd}{\dd \eta_2} \, \left (\{\e^{\ii m \eta_2}
        \hat{a}_{m,n}\}, \varphi, S, T  \right ) \right |_{\eta_2=0}  
        & = &   \left ( \{\ii m \: \hat{a}_{m,n}\}, 0, 0, 0 \right ),   
        \label{eq:nullvec2}  \\  
    \mmbf{\hat{v}}_3  \ := \
        \left. \frac{\dd}{\dd \tau_{\ }} \, \left (\{\e^{\ii n \tau} 
        \hat{a}_{m,n}\}, \varphi, S, T  \right ) \right |_{\tau=0\ } 
        & =  &  \left (\{\ii n \: \hat{a}_{m,n}\}, 0, 0, 0 \right ).
        \label{eq:nullvec3}
\end{eqnarray}
Split into real and imaginary parts, the above vectors $\mmbf{\hat{v}}_1$, $\mmbf{\hat{v}}_2$, and
$\mmbf{\hat{v}}_3$ are in \kernel{$J$} at a solution $(\nleargs)$ of \Feqzero.

As previously pointed out, we will augment the system \Feqzero\ with additional equations and look
for solutions of the augmented system.  The additional equations should be defined so that
the vectors in~\refEqns{eq:nullvec1}{eq:nullvec3} are not in the kernel of the augmented Jacobian matrix
at a solution, since a singularity of the Jacobian is usually problematic for solvers of systems of
nonlinear algebraic equations.  One must then confront the question of whether, at a solution of
\Feqzero, the vectors in~\refEqns{eq:nullvec1}{eq:nullvec3} span \kernel{$J$}, so that the Jacobian of
the augmented system will be nonsingular at solutions.  Typically this will be true, as explained next.

For a $(g_{\sss 0},T)$-\relsoln\ $\gamma(t) \subset \mathbb{C}^N$ of a $G$-equivariant system of ODEs,
where $g \in G$ acts on $\mathbb{C}^N$ via multiplication by a matrix $N_{g}$, define the
\emph{\relmon\ matrix} $M$ to be 
\[  M = N_{g_{\sss 0}} \tM,  \]
where $\tM$ is the time $T$ monodromy matrix along the solution.  (For our problem $N_{g_{\sss 0}} =
e^{L_{g_{\sss 0}}}$, where $L_{g_{\sss 0}}$ is the matrix $\diag(\ii\varphi + \ii mS)$.)  Thus, $M$ is the
derivative at $\gamma(0)$ of the transformation $g_{\sss 0} \cdot \phi_T $ that sends $\gamma(0)$ to
itself by flowing along the \relsoln\ for time $T$ and then moving as dictated by the drift $g_{\sss 0}$.
Say that $\gamma$ is a \de{regular \relsoln} if the only eigenvalues of $M$ equal to 1 are those coming
from the continuous symmetries (including time translation) of the system of ODEs.  Let \mbox{F $=0$}
denote the system of nonlinear algebraic equations whose solutions give $(g_{\sss 0},T)$-\relsolns\ of
the system of ODEs.  If $\gamma$ is a $(g_{\sss 0}, T)$-\relsoln, then one can show that each eigenvalue
$\lambda$ of the \relmon\ $M$ corresponds to an eigenvalue $-\log(\lambda)/T$ of the Jacobian matrix $J$
of F at the solution to \mbox{F $=0$}~\cite{inprep1}. Now, $J$ will have many eigenvalues that are not
related to eigenvalues of $M$, but every zero eigenvalue of $J$ does correspond to a 1 in the spectrum
of $M$. Thus, if $\gamma$ is a regular \relsoln, then the kernel of the Jacobian matrix of F is spanned by 
the infinitesimal generators of the group action.  Furthermore, among ODEs with a given continuous
symmetry, it is $C^k$-generic for all $k>1$ that all \relpers\ are regular~\cite{Field1}.

\subsection{Additional Equations}
\label{sec:additional_eqns}

When defining additional equations to augment the system \Feqzero, we want to accomplish two objectives:
first, to guarantee that solutions are not lost as a result of adding new equations; second, 
to derive conditions necessary so that the vectors given in~\refEqns{eq:nullvec1}{eq:nullvec3} are
not in the kernel of the Jacobian matrix of the augmented system at a solution.  There are a variety of
options available to accomplish these two goals. The construction we found to work best for our
problem is to fix the sum of the arguments of some given sets 
$\Lambda_l = \{\hat{a}_{m_j^l,n_j^l}\}_{j=1}^{K}$, $l = 1,2,3$, of $K$ nonzero coefficients by imposing
the constraints
\begin{equation}   \label{eq:constraints}
    \sum_{j=1}^{K} \arg(\hat{a}_{m_j^l,n_j^l})  =  c_l,
\end{equation}
where $c_1, c_2,$ and $c_3$ denote some given constants.  Note that in the indexing notation used above,
the superscript $l$ is used to indicate membership of a given coefficient $\hat{a}_{m_j^l,n_j^l}$ into
one of the sets $\Lambda_1$, $\Lambda_2$, or $\Lambda_3$.  To verify that imposing the
constraints~\refEqn{eq:constraints} does not result in a loss of solutions to \Feqzero, recall that,
because of the symmetries \refEqns{Fsym1}{Fsym3}, if $(\{\hat{b}_{m,n}\}, \varphi,S, T)$ is a solution 
to \Feqzero, then so is $(\{\hat{a}_{m,n}\}, \varphi, S, T)$, where
\begin{equation*}
    \hat{a}_{m,n}  =  \e^{\ii\eta_1} \e^{\ii m \eta_2} \e^{\ii n \tau} \hat{b}_{m,n}
\end{equation*}
and the element $(\eta_1,\eta_2, \tau) \in \torus^3$ is arbitrary.  Therefore, the
constraints~\refEqn{eq:constraints} are well defined if there exist some $\eta_1, \eta_2, \tau$ such
that (for $l=1,2,3$)
\begin{equation*} 
    \sum_{j=1}^{K} \arg(\exp(\ii\eta_1) \exp(\ii m_j^l \eta_2)
	 \exp(\ii n_j^l \tau) \hat{b}_{m_j^l,n_j^l})  =  c_l
\end{equation*}
or, equivalently, if a solution to the linear system 
\begin{equation}   \label{eq:linsys} 
     \left [ \begin{array}{ccc}
	      K\ \  &  \sum_{j} m_j^1\ \  &  \sum_{j} n_j^1    \\
	      K\ \  &  \sum_{j} m_j^2\ \  &  \sum_{j} n_j^2    \\
	      K\ \  &  \sum_{j} m_j^3\ \  &  \sum_{j} n_j^3  
              \end{array}
      \right ]
      \left [ \begin{array}{c}
	      \eta_1  \\ \eta_2  \\ \tau
              \end{array}
      \right ] 
      = 
      \left [ \begin{array}{c}
	      c_1 - \sum_{j} \arg(\hat{b}_{m_j^1,n_j^1})   \\
	      c_2 - \sum_{j} \arg(\hat{b}_{m_j^2,n_j^2})   \\
	      c_3 - \sum_{j} \arg(\hat{b}_{m_j^3,n_j^3})
              \end{array}
       \right ]  \qquad 
\end{equation}
exists.  But one can always assign coefficients to each set $\Lambda_l$ so that the matrix in the above
system of linear equations is nonsingular and therefore guarantee the existence of a unique solution to
\refEqn{eq:linsys}.  Thus, there always exists a symmetry transformation by which it is possible to
transform any solution of \Feqzero\ so that it satisfies the constraints~\refEqn{eq:constraints}.
Hence, the constraints are well defined.

Now let $(\nleargs)$ be a solution of \Feqzero.  Let $J$ denote the Jacobian matrix of $\F$
augmented with the equations~\refEqn{eq:constraints} and evaluated at $(\nleargs)$, and let the vectors
$\mmbf{\hat{v}}_1$, $\mmbf{\hat{v}}_2$, and $\mmbf{\hat{v}}_3$ be as in~\refEqns{eq:nullvec1}{eq:nullvec3}.
Then it follows that
\begin{equation*}
\begin{array}{ccc}
    J \mmbf{\hat{v}}_1 =  \left [ \begin{array}{c}
	\mmbf{0}  \\  K  \\  K  \\  K
        \end{array} \right ],  & \quad
    J \mmbf{\hat{v}}_2 =  \left [ \begin{array}{c}
	\mmbf{0}  \\  \sum_{j} {m_j^1}  \\ 
	\sum_{j} {m_j^2}  \\  \sum_{j} {m_j^3}
        \end{array} \right ],  & \quad
    J \mmbf{\hat{v}}_3 =  \left [ \begin{array}{c}
	\mmbf{0}  \\  \sum_{j} {n_j^1}  \\
	\sum_{j} {n_j^2}  \\  \sum_{j} {n_j^3}
        \end{array} \right ],
\end{array}
\end{equation*}
where $\mmbf{0} \in \reals^{2(N_x-1)(N_t-1)}$.  Thus, if the sets $\Lambda_1$, $\Lambda_2$, and
$\Lambda_3$ are nonempty, and if at least one of the sums $\sum_{j} {m_j^l}$ and $ \sum_{j} {n_j^l}$ are
nonzero, then the vectors $\mmbf{\hat{v}}_1$, $\mmbf{\hat{v}}_2$, and $\mmbf{\hat{v}}_3$ will not be in
the null space of $J$.

\subsection{Identifying Minimum Time Period and Relative Equilibria}
\label{sec:mintimeperiod}

It can happen that a computed solution $(\nleargs)$ to \Feqzero\ represents a \relsoln\ of the CGLE
whose period $T$ is not minimal.  In such a case, there exists an integer $p > 1$ and rational numbers
$j$ and $l$ so that
\[
    \tilde{T} \ = \ T/p, \qquad
    \tilde{S} \ = \ (S+2\pi j)/p, \qquad
    \tilde{\varphi} \ = \ (\varphi+2\pi l)/p
\]
satisfy
\begin{equation}   \label{eq:solminT}
    \e^{\ii\tilde{\varphi}} A(x+\tilde{S}, t+\tilde{T}) \ = \ A(x,t).
\end{equation}
Using the expansions \refEqn{eq:xFseries} and \refEqn{eq:am_ansatz} one has that
\[
    \e^{\ii\tilde{\varphi}} A(x+\tilde{S}, t+\tilde{T}) \ = \
        \sum_{m} \left ( \e^{-\ii\frac{\varphi}{T}t} \e^{-\ii m\frac{S}{T}t}
        \sum_{n} \hat{a}_{m,n} \e^{\ii n\frac{2\pi}{T}t}
	\e^{\ii\frac{2\pi(n + l + jm)}{p}} \right ) \e^{\ii m x}, 
\]
so \refEqn{eq:solminT} holds only if $\hat{a}_{m,n} = 0$ whenever $p$ does not divide $n + l + jm$.  The
nonzero coefficients $\hat{a}_{m,n}$ define a set $\{ \tilde{a}_{m,n^{\prime}} \}$, where
\[
    \tilde{a}_{m, (n+l+jm)/p} \ = \ \hat{a}_{m,n}
\]
if $p$ divides $n + l + jm$, giving a point $(\tilde{\mmbf{a}},\tilde{\varphi},\tilde{S},\tilde{T})$ that
corresponds to a relative periodic solution of the CGLE with minimal time period $\tilde{T}$.

Now, a \releq\ $A(x,t) = \e^{\ii\omega_1 t} \alpha(x+\omega_2 t)$ of the CGLE satisfies the
equation \refEqn{eq:Atpsol} defining \relsolns\  with $(\varphi,S,T)  = ( -\omega_1u, -\omega_2 u, u)$,
for any $u > 0$. Using the expansion $\alpha(x) = \sum c_m \e^{\ii m x}$ and such values of $(\varphi,S,T)$
in the ansatz~\refEqn{eq:am_ansatz} yields that for all $m$, $\hat{a}_{m,0} = c_m$, and when $n \not = 0$,
$\hat{a}_{m,n} = 0$.  More generally, a solution to \Feqzero\ representing a \releq\ will have
$\hat{a}_{m,n^{\prime}} \ne 0$ for one index $n^{\prime}$ (not necessarily $n^{\prime}=0$) and
$\hat{a}_{m,n} = 0$ when $n \ne n^{\prime}$.  Thus, from the computed values of $(\nleargs)$ it is easy
to detect when a solution is a \releq\ and not a \relsoln.

\subsection{Starting Values for Solving \Feqzero}
\label{sec:initialguess}

As with almost any system of nonlinear algebraic equations, the augmented system \Feqzero\ must be
solved using a method that involves iteration.  This requires us to supply an initial guess as a
starting point for the iterative procedure.  In general, it is very important to supply initial guesses
that in some sense are ``close'' to actual solutions of the system being solved.  For our problem, we
adapted the idea of searching for close returns (or recurrences) of a chaotic dynamical system and
sought \emph{relative close returns} of the truncated system of ODEs~\refEqn{eq:odes} to generate the
required initial guesses.

When computing solutions $\mmbf{u}(t)$ of a chaotic dynamical system that are truly periodic in time
(i.e., $\mmbf{u}(t) = \mmbf{u}(t+\tilde{T})$ for some time period $\tilde{T}$) it is common to search
for close returns of the system, that is, integrate the system until $\mmbf{u}(t_1) \approx
\mmbf{u}(t_2)$ for some times $t_1 < t_2$, to within some tolerance, and use the close return as an
initial approximation to a time-periodic solution. Since for a $(g,T)$-\relsoln\ of the truncated
system of ODEs~\refEqn{eq:odes} one has that $|a_m(t)| = |a_m(t+T)|$ (cf. equation~\refEqn{eq:am_rpo}),
we sought sets of coefficients $\{a_m(t)\}$ for which
\[  |a_m(t_1)|   \,  \approx \, |a_m(t_2)|  \]
for some times $t_1 < t_2$.  Given a measure $\delta$ of desired closeness, the truncated system of 
ODEs~\refEqn{eq:odes} was integrated starting with some random set $\{a_m(0)\}$ of coefficients until
times $t_1 < t_2$ were found for which 
\[  \underset{m}{\max}\,  \left | 1 - \frac{|a_m(t_2)|}{|a_m(t_1)|} \right | \ \le \ \delta. \]
The initial value $T_0$ for the time period was then set to $T_0 = t_2 - t_1$ and a least squares
solution over $\varphi$ and $S$ to the system
\[  a_m(t_1) \, = \, \e^{\ii\varphi} \e^{\ii mS} a_m(t_2) \]
was taken as an initial value $g_0 = (\varphi_0, S_0)$ for the drift.  Once such a relative close return
was identified, initial values for the Fourier coefficients $\{\hat{a}_{m,n}\}$ were generated.  We set 
(cf. equation~\refEqn{eq:am_ansatz})
\[  \e^{\ii (\varphi_0/T_0)t} \e^{\ii m (S_0/T_0)t} a_m(t) \ = \ 
    \sum_{n} \hat{a}_{m,n} \e^{\ii n (2\pi/T_0) t} \]
and used the Fast Fourier Transform (FFT) algorithm to compute the initial guess for the coefficients
$\{\hat{a}_{m,n}\}$.

\subsection{Numerical Solution of \Feqzero}
\label{sec:solveF}

The augmented system \Feqzero\ was solved using the nonlinear least squares solver \texttt{lmder} from
the MINPACK software package~\cite{minpack}.  This solver is an implementation of the \LM
method~\cite{levenberg44, marquardt63, more77} for solving nonlinear least squares problems.  The solver
\texttt{lmder} was chosen because it performed better on the CGLE problem addressed in this study than
the other alternatives considered, namely \newtons method with a line search~\cite{bk:dennis} and a
modification of \mbox{Powell's} hybrid method~\cite{minpack, powell70} for solving nonlinear equations. 
A brief discussion on the performance of these other solvers on our problem is given at the end of this
section.

To use the solver \texttt{lmder} we need to provide a routine that returns the value of $\F$
(augmented) evaluated at a given point $(\nleargs)$.  Computing \FLargs, as defined in \refEqn{eq:Feq0},
is straightforward.  As for \FNLargs, note that since it is a vector with components given by the
coefficients in the truncated (space-time) Fourier series expansion of $(1+\ii\mu) A|A|^2$, one can
compute it using the FFT algorithm.  This is a standard procedure for computing nonlinear terms in
spectral discretizations, see for example~\cite{bk:canuto}, and is briefly detailed here.  Let $x_j$ and
$t_l$ denote grid-point values of the spatial variable $x$ and time variable $t$, respectively, in
physical space.  To obtain the Fourier coefficients of $A|A|^2$, the inverse FFT algorithm was used to
compute the values $A(x_j,t_l)$, given the coefficients $\{\hat{a}_{m,n}\}$.  In order to avoid aliasing
error when computing the convolution sums in~\refEqn{eq:cgle_nleqns} via the FFT algorithm, the
coefficients $\{\hat{a}_{m,n}\}$ were padded with zeros to obtain a set $\{\tilde{a}_{m^{\prime},
n^{\prime}}\}$, where \mbox{$m^{\prime} = -N_x,\ldots,N_x$}, \mbox{$n^{\prime} = -N_t,\ldots,N_t$},
and $\tilde{a}_{m^{\prime},n^{\prime}}$ was equal to  $\hat{a}_{m^{\prime},n^{\prime}}$ if
$|m^{\prime}| < N_x/2$ and $|n^{\prime}| < N_t/2$, and $0$ otherwise.  
(Based on the $3/2$ rule for dealiasing quadratic products~\cite{bk:canuto}, the number of modes  $\{\tilde{a}_{m^{\prime},n^{\prime}}\}$ used are sufficient for dealiasing a cubic nonlinearity.)
The inverse FFT was done using
the coefficients $\{\tilde{a}_{m^{\prime},n^{\prime}}\}$.  Then, the pointwise products
$A(x_j,t_l)|A(x_j,t_l)|^2$ were computed in physical space.  Finally, the FFT algorithm was used to
compute the Fourier coefficients of $A|A|^2$.  At this point only those coefficients whose index
$(m^{\prime},n^{\prime})$ satisfied $|m^{\prime}| < N_x/2$ and $|n^{\prime}| < N_t/2$ were retained.
The FFTW software package~\cite{fftw} was used in our computations.

Some points are worth mentioning regarding the additional equations~\refEqn{eq:constraints}. First,
pairs of indices $\{(m_j^l,n_j^l)\}_{j=1}^{K}$, $l=1,2,3$, were associated with the sets $\Lambda_1$,
$\Lambda_2$, and $\Lambda_3$ so that the linear system in~\refEqn{eq:linsys} was guaranteed to have a
solution.  Second, the values of $c_1$, $c_2$, and $c_3$ in \refEqn{eq:constraints} were set to the sum
of the arguments of the coefficients corresponding to the current approximation to a solution.  That is,
each time $\F$ augmented was evaluated, the equations~\refEqn{eq:constraints} were satisfied
exactly.  This is appropriate since the justification for using these additional equations does not
depend on the specific values of $c_1$, $c_2$, and $c_3$.  Finally, which elements were in the sets
$\Lambda_1$, $\Lambda_2$, and $\Lambda_3$ depended not only on the pairs of indices associated with
them, but also on the value of the coefficients corresponding to the current approximation to a
solution.  Coefficients whose absolute value was below a given tolerance, usually $10^{-5}$, were
excluded from the sets.  

The MINPACK software package makes use of the Jacobian matrix of the system being solved (or some
approximation to it) and provides the option of computing the Jacobian using finite differences or
(analytically) via a user supplied routine.  We chose the latter option.  Note that it is
straightforward to compute the derivatives of $\FLargs$ with respect to $\varphi$, $S$, $T$, and the
real and imaginary parts of the coefficients $\{\hat{a}_{m,n}\}$.  On the other hand, computing the
derivatives of $\FNLargs$ with respect to the real and imaginary parts of the coefficients
$\{\hat{a}_{m,n}\}$ directly from the convolution sums in~\refEqn{eq:cgle_nleqns} is potentially
an error-prone process.  However, one can use the directional derivative $A^2V^{*} + 2|A|^2V$ of the
function $A|A|^2$ in the direction of a perturbation $V(x,t)$ to $A(x,t)$ to compute the product of a
given vector $\mmbf{\hat{v}}$ (whose components are the real and imaginary parts of the Fourier
coefficients $\{\hat{v}_{m,n}\}$ of $V(x,t)$) and the Jacobian matrix of $\FNL$ 
evaluated at a point $\mmbf{\hat{a}}$.  By virtue of the discretization used, this matrix-vector product
has components given by the coefficients in the truncated Fourier series expansion (in both space and
time) of the function $(1 + \ii\mu) (A^2V^{*} + 2|A|^2V)$.  Hence, the matrix-vector product can be 
computed using a procedure analogous to that for computing \FNLargs, which was outlined earlier in this
section.  The difference here is that one has two sets of Fourier coefficients, $\{\hat{a}_{m,n}\}$ and
$\{\hat{v}_{m,n}\}$, and that the pointwise products to be computed in physical space are for the
function \mbox{$A^2V^{*} + 2|A|^2V$}.  The Jacobian matrix of $\FNL$ can thus be computed by
substituting for $\mmbf{\hat{v}}$ the standard basis vectors in $\reals^{2(N_x-1)(N_t-1)}$.

As for the measure of closeness $\delta$ used to identify relative close returns, setting 
$\delta \ge 0.5$ worked well, but choosing smaller values of $\delta$ usually proved unsuccessful in
identifying relative close returns.  We used the value $\delta = 0.5$, chosen experimentally, and the
initial guesses generated from the resulting relative close returns were frequently good starting points 
for the solver \texttt{lmder}.  When searching for relative close returns, we also specified a range
$[p_1,p_2]$ for the initial time period $T_0$.  If $\delta$ was chosen to be too large, then most relative
close returns would end up having period $T_0 = p_1$.  With $\delta = 0.5$, the search for relative close
returns was always successful, but also resulted in relative close returns with different values within the
interval $[p_1,p_2]$ for the initial time period $T_0$.  We remark that the use of random initial data
(i.e., not resulting from a relative close return), even with the spectral profile of a typical trajectory
of the truncated system of ODEs, was unsuccessful in generating solutions with any of the nonlinear
equations solvers.

Initially, the system of nonlinear algebraic equations was solved using values of $N_x=32$ and $N_t=48$
(giving a total of 2,917 real variables).  For this value of $N_x$, the decay in the spatial spectra of
the relative close returns was around five orders of magnitude.  This proved to be sufficient in order
to have well defined solutions in the sense that once a solution was found, continuing from it with
$N_x > 32$ posed no challenge for the nonlinear equations solver and in a very small number of
iterations it converged to the same solution.  The value $N_t = 48$ was chosen to allow for a relatively
large range of time periods.  If the solver \texttt{lmder} converged to a solution of \Feqzero, then we
verified whether the solution had minimal time period using the criteria from
\S~\ref{sec:mintimeperiod}.  If not, a new set of Fourier coefficients was defined and the values
of $T$, $S$, and $\varphi$ modified in order to obtain a solution with minimal time period.  Then the
values of $N_x$ and $N_t$ were increased, if necessary, to have a good decay (around eight or more orders
of magnitude) in the spectra of the solution.  At this point, using \newtons method to solve the system
of nonlinear algebraic equations worked well: only \mbox{$2$--$4$} iterations were required for
convergence.  Finally, the solution to \Feqzero\ was used as an initial condition and the truncated
system of ODEs~\refEqn{eq:odes} was integrated to verify that we indeed had a \relsoln\ of the system of
ODEs (and thus of the CGLE).

As for the other solvers considered,  the performance of \newtons method with a line
search~\cite{bk:dennis} was, in general, unsatisfactory for our problem.  The typical behavior was that,
as the iterations progressed, the line search parameter became very small and no significant change in
the iterates nor reduction in the value of $||\F||_2^2$ was achieved.  The MINPACK solver for
systems of nonlinear algebraic equations, \texttt{hybrj}, was also tried on our problem.  It is an
implementation of a modification of \mbox{Powell's} hybrid method~\cite{minpack, powell70} for solving
nonlinear equations.  The solver \texttt{hybrj} often terminated without converging to a solution of
\Feqzero\ nor to a minimum of $\F^{\mathrm{T}} \F$ for which \mbox{$\F \ne \mmbf{0}$}.

%%% new section %%%
\section{The Relative Periodic Solutions}
\label{sec:results}

\subsection{Dynamical Properties}

\begin{figure}[hbt!]
   \begin{center}
   \resizebox{2.50in}{!} 
       {\includegraphics{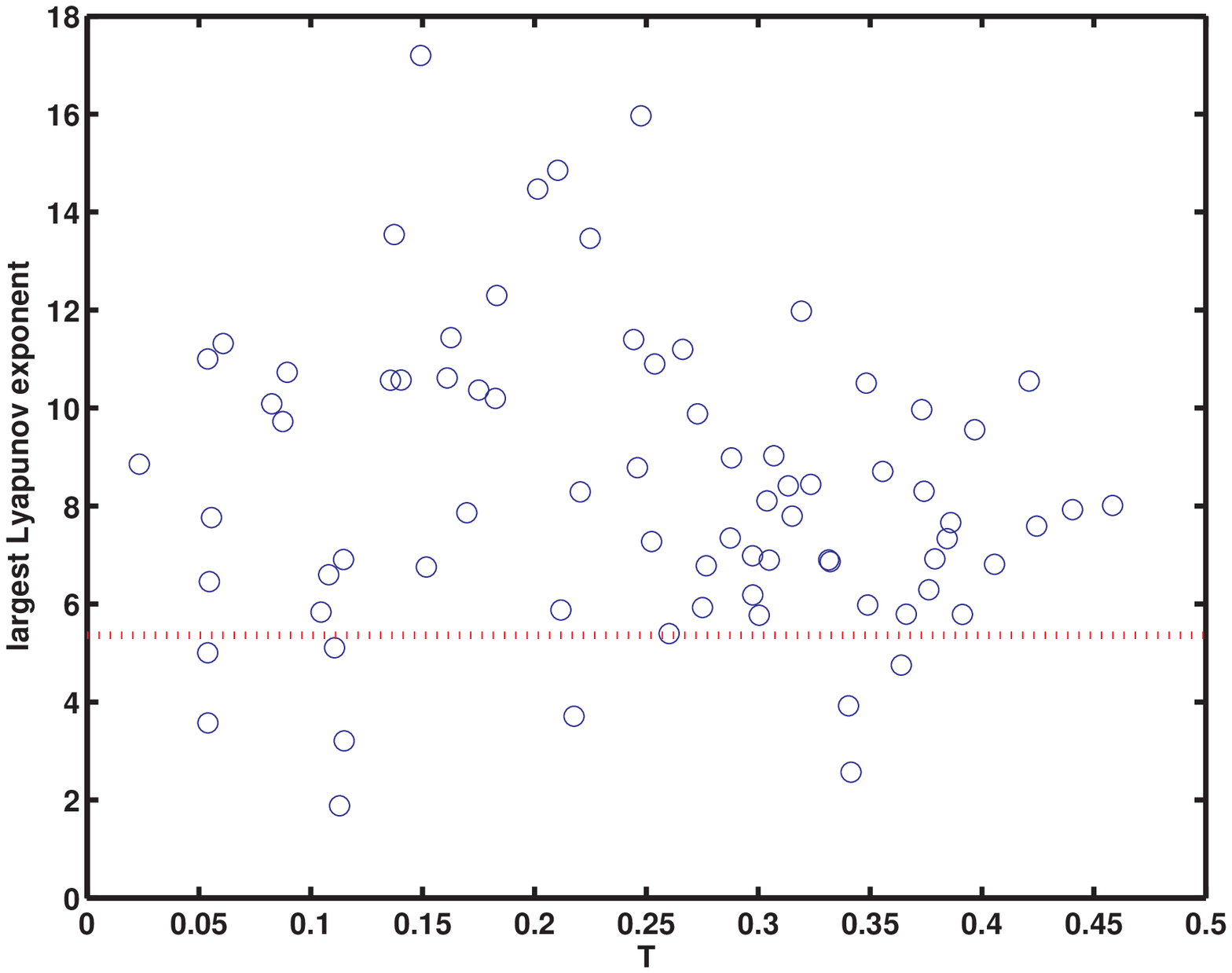}}
   \resizebox{2.50in}{!} 
       {\includegraphics{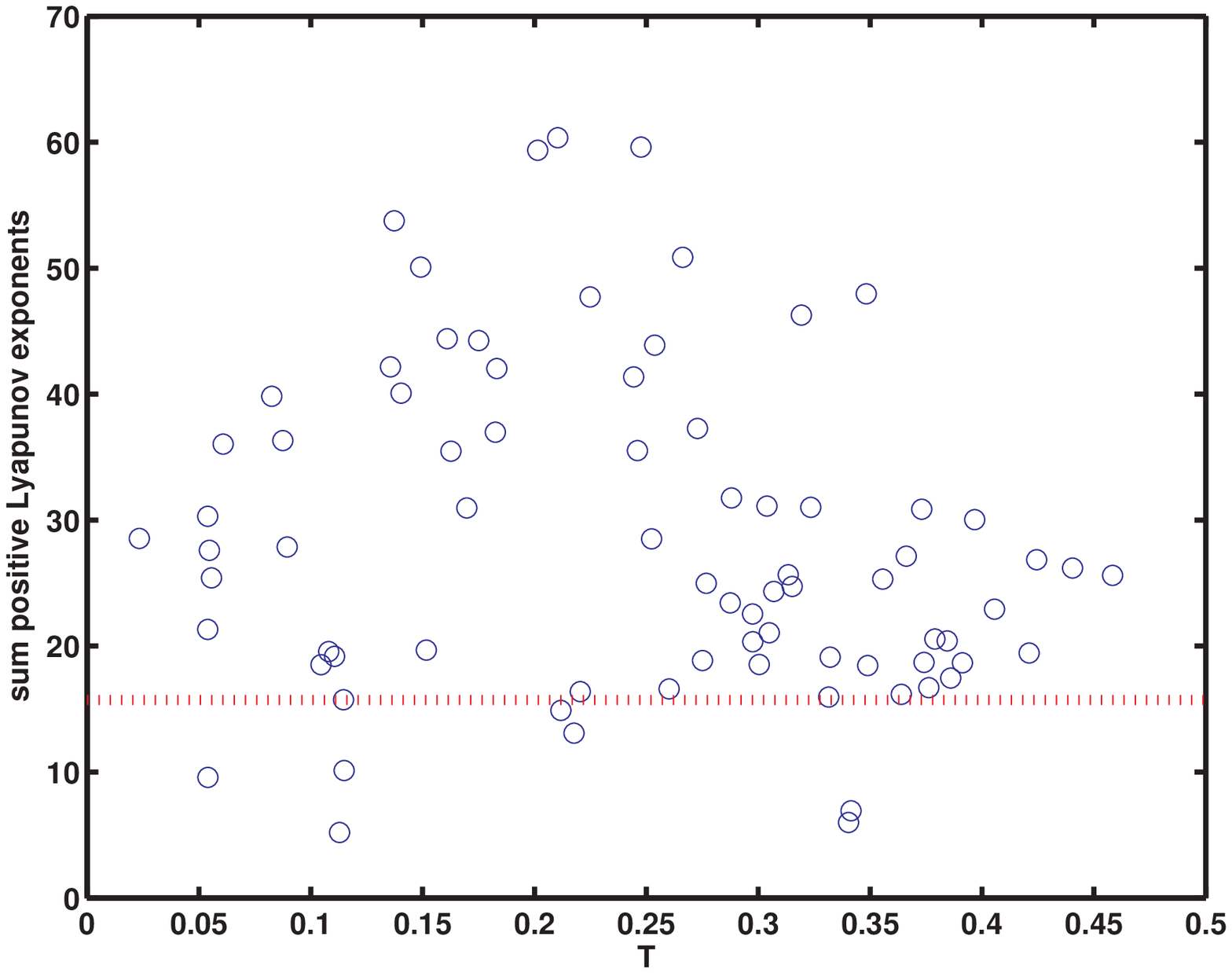}}
   \makebox[2.50in]{(a)}  \makebox[2.50in]{(b)}
   \caption{(a) Largest Lyapunov exponent vs.~period and (b) total instability vs.~period for \relsolns.
       Mean value for typical solutions denoted by horizontal dotted line.}
   \label{fig:TvsLyap}
   \end{center}
\end{figure}

\begin{figure}[hbt!]
   \begin{center}
   \resizebox{2.55in}{!} 
       {\includegraphics{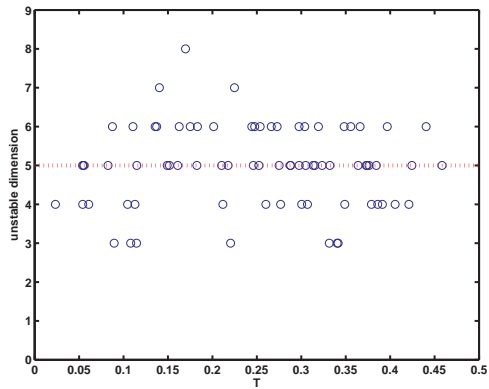}}
   \caption{Unstable dimension vs.~period for \relsolns. Value for typical solutions denoted by 
       horizontal dotted line.} 
   \label{fig:unstdim}
   \end{center}
\end{figure}

Using the procedure described in \S~\ref{sec:compute_relsolns}, 77 distinct \relsolns\ were found
for the CGLE \refEqn{eq:cgle_pde} with the parameter values $R=16$, $\nu=-7$, and $\mu=5$.  The solutions
display a wide variety of dynamical and spatial properties, some of which are summarized in 
Appendix~A, Table~\tblrelsolns, and illustrated in Figures~\ref{fig:TvsLyap} and
\ref{fig:unstdim}.  In Table~\tblrelsolns\ the solutions are sorted by increasing period, with the
periods in the range $0.02 < T < 0.46$. The solutions are given an identifying number in the left column
corresponding to their position in the list.

The Lyapunov exponents of the solutions were computed from the eigenvalues of the \relmon\ matrix defined
in \S~\ref{sec:jacobian}.  As a consequence of the continuous symmetries and the flow direction, each
\relsoln\ has three zero exponents~\cite{zerolyap}. This was confirmed numerically. 
Figure~\ref{fig:TvsLyap}(a) shows the largest Lyapunov exponent of each solution plotted against the
corresponding time period, while Figure~\ref{fig:TvsLyap}(b) shows the total instability.  The horizontal
dotted line on each graph gives the value of the corresponding dynamical quantity computed for a typical
trajectory, which is assumed to be characteristic of the main attractor (cf. \S~\ref{sec:dynatparam}).

Figure~\ref{fig:unstdim} shows the unstable dimension of each solution plotted against the period.  As
noted in \S~\ref{sec:cglerelper}, each \relsoln\ represents an invariant three-torus in the CGLE
semiflow, and so, for example, a \relsoln\ with five positive Lyapunov exponents corresponds to a flow
invariant three-torus with a $5$-dimensional unstable manifold. The unstable dimension variability of
the system is clearly illustrated in Figure~\ref{fig:unstdim}.  We have been unable to associate the
various values of unstable dimension with specific regions of the main attractor, but the high
dimensionality makes a clear geometric picture of the attractor very difficult to obtain.

As seen in Table~\tblrelsolns, the solutions possess a wide variety of space shifts $S$ and phase
shifts $\varphi$.  A \relsoln\ can be the ``$k^{th}$ root'' of true periodic solution if $A(x, kT) = A(x, 0)$
for some integer $k$. This happens exactly when $k \varphi$ and $k S$ are both integer multiples of $2 \pi$ 
and corresponds to the dynamical situation of a rational flow (every orbit is periodic) on the invariant
three-torus.  To test for this possibility we computed the best rational approximation to each
$\varphi/(2 \pi)$ and $S/(2 \pi)$ using the MATLAB function {\tt rationalize}. For no solutions did the
rational approximations of $\varphi/(2 \pi)$ and $S/(2 \pi)$ share the same denominator, and in most
cases the denominators of the rational approximates were on the order of several hundred. These results
indicate that, at least for moderate size $k$, the computed \relsolns\ are not $k^{th}$ roots of true
periodic solutions and that the dynamics on the invariant three-tori are either an irrational flow or
else a periodic flow with long period.

The semilog plots of the temporal and spatial power spectra of all the computed \relsolns\ are
approximately linear, indicating near exponential decay in these spectra. Figure~\ref{fig:sol14_spectra}
in \S~\ref{sec:specific_solns} shows plots of the spectra for solution 14.  The exponential decay
reflects the fact that solutions on the CGLE attractor are $C^\omega$ in time~\cite{promislow1}
and space~\cite{Doelman}. In addition, the temporal spectra indicate that none of the 77 computed
\relsolns\ are coherent structures.

\subsection{Properties of Specific Solutions}
\label{sec:specific_solns}

\begin{figure}[hbt!]
   \begin{center}
   \resizebox{1.25in}{!} 
       {\includegraphics{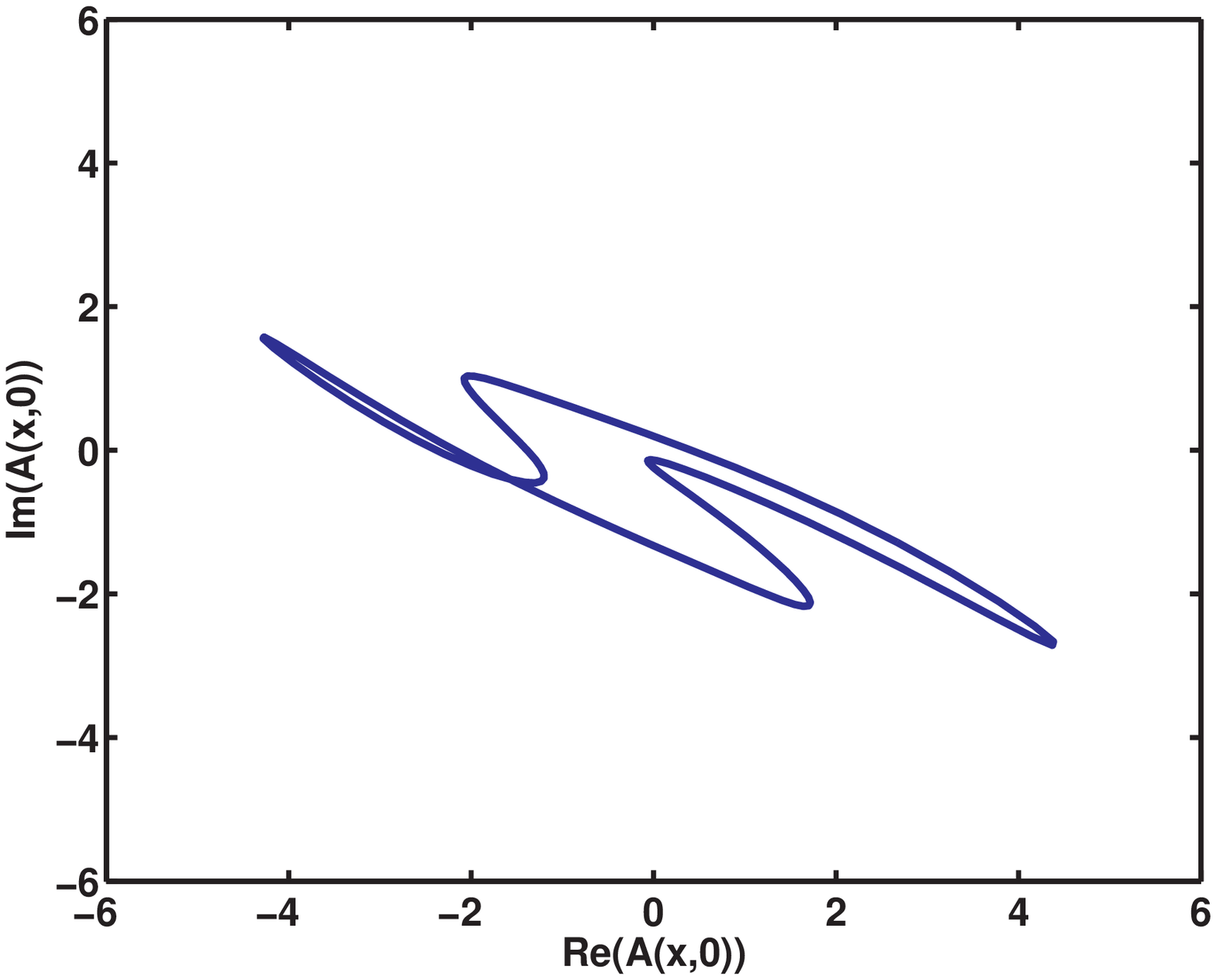}}
   \resizebox{1.25in}{!} 
       {\includegraphics{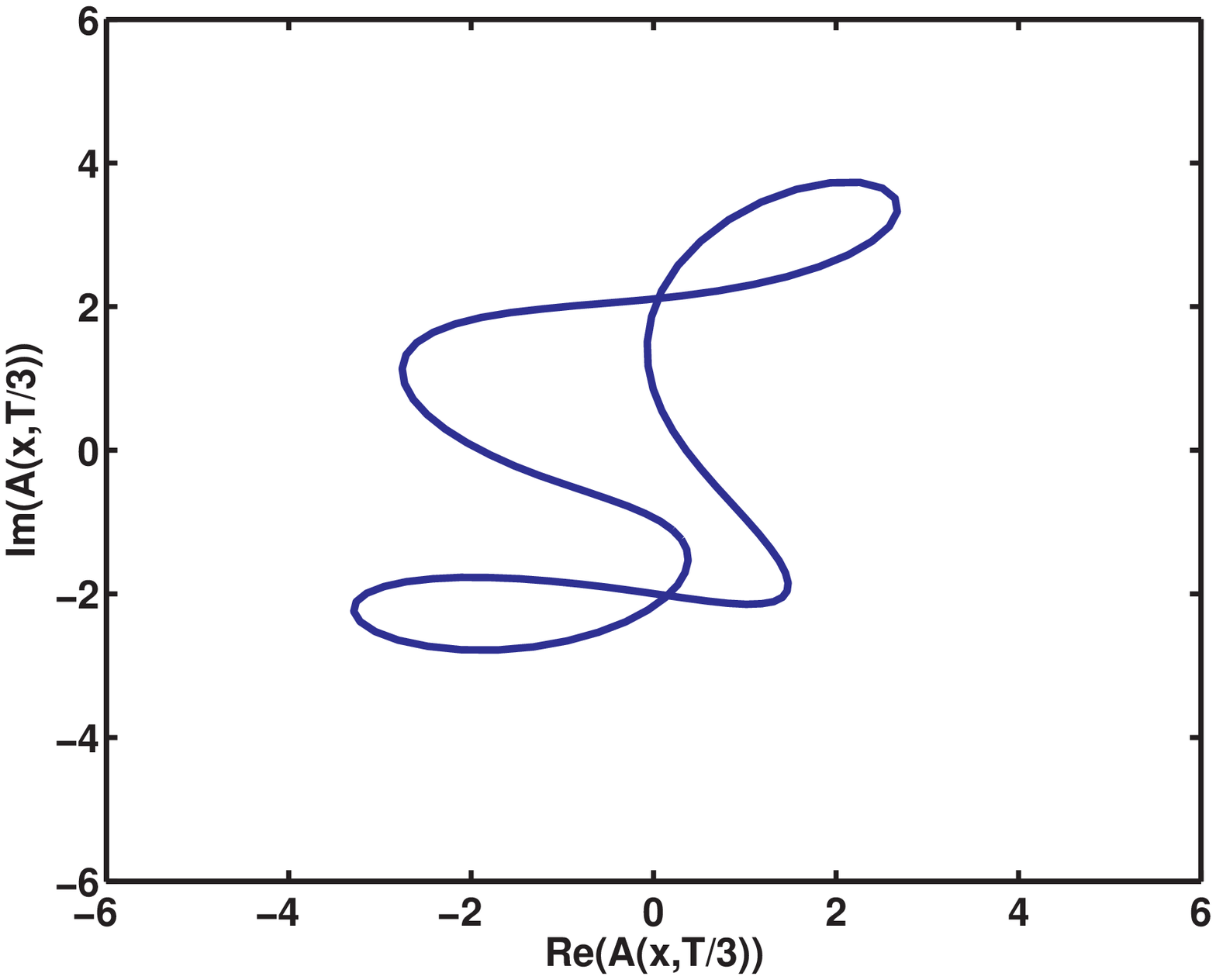}}
   \resizebox{1.25in}{!} 
       {\includegraphics{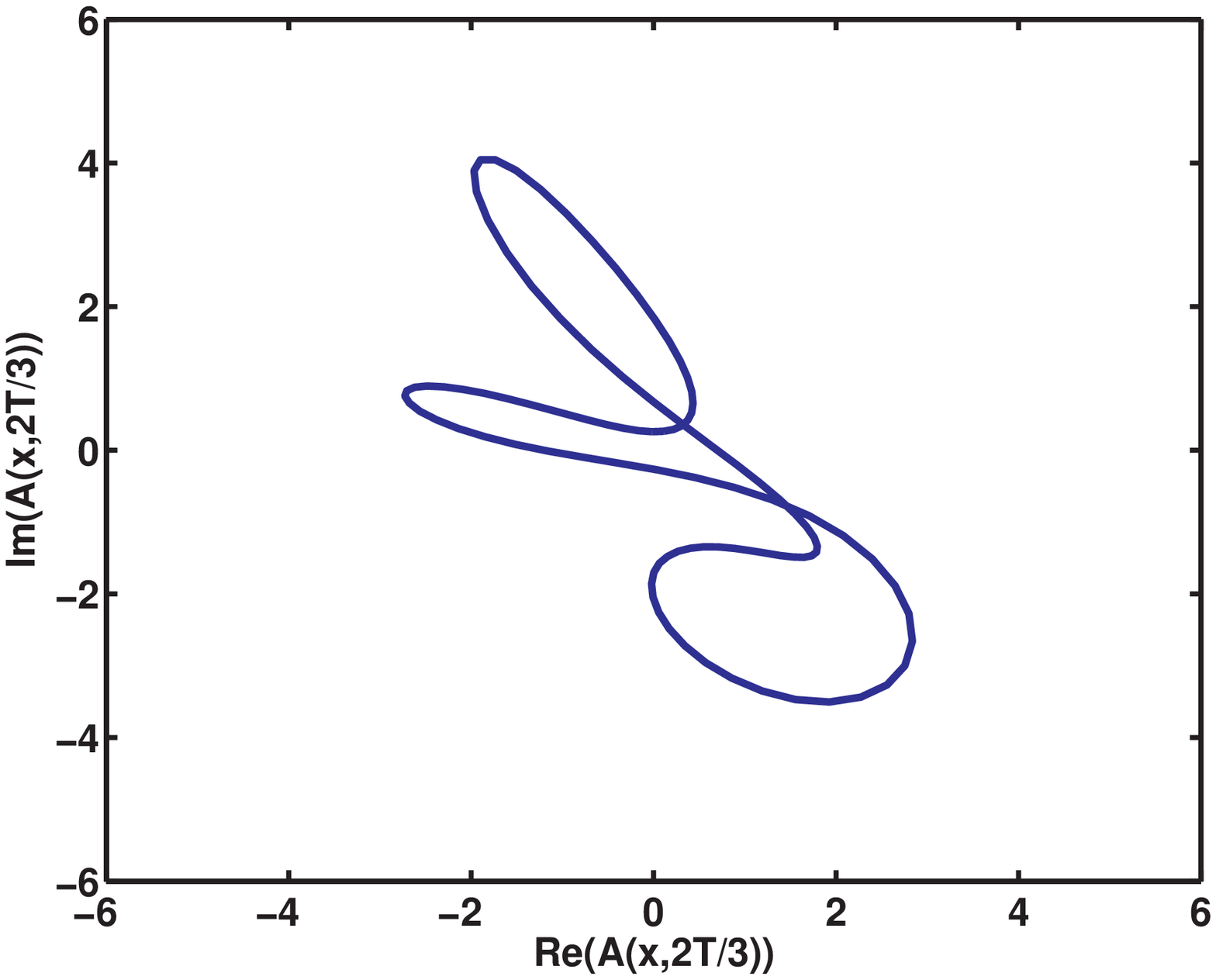}}
   \resizebox{1.25in}{!} 
       {\includegraphics{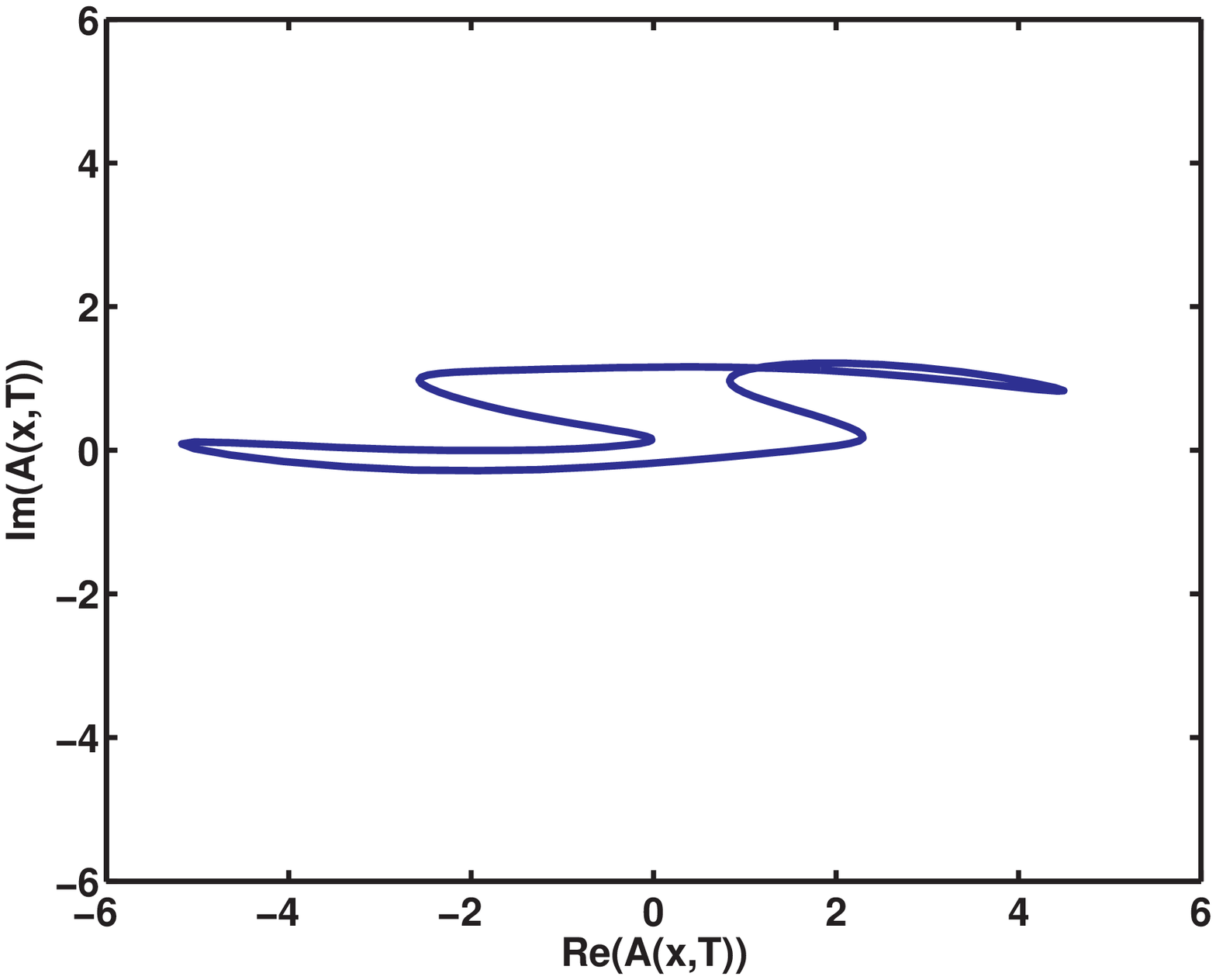}}
   \end{center}
   \caption{Imaginary vs.~real part of solution 14 at times $t = 0, T/3, 2T/3, T$.  Solution has phase
       shift \mbox{$\varphi=2.61$}. 
%%       \href{run:sol14.avi}{\textcolor{red}{Click here}} for movie of time evolution.}
       {Click here} for movie of time evolution.}
   \label{fig:sol14_realvsimag}
\end{figure}

\begin{figure}[hbt!]
   \begin{center}
   \resizebox{2.25in}{!} 
       {\includegraphics{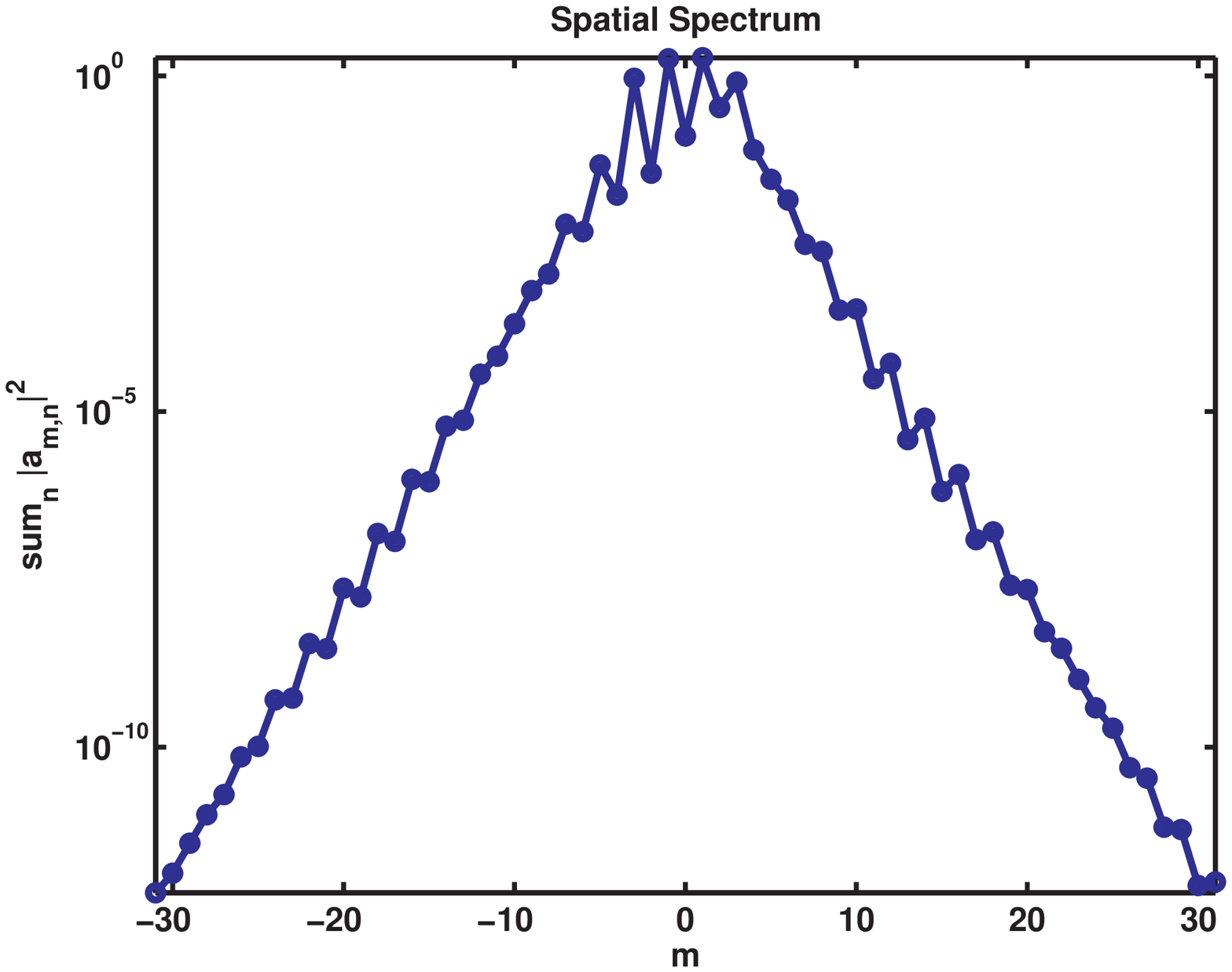}}
   \resizebox{2.25in}{!} 
       {\includegraphics{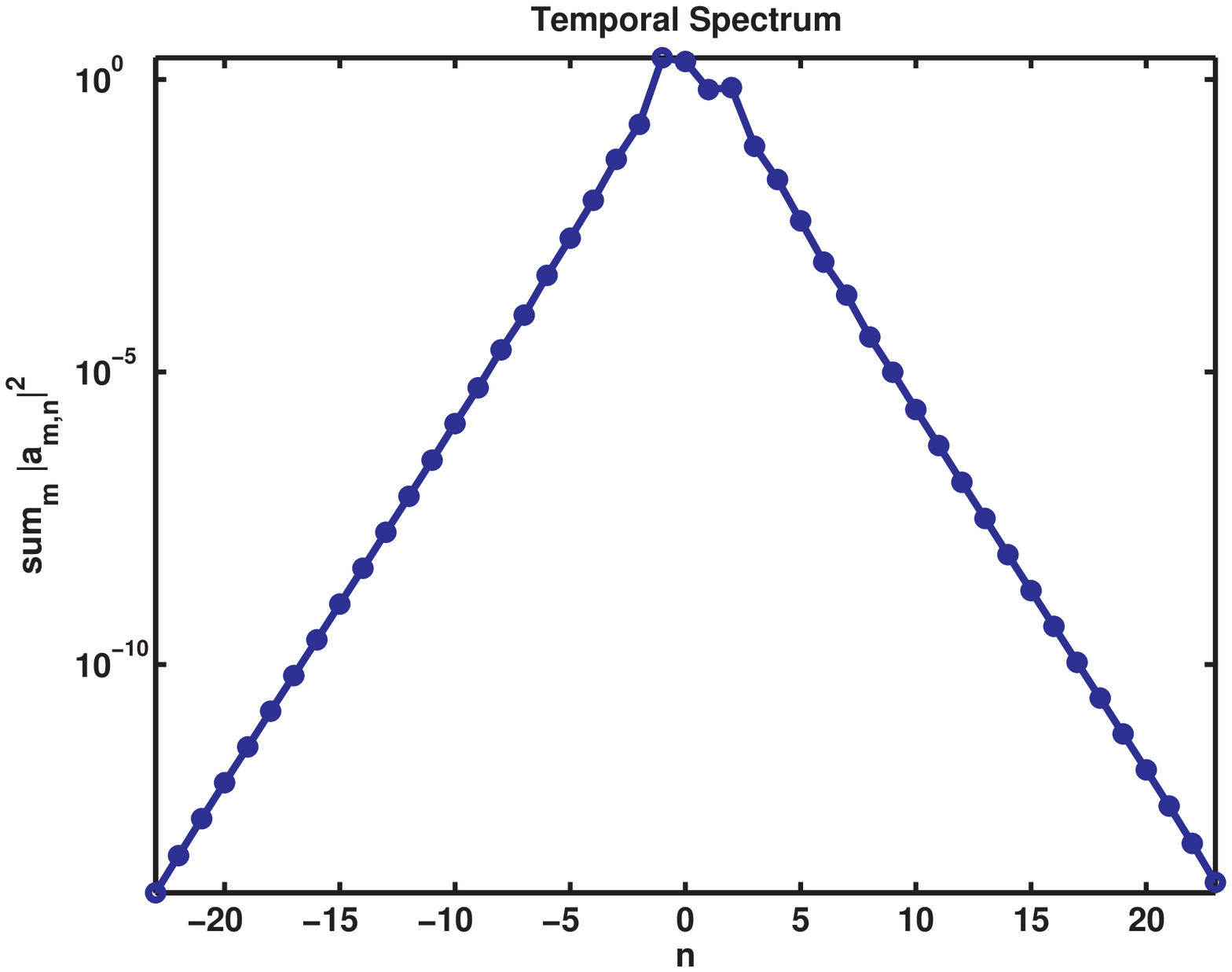}}
   \caption{Spectra of solution 14.}
   \label{fig:sol14_spectra}
   \end{center}
\end{figure}

To give an idea of the character of the solutions we describe a few in some detail.  The first \relsoln\
found, identified as solution 14 in Table~\tblrelsolns, is the least unstable among our collection of 
\relsolns.  The time evolution of solution 14, represented as curves on the plane with coordinates defined 
by the real and imaginary parts of $A(x,t)$ at different times in the interval $[0,T]$, is displayed
in Figure~\ref{fig:sol14_realvsimag}.
The nontrivial time behavior of the solution (i.e., the excitation
of multiple temporal frequency modes) is apparent from these curves.  For single-frequency solutions
$A(x,t) = B(x) \e^{\ii\omega t}$ and generalized traveling waves $A(x,t) = \rho(x-vt) \e^{\ii\phi(x-vt)}
\e^{\ii\omega t}$, where $\omega$ is some single frequency, plots of this kind at different times would
show, except for a rotation, the same curve.  It is clear, then, that solution 14 is not of either
of these types.  Note also that the curve at time $t=T$ differs from that at time $t=0$ because of the
rotation of the complex field $A(x,0) \rightarrow \e^{\ii\varphi}A(x+S,T)$.  The spectra of solution 14 are
displayed in Figure~\ref{fig:sol14_spectra}, where one can also observe the multiple modes active in the
temporal spectrum of the solution.

\begin{figure}[hbt!]
   \begin{center}
   \resizebox{1.25in}{!} 
       {\includegraphics{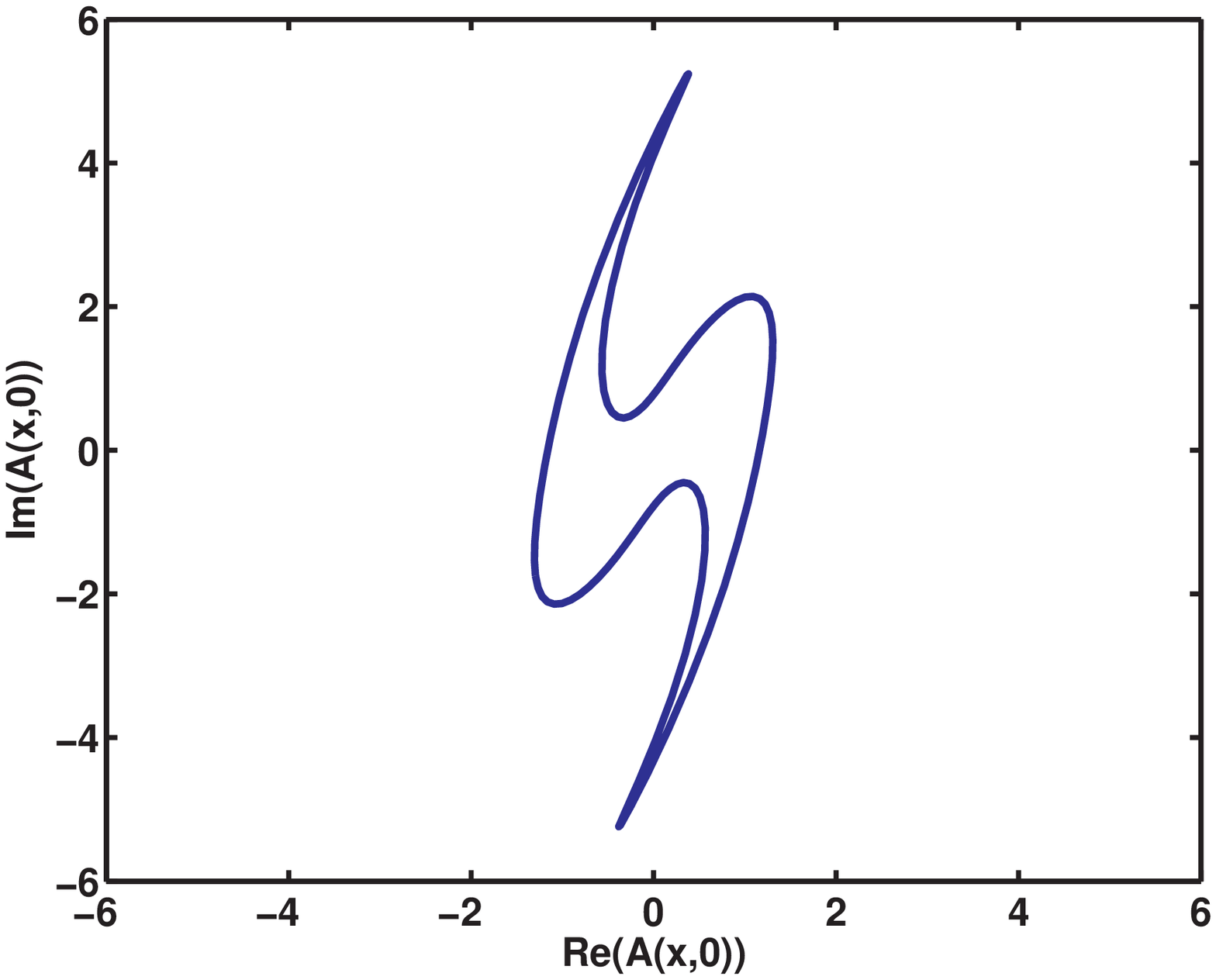}}
   \resizebox{1.25in}{!} 
       {\includegraphics{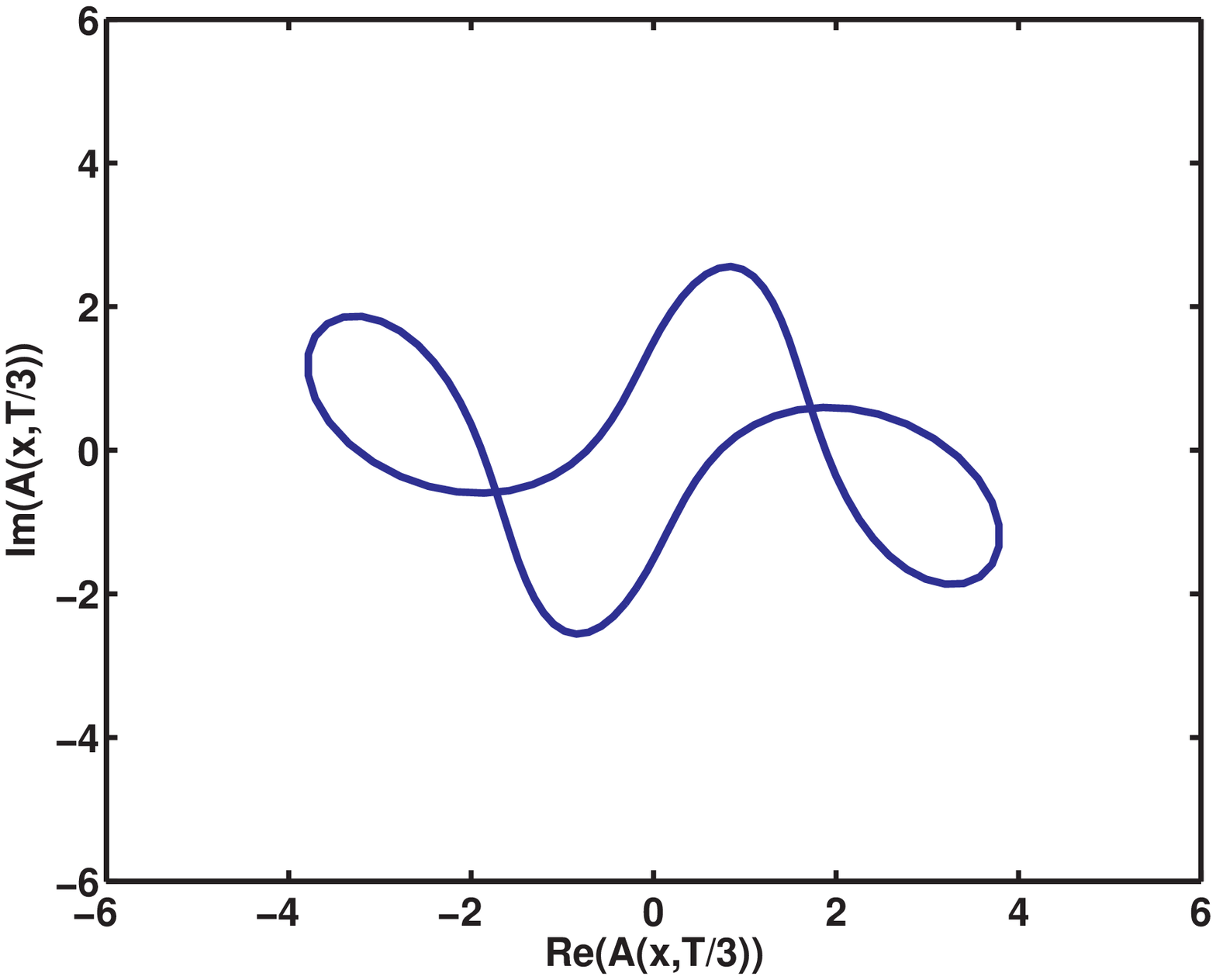}}
   \resizebox{1.25in}{!} 
       {\includegraphics{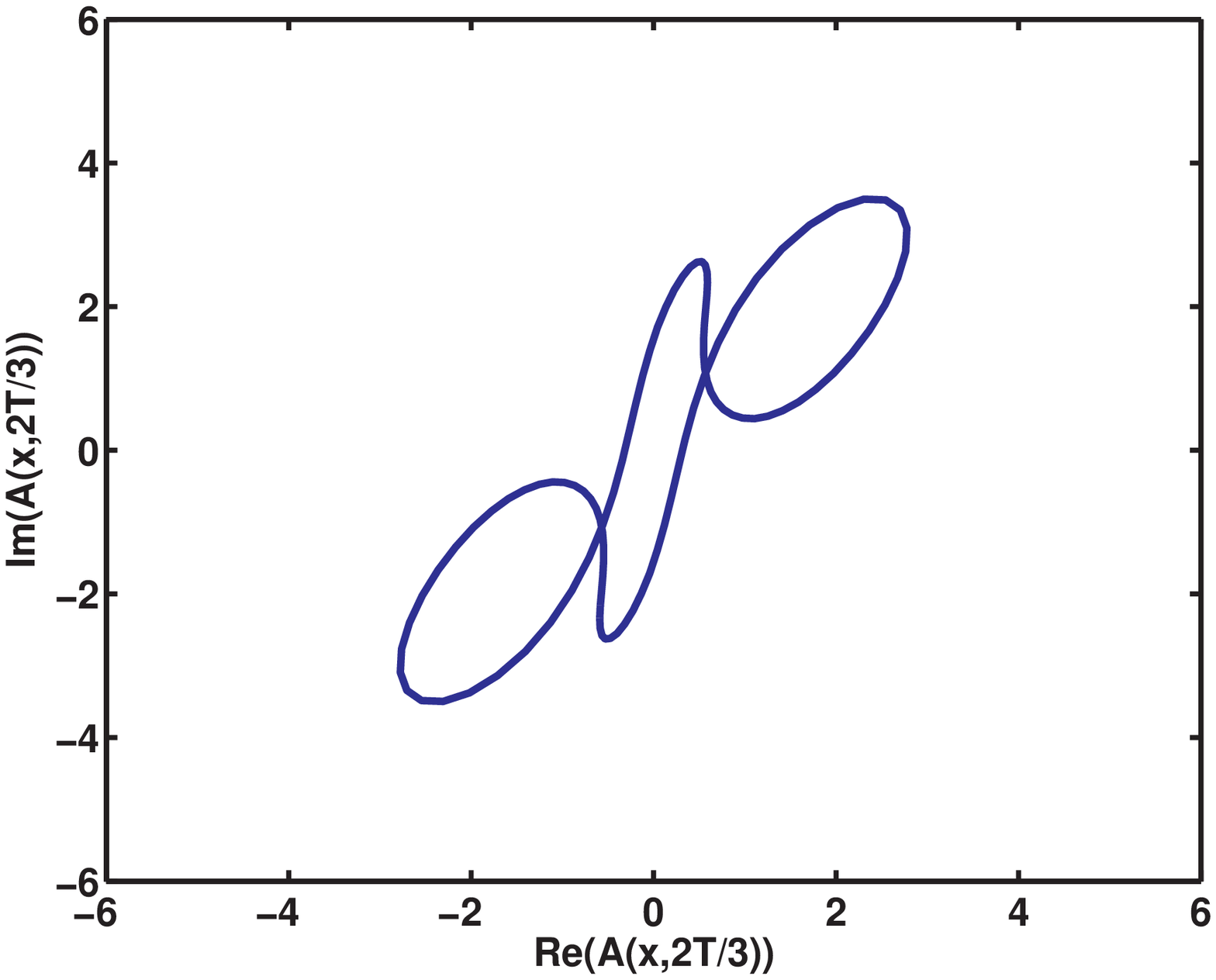}} 
   \resizebox{1.25in}{!} 
       {\includegraphics{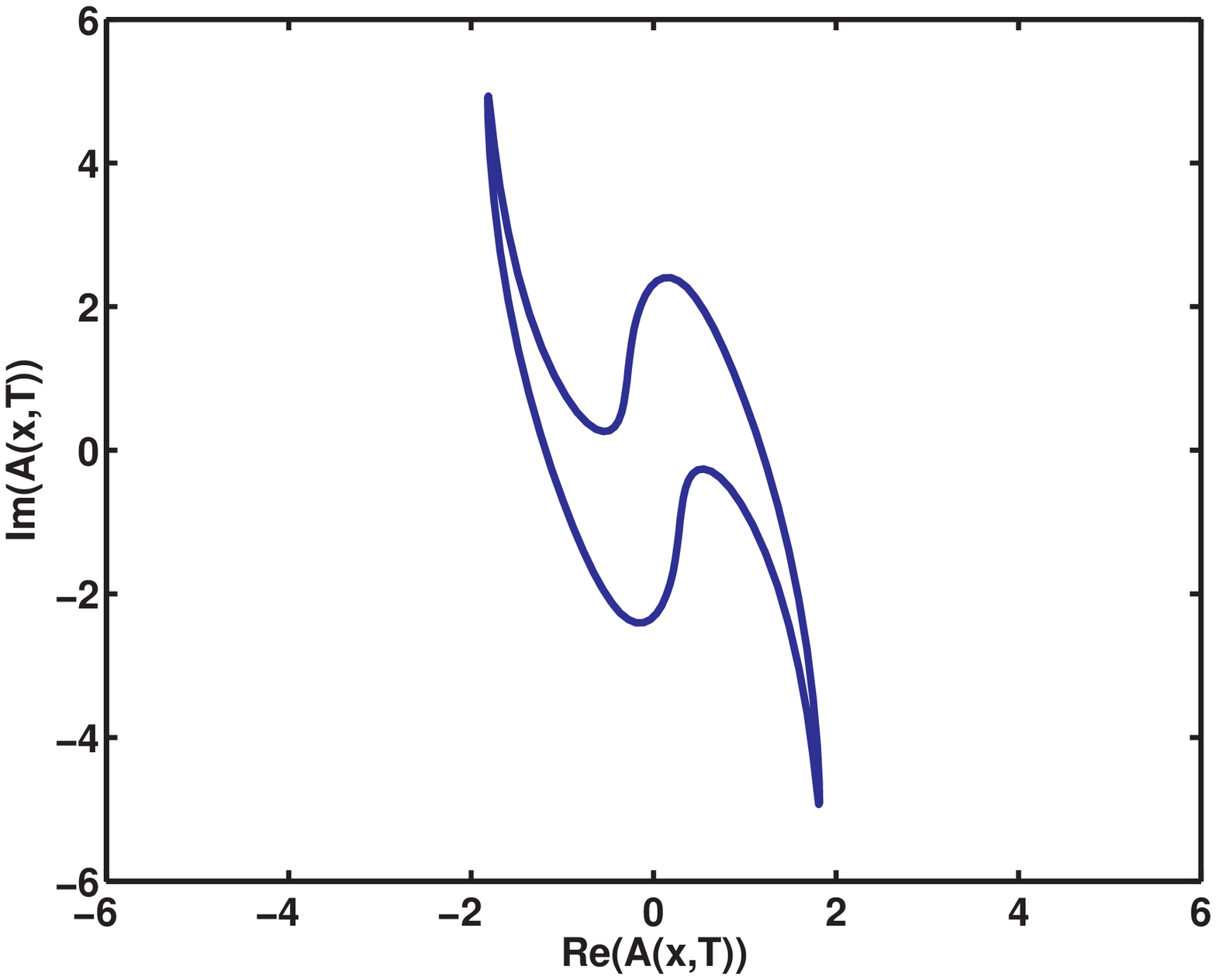}}  
   \caption{Imaginary vs.~real part of solution 16 at times $t = 0,T/3,2T/3,T$.  Solution has phase
       shift \mbox{$\varphi=2.72$}.  It satisfies $A(x,t)=-A(x+\pi,t)$.
%%       \href{run:sol16.avi}{\textcolor{red}{Click here}} for movie of time evolution.}
       {Click here} for movie of time evolution.}
   \label{fig:sol16_realvsimag}
   \end{center}
\end{figure}

Another of the solutions with total instability less than typical, listed as solution 16 in 
Table~\tblrelsolns, has period and drift close to those of solution 14.  However, solution 16 has
stabilizer $C(1,2)$ (cf. \S~\ref{sec:cglestabs}), that is, it satisfies $A(x,t) = -A(x+\pi,t)$.  This
property is apparent from the symmetry in each of the curves representing the time evolution of solution 16,
displayed in Figure~\ref{fig:sol16_realvsimag}.  We will come back to solutions 14 and 16 in 
\S~\ref{sec:family_solns}, where an apparent relationship between these and several other solutions
is explored.

\begin{figure}[hbt!]
   \begin{center}
   \resizebox{3.80in}{1.80in}
       {\includegraphics{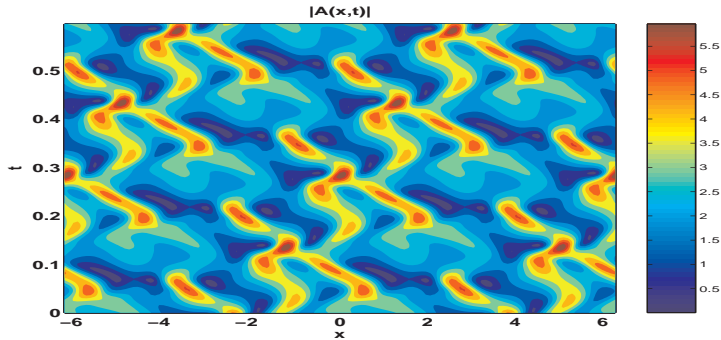}}
   \caption{Contours of $|A(x,t)|$ for solution 20, $x \in [-2\pi,2\pi]$, $t \in [0, 4T]$.  Solution has
       space shift $S=1.27$.}
   \label{fig:sol20_abs_cntr}
   \end{center}
\end{figure}

\begin{figure}[hbt!]
   \begin{center}
   \resizebox{1.25in}{!} 
       {\includegraphics{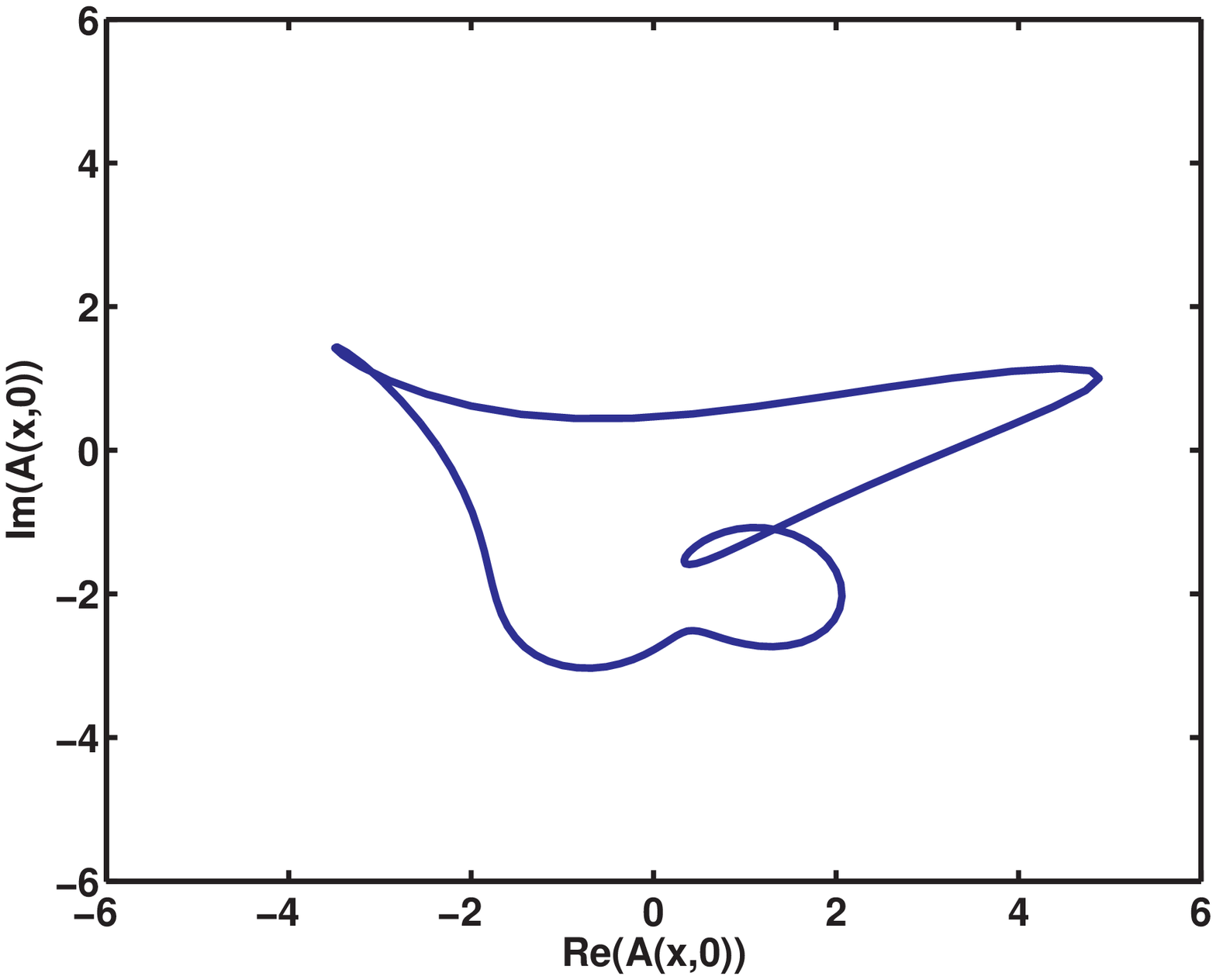}}
   \resizebox{1.25in}{!} 
       {\includegraphics{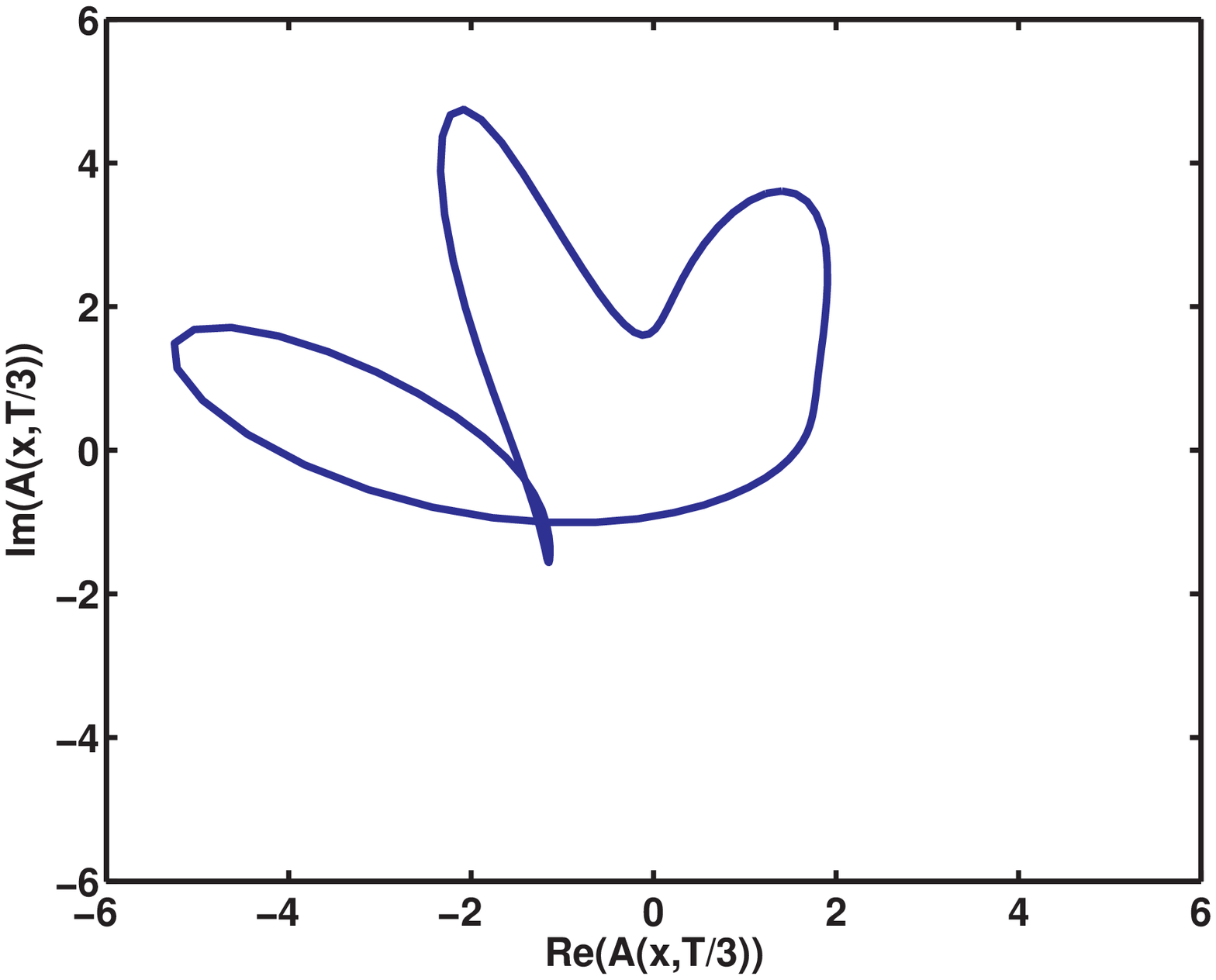}}
   \resizebox{1.25in}{!} 
       {\includegraphics{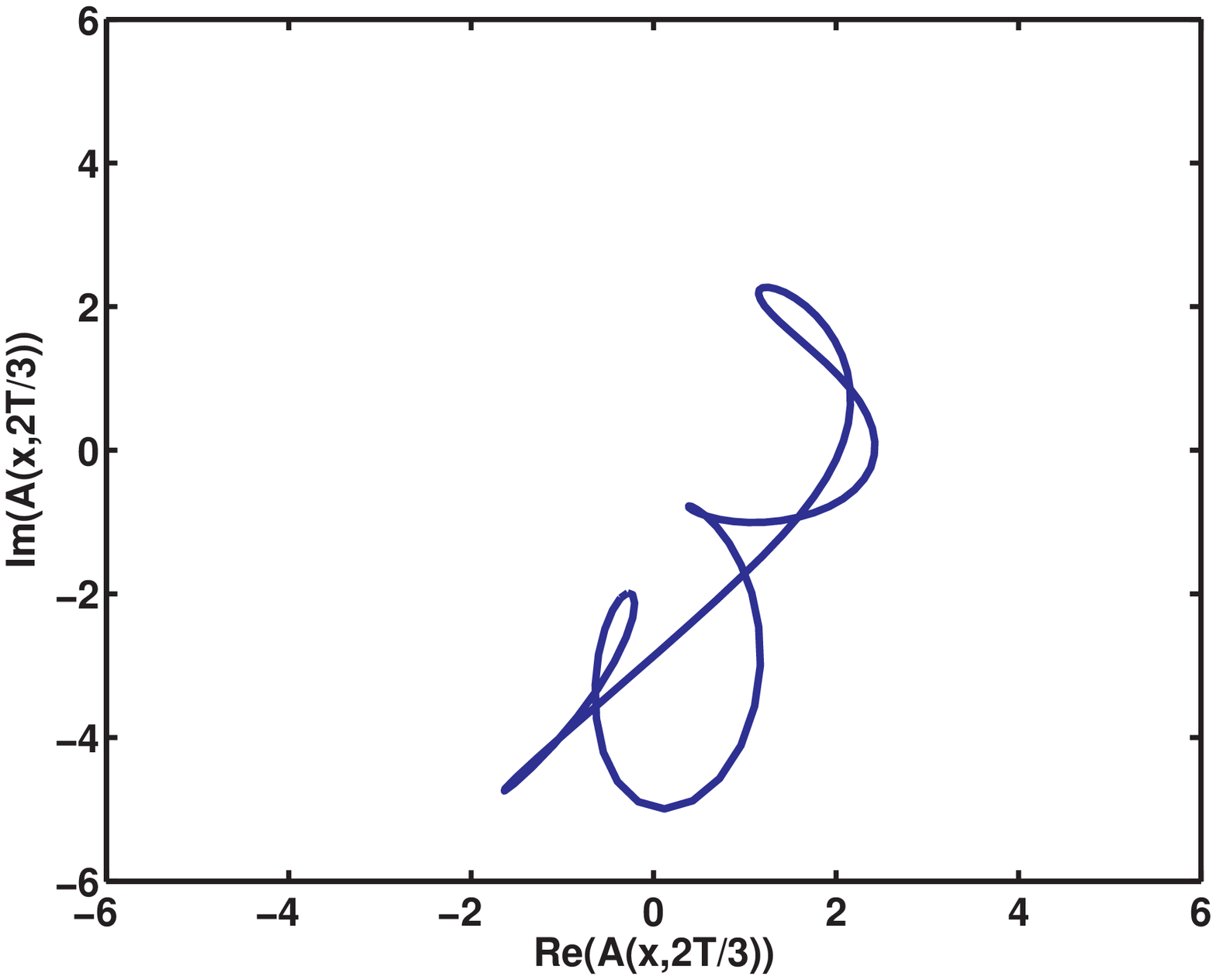}}
   \resizebox{1.25in}{!} 
       {\includegraphics{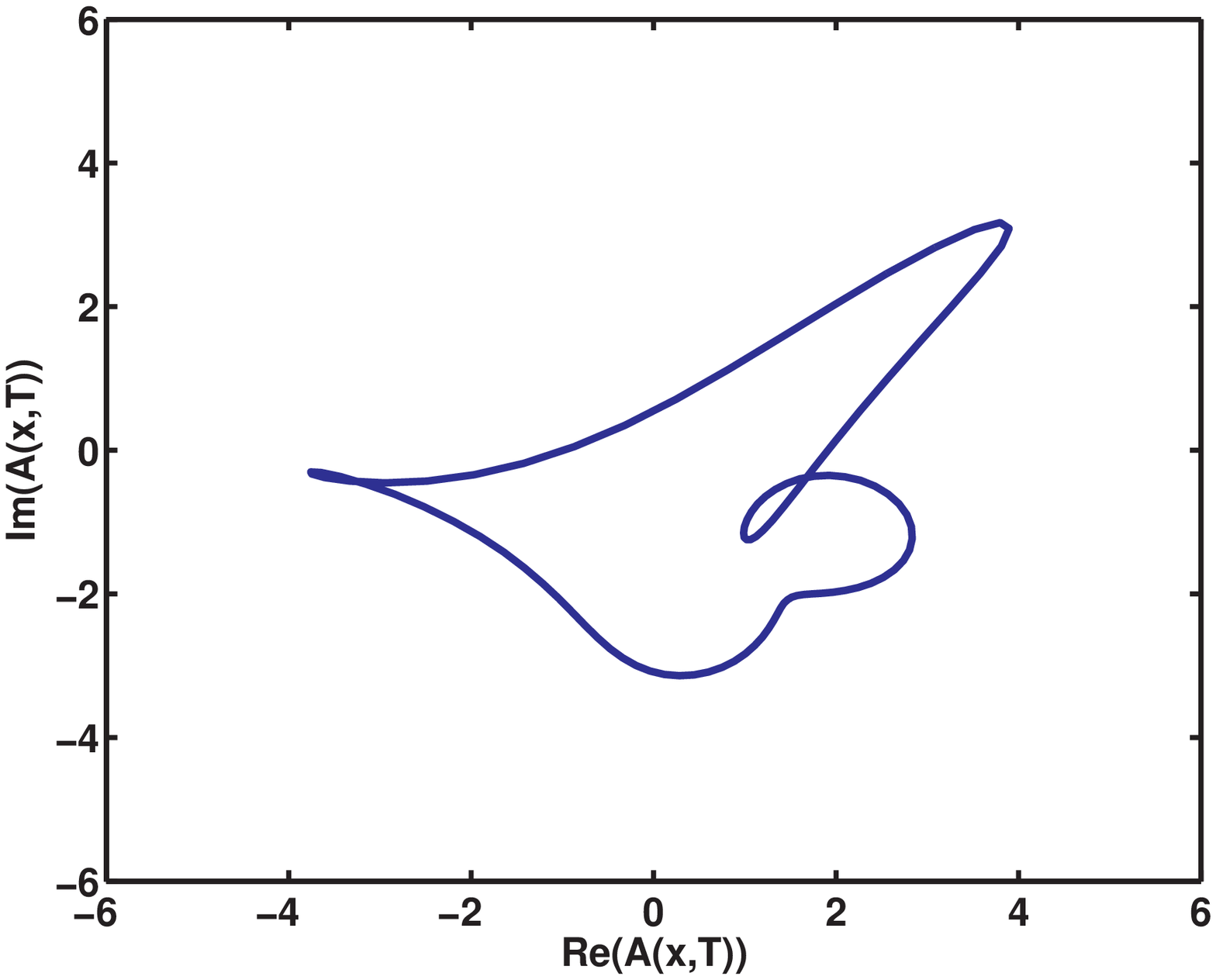}}  
   \caption{Imaginary vs.~real part of solution 20 at times $t = 0,T/3,2T/3,T$.  Solution has phase
       shift \mbox{$\varphi=5.81$}.
%%       \href{run:sol20.avi}{\textcolor{red}{Click here}} for movie of time evolution.}
       {Click here} for movie of time evolution.}
   \label{fig:sol20_realvsimag}
   \end{center}
\end{figure}

Contour plots of the absolute value of a \relsoln\ provide a way of visualizing periodicity in time
coupled with the shift $S$ in space.  Figure~\ref{fig:sol20_abs_cntr} shows one such plot for one of the
most unstable among the \relsolns\ found, listed as solution 20 in Table~\tblrelsolns.
The absolute value of the solution has been plotted over two space and four time intervals.  Moving
vertically (in time) and horizontally (in space) allows us to observe periodicity in time, after taking
into account the shift $S$ in space.  From the contours of $|A|$ one can also locate points in space-time
where the magnitude of $A$ vanishes (the darkest shaded regions (in blue) in
Figure~\ref{fig:sol20_abs_cntr}).
The time evolution of solution 20 is shown in Figure~\ref{fig:sol20_realvsimag}.
To view the time evolution of several other \relsolns, click on the following red links:
%%\href{run:sol8.avi}{\textcolor{red}{solution 8}},
%%\href{run:sol24.avi}{\textcolor{red}{solution 24}}, and
%%\href{run:sol26.avi}{\textcolor{red}{solution 26}}.
{solution 8},
{solution 24}, and
{solution 26}.

\begin{figure}[hbt!]
   \begin{center}
   \resizebox{4.50in}{!}
       {\includegraphics{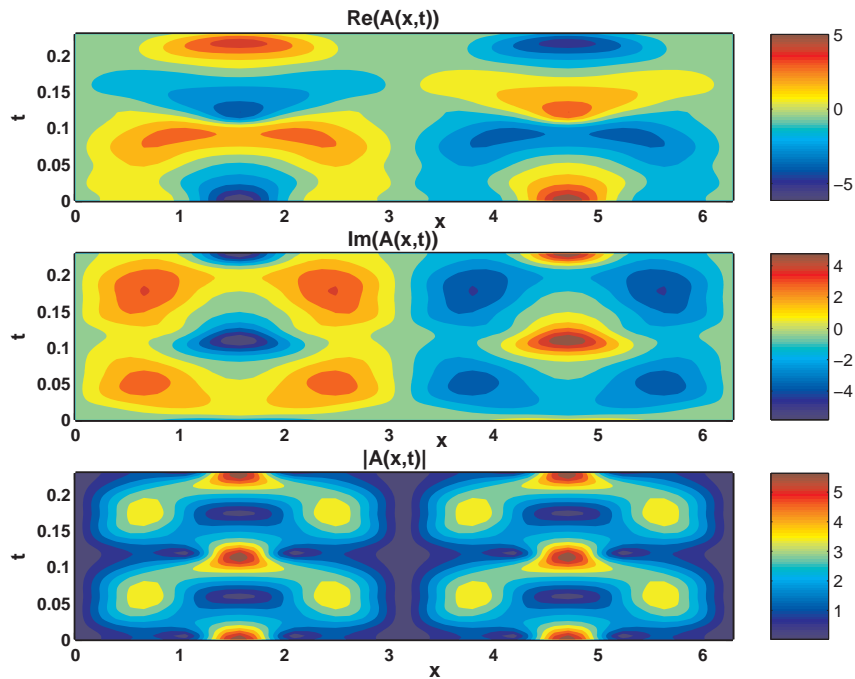}}
   \caption{Contour plots for solution 15, $x \in [0,2\pi]$, $t \in [0, 2T]$.  Solution satisfies
       $A(x,t) = -A(x+\pi,t)$ and $|A(x,t)| = |A(x,t+T)|$.  It is odd about $\pi$ for $x \in [0,2\pi]$, 
       even about $\pi/2$ for $x \in [0,\pi]$, and even about $3\pi/2$ for $x \in [\pi,2\pi]$.}
   \label{fig:sol15_cntrs}
   \end{center}
\end{figure}

Finally, several other \relsolns\ have nontrivial \mbox{$\torus^2$-stabilizers} and some are even or odd
about certain points in the spatial domain.  For instance, Figure~\ref{fig:sol15_cntrs} displays contour
plots of the real and imaginary parts and the absolute value of a \relsoln\ having all these properties.
The solution, identified as solution 15 in Table~\tblrelsolns, has total instability close to that of
a typical trajectory. The drift and period of solution 15 are close to those of solutions 14 and 16, but the
unstable dimension of each of these solutions is different.  Like solution 16, solution 15 has stabilizer
$C(1,2)$.  In addition, solution 15 is odd about $x=\pi$ and even about $x=\pi/2, 3\pi/2$. Since
solution 15 has stabilizer $C(1,2)$, it is also a \relsoln\ with period $T$ and drift
$(\tilde{\varphi},\tilde{S}) = (\varphi+\pi, S+\pi)$.  The space shift $S$ of solution 15 is 
(numerically) equal to $\pi$, so it follows that its absolute value is (truly) periodic in time, with
time period $T$.  All these properties of solution 15 can be observed from the contour plots in
Figure~\ref{fig:sol15_cntrs}.  

\begin{figure}[hbt!]
   \begin{center}
   \resizebox{1.33in}{!} 
       {\includegraphics{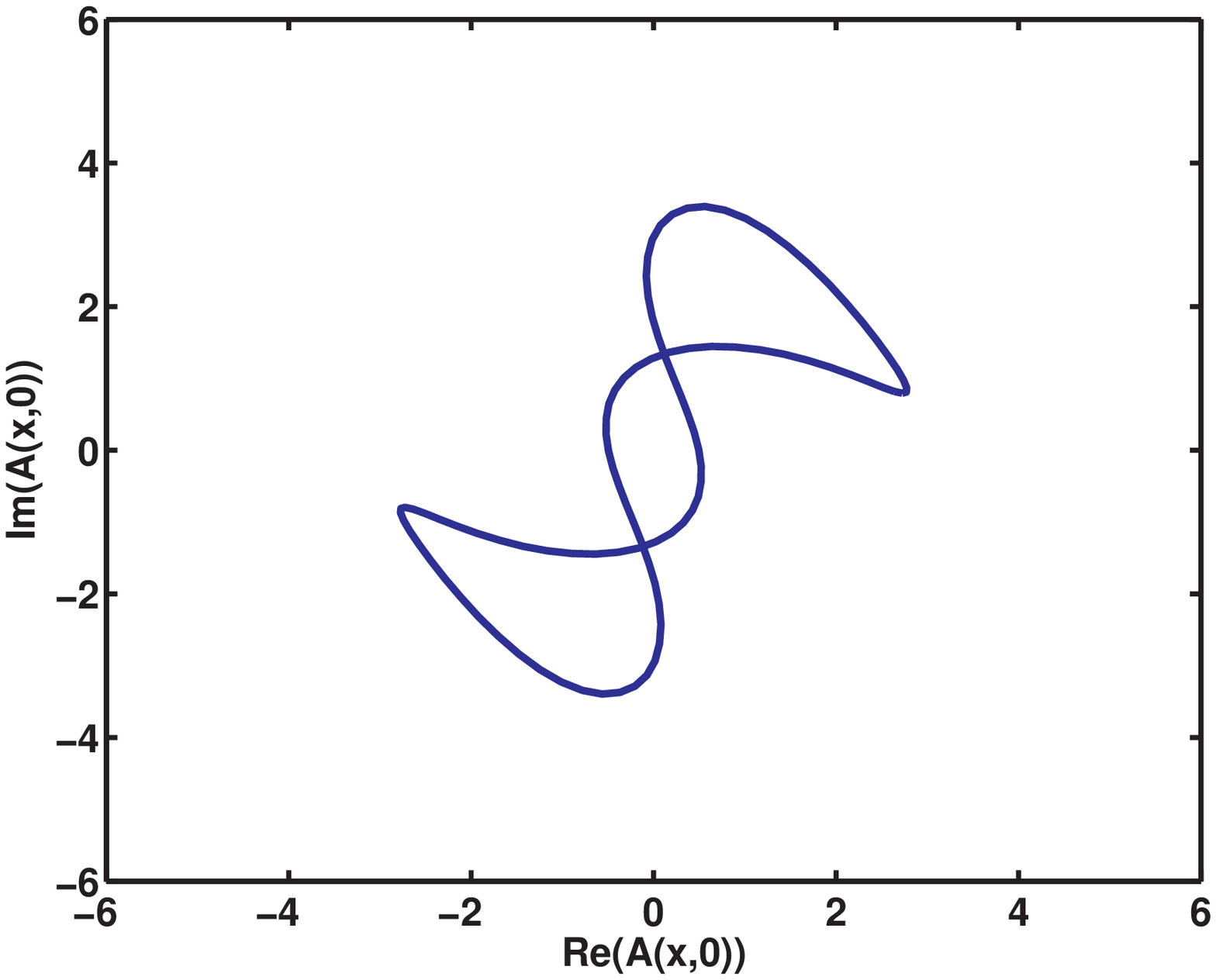}}   
   \resizebox{1.33in}{!} 
       {\includegraphics{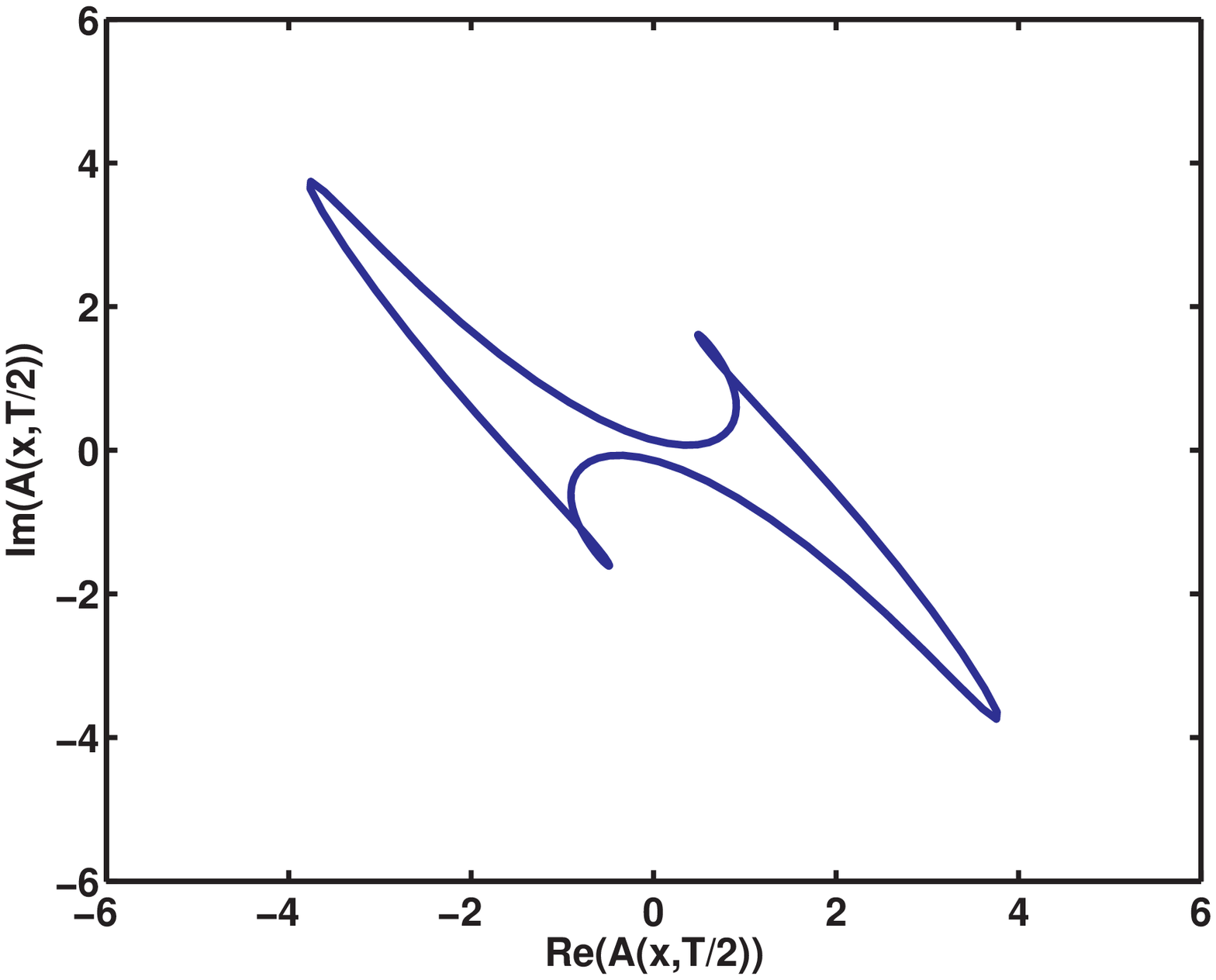}}
   \resizebox{1.33in}{!} 
       {\includegraphics{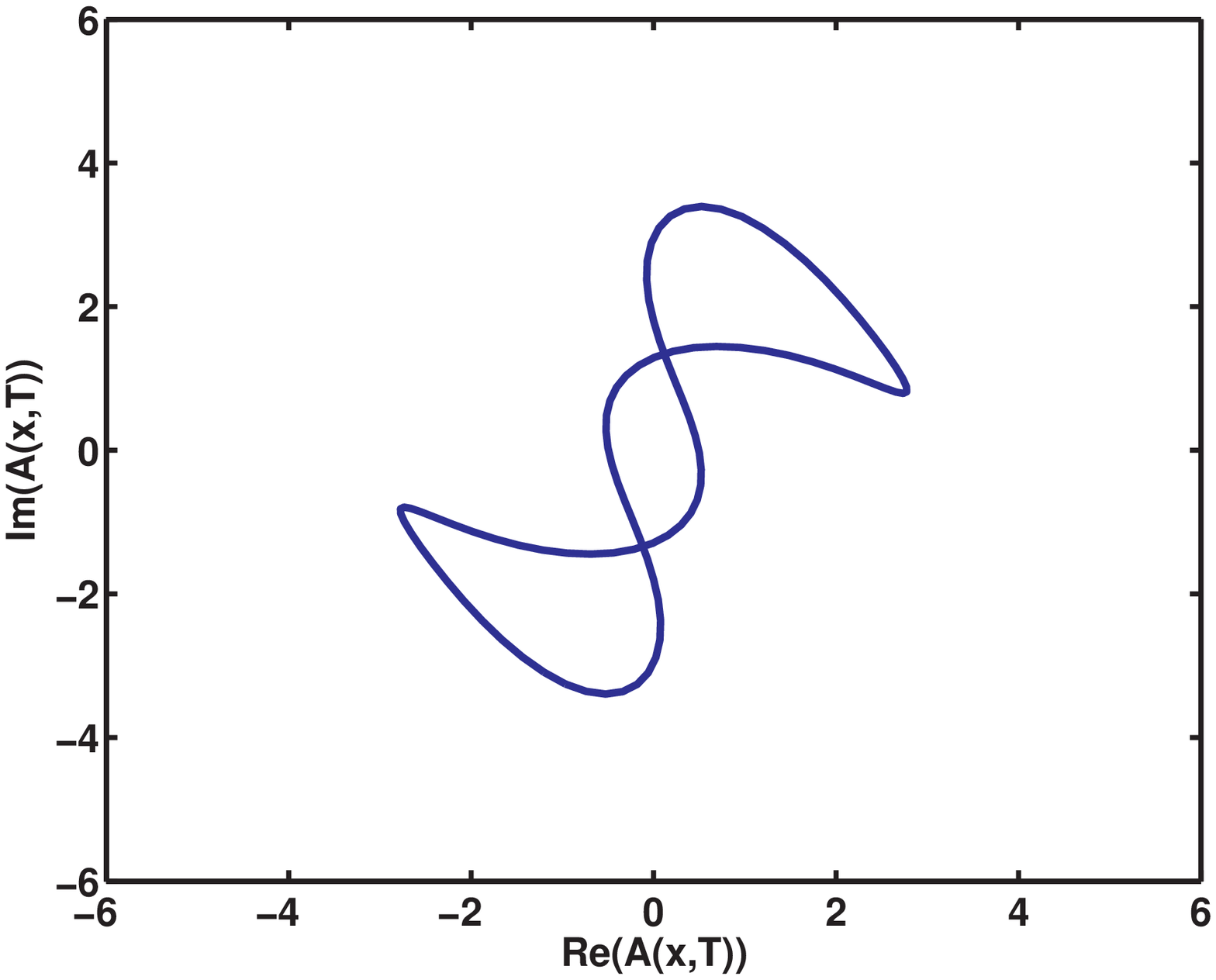}}
   \caption{Imaginary vs.~real part of solution 3 at times $t = 0,T/2,T$.  Solution satisfies
       $A(x,t) = -A(x+\pi,t)$.
%%       \href{run:sol3.avi}{\textcolor{red}{Click here}} for movie of time evolution.}
       {Click here} for movie of time evolution.}
   \label{fig:sol3_realvsimag}
   \end{center}
\end{figure}

\begin{figure}[hbt!]
   \begin{center}
   \resizebox{1.33in}{!} 
       {\includegraphics{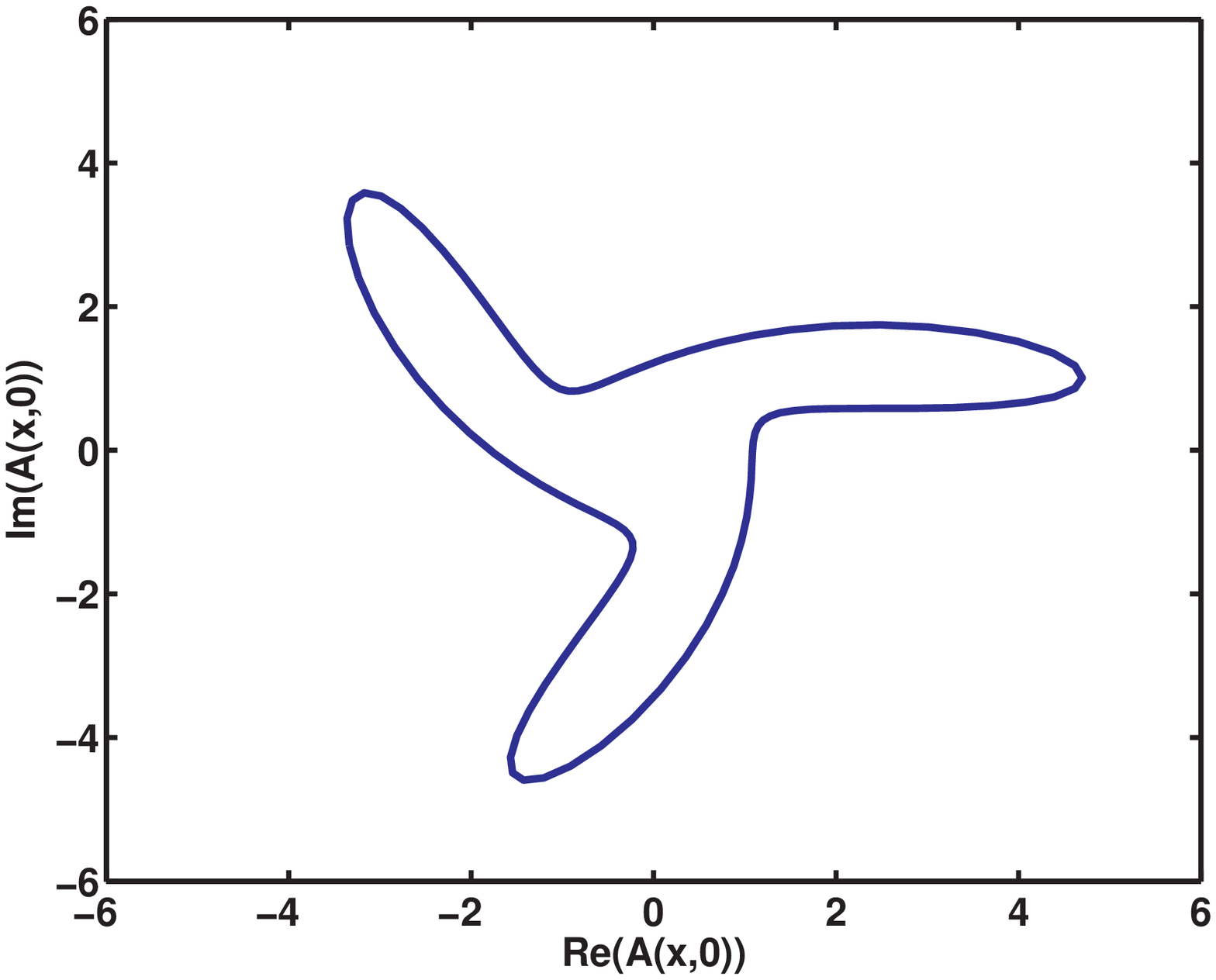}}   
   \resizebox{1.33in}{!} 
       {\includegraphics{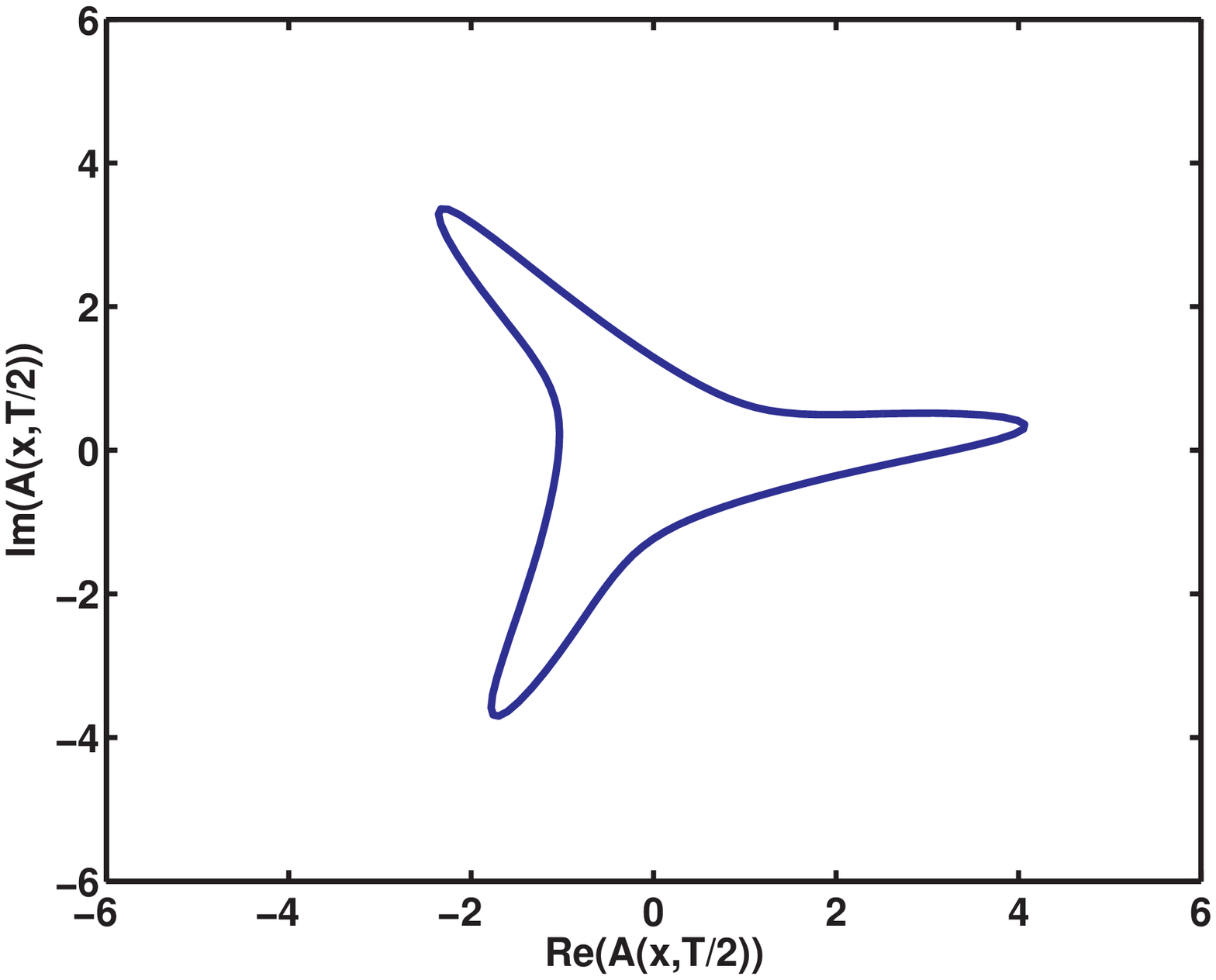}}
   \resizebox{1.33in}{!} 
       {\includegraphics{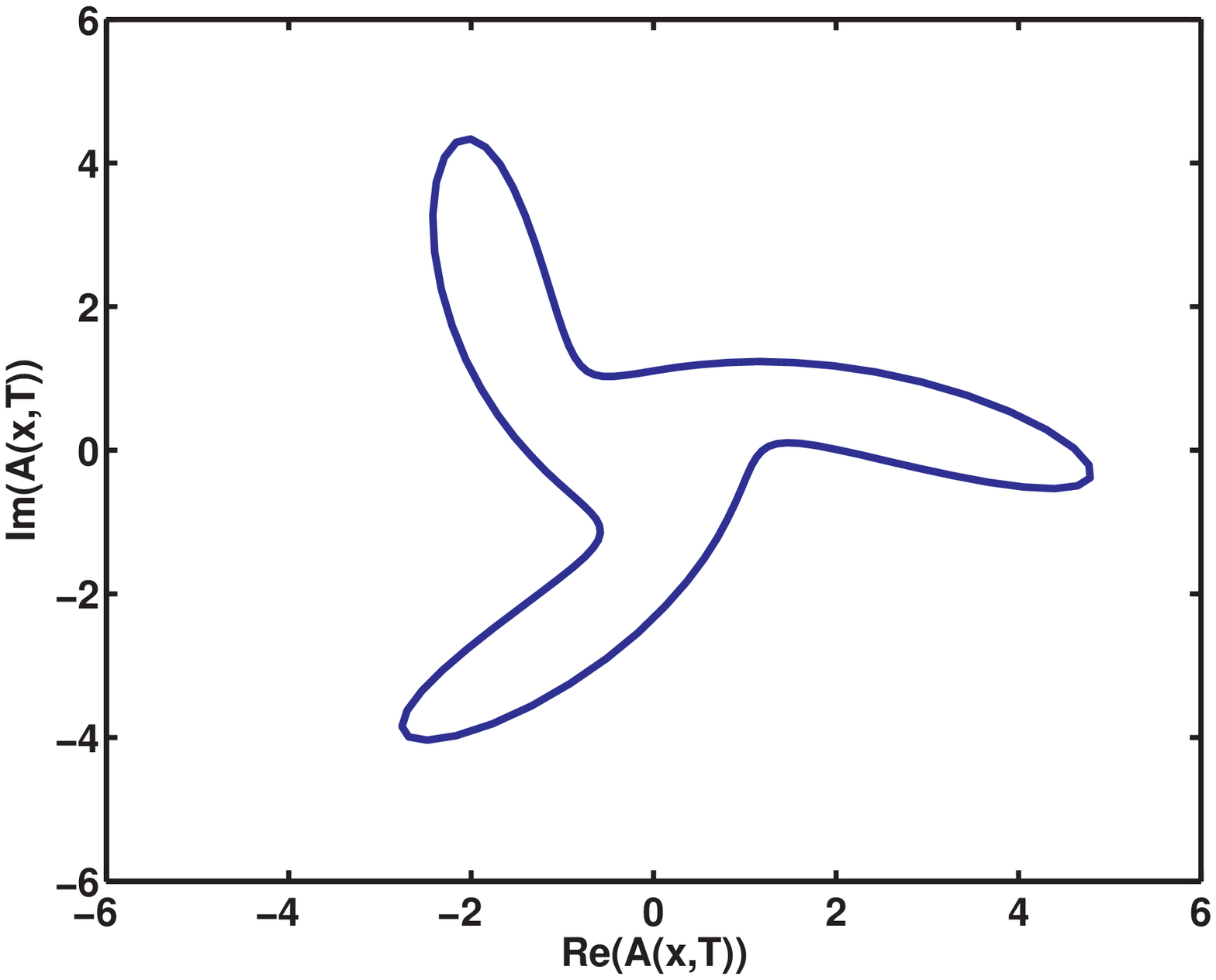}}
   \caption{Imaginary vs.~real part of solution 9 at times $t = 0,T/2,T$.  Solution satisfies
       $A(x,t) = \e^{\ii 2\pi/3}A(x+2\pi/3,t)$.
%%       \href{run:sol9.avi}{\textcolor{red}{Click here}} for movie of time evolution.}
       {Click here} for movie of time evolution.}
   \label{fig:sol26_realvsimag}
   \end{center}
\end{figure}

\begin{figure}[hbt!]
   \begin{center}
   \resizebox{2.50in}{1.50in}
       {\includegraphics{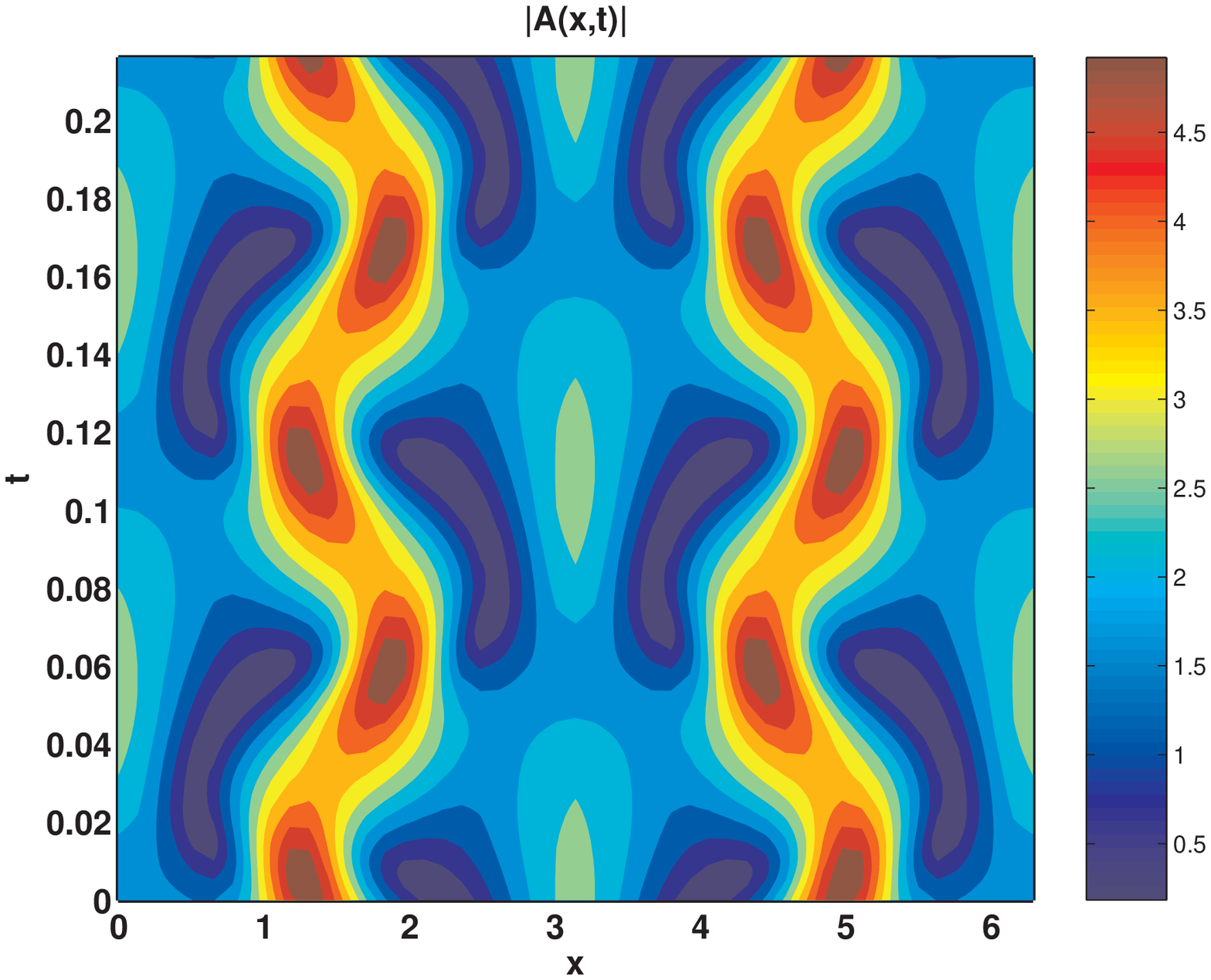}}
   \resizebox{2.50in}{1.50in}
       {\includegraphics{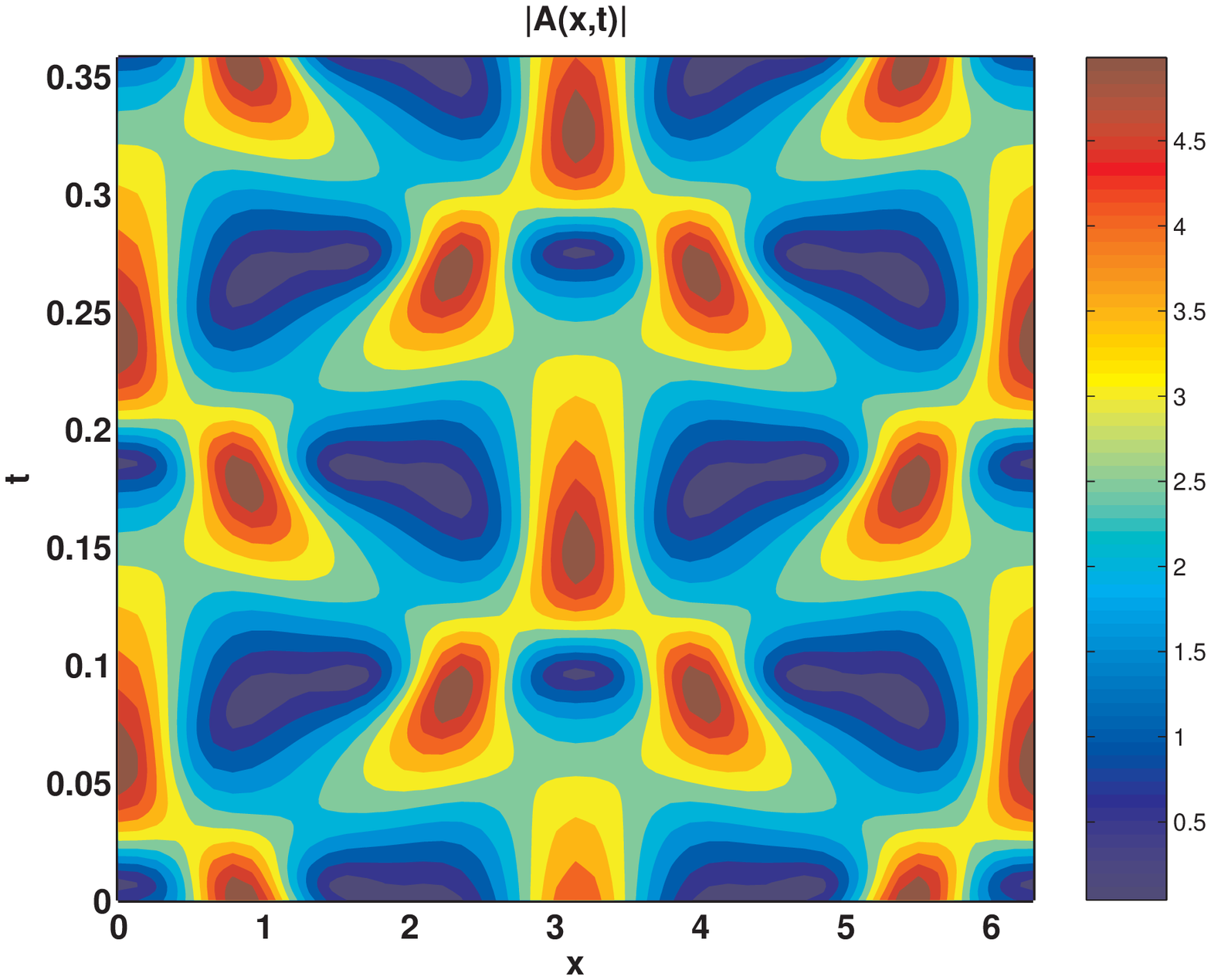}}
   \makebox[2.50in]{(a)}  \makebox[2.50in]{(b)}
   \caption{Contours of $|A(x,t)|$ for solutions (a) 4 and (b) 10, $x \in [0,2\pi]$, $t \in [0, 4T]$.
       Each solution is even about $x=\pi$ and satisfies \mbox{$|A(x,t)| = |A(x,t+2T)|$}.}
   \label{fig:sol4_10_cntrs}
   \end{center}
\end{figure}

The other \relsolns\ with nontrivial \mbox{$\torus^2$-stabilizers} are solutions 3 and 9, which have
stabilizers $C(1,2)$ and $C(1,3)$, respectively.  Solutions 4 and 10 are even about $x=\pi$ and have a
space shift $S$ (numerically) equal to $\pi$.  The latter results in the absolute value of each of these
solutions being periodic in time with period $2T$. Such properties of solutions 3, 4, 9, and 10 can be
seen from Figures~\ref{fig:sol3_realvsimag}--\ref{fig:sol4_10_cntrs}.  We remark that solutions
3, 9, and 10 are more unstable than typical trajectories, whereas solution 4 is among the least unstable
of the solutions found.

\subsection{Instability and a Family of Solutions}
\label{sec:family_solns}

There are several features of the instability of the collection of \relsolns\ that are apparent from 
Figures~\ref{fig:TvsLyap} and \ref{fig:unstdim}: the largest Lyapunov exponent and total instability of
most of the \relsolns\ are larger than those of a typical trajectory, and  in contrast, the unstable
dimensions are nearly evenly distributed about the typical value.  In addition, the periods of the
\relsolns\ with instability less than typical in Figure~\ref{fig:TvsLyap} are clustered around what
appear to be multiples of a single fundamental period.

More specifically, if $T_f = 0.054$, then the seven \relsolns\ whose largest Lyapunov exponent and total
instability are less than typical have periods close to one of $T_f$, $2 T_f$, $4 T_f$, and $6 T_f$. 
As will be discussed below, the drifts $(\varphi, S)$ of the corresponding solutions are also
approximate integer multiples.  This suggests that these least unstable \relsolns\ arose via a sequence
of period-multiplying bifurcations and that the resulting collection of solutions are members of a low
instability (and thus high traffic) region of the main attractor, which is structured around a
three-torus that projects in the reduced flow to a loop with the fundamental period $T_f$. Further
evidence for this proposed structure is the observed repeated appearance in typical trajectories of
sequences of patterns from the low instability solutions.  The role of $R$ in the equations 
\refEqn{eq:odes} makes it clear that, in general, instability will increase with increasing $R$. Thus,
low instability orbits could be continuations of solutions present for smaller values of $R$. This
conjectured attractor structure and bifurcation scheme should certainly be investigated in further
detail. 

In the initial phases of this investigation we have found a collection of solutions that clearly fit
together in a dynamical family. The family contains four of the seven solutions with total instability
less than typical and, in fact, consists of four of the five solutions with least total instability.
The family splits into the pair 14/16, which have periods approximately $2 T_f$, and the pair 58/59, which
have periods approximately $6 T_f$. The elements of a pair have very similar $(\varphi, S, T)$, and all 
these values for the pair 58/59 are close to three times those of pair 14/16. Solutions 14 and 16 are
distinguished from each other by their unstable dimensions of $4$ and $5$, respectively. Solutions 58
and 59 both have unstable dimension of $3$, but they differ in their unstable Floquet spectrum (i.e.,
the eigenvalues of the \relmon\ matrix with magnitude greater than one).  In this portion of the spectrum, 
both solutions have a conjugate pair of complex eigenvalues, but the remaining real eigenvalue is
positive for solution 58 and negative for solution 57. Thus, they have opposite Lefschetz indices as
fixed points of their Poincar\'e map and hence the available information is consistent with solutions
58 and 59 arising together in an equivariant period-tripling bifurcation.

\begin{figure}[hbt!]
   \begin{center}
   \resizebox{2.45in}{!}
       {\includegraphics{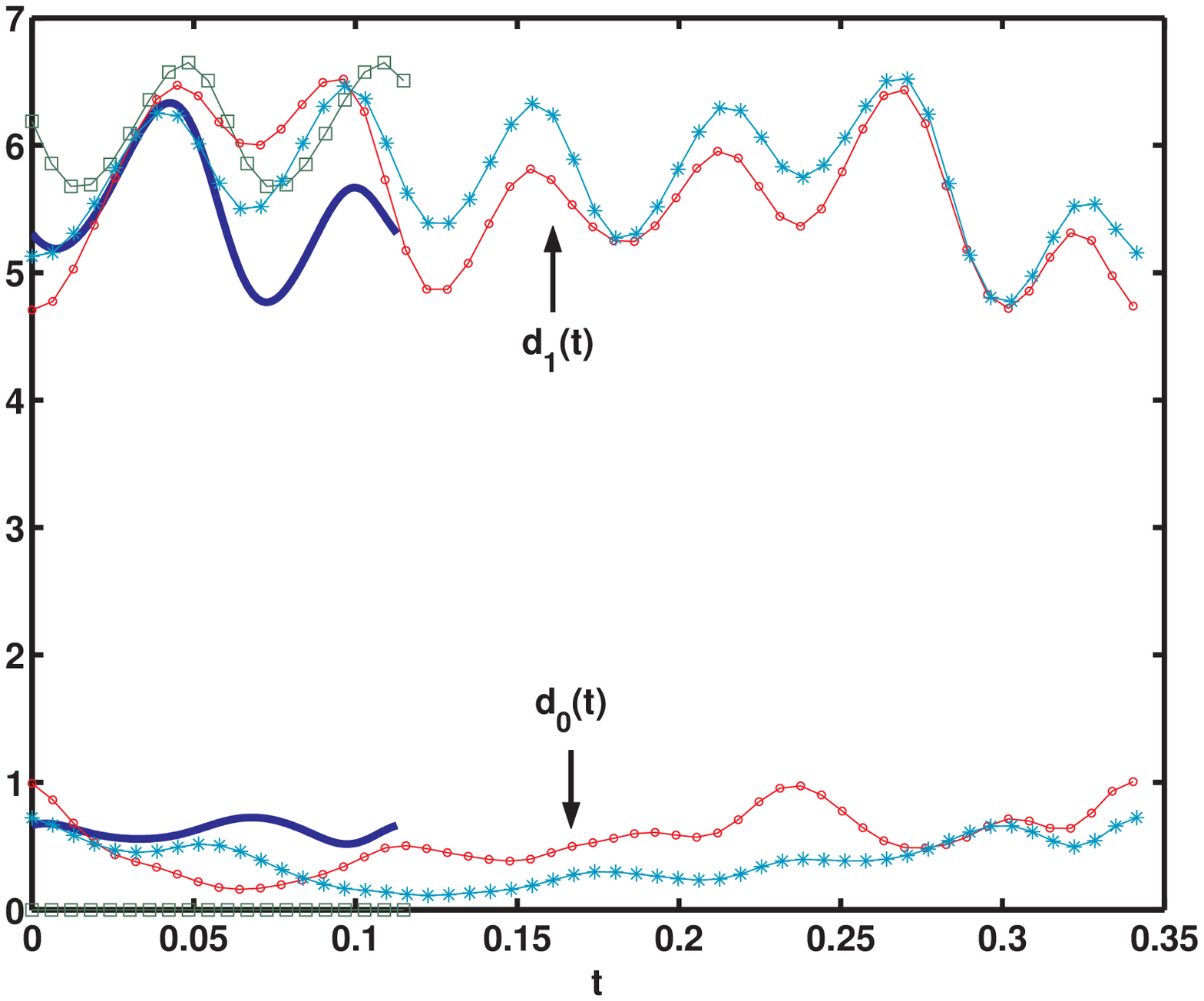}}
   \hspace*{0.1in}
   \resizebox{2.45in}{!}
       {\includegraphics{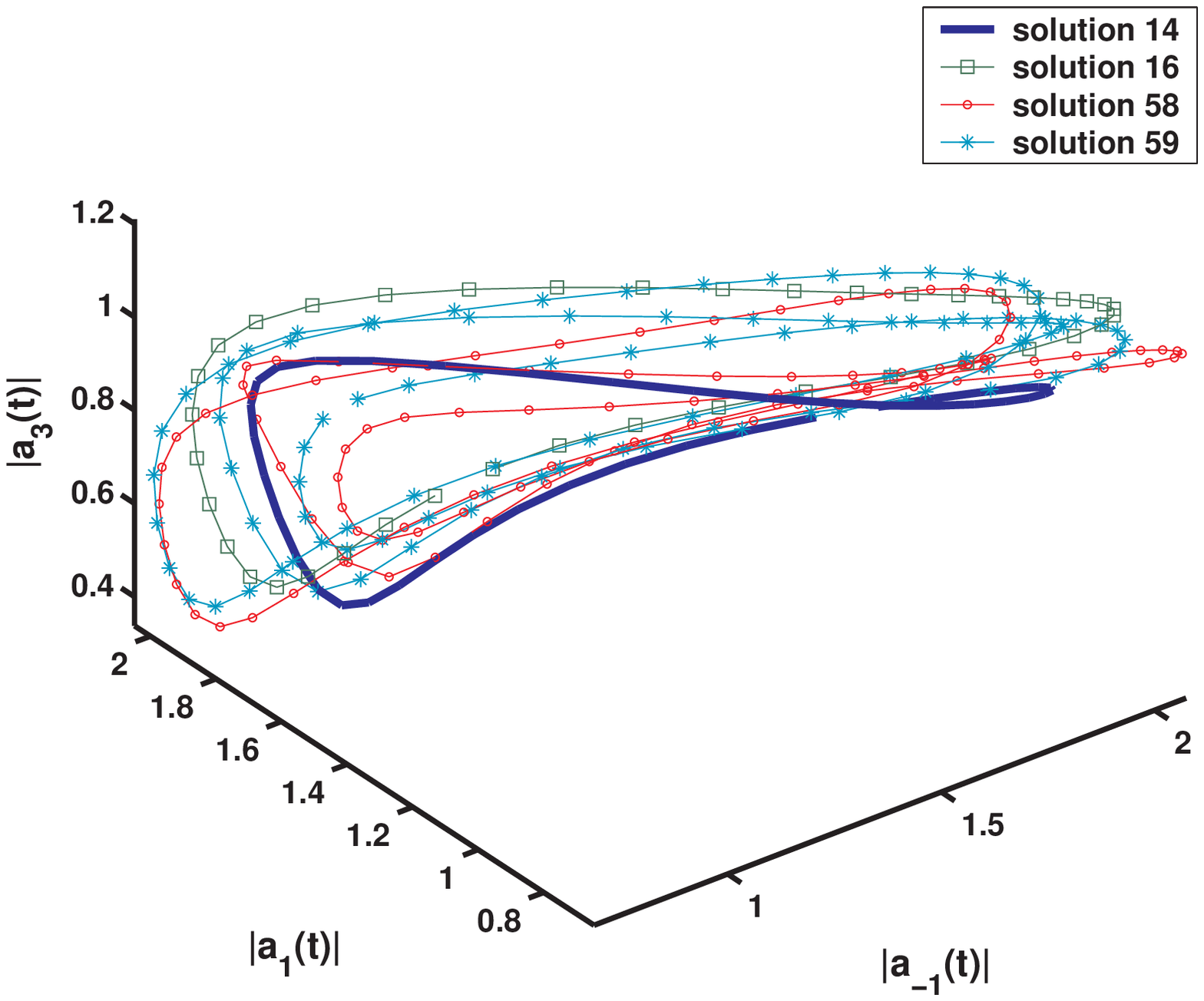}}
   \makebox[2.45in]{(a)}  \makebox[2.45in]{(b)}
   \caption{(a) Magnitude of even ($d_0(t)$) and odd ($d_1(t)$) parts of power spectrum for
       solutions 14, 16, 58, 59.  (b) Projection of time evolution onto three-dimensional space with
       coordinates $(|a_{-1}(t)|, |a_1(t)|, |a_3(t)|)$.}
   \label{fig:family}
   \end{center}
\end{figure}

All four of the solutions in the family have similar appearing pattern evolution, and indeed, this is
one of the features that brought them to our attention. In particular, solution 16 has stabilizer
$C(1,2)$ and thus satisfies $A(x + \pi, t) = -A(x, t)$, while the other three solutions approximately
have this symmetry.  To investigate these ``near symmetries'', recall from \S~\ref{sec:cglestabs}
that a solution has a $C(1,2)$ stabilizer exactly when its spatial Fourier coefficients $(a_m)$ satisfy
$a_m = 0$ for all even $m$. Thus, for each of the solutions we plotted the magnitude of the even and odd
parts of the power spectrum using
\[  d_0(t) = \sqrt{ \sum |a_{2k}(t)|^2},  \qquad d_1(t) = \sqrt{ \sum |a_{2k+1}(t)|^2}.  \]
The results are shown in Figure~\ref{fig:family}(a), which clearly shows the strong similarities in the
evolution of $d_1(t)$ and the differences in $d_0(t)$ among the members of the family.
Furthermore, the magnitude of $d_1(t)$ is significantly larger than that of $d_0(t)$, confirming the presence
of the approximate stabilizer.  As further confirmation of the similarities in the evolution of the
coefficients $a_{2k + 1}(t)$ for members of the family, we show in Figure~\ref{fig:family}(b) the projection
of the solutions onto the three-dimensional space with coordinates $(|a_{-1}(t)|, |a_1(t)|, |a_3(t)|)$
(these are the three $a_{2k + 1}(t)$ with largest magnitude).  We confirmed comparable similarities
among the evolution of the other $|a_{2k + 1}(t)|$.  Figure~\ref{fig:family}(b) also shows the pair
58/59 traversing the other solutions three times.

To put these findings in a more geometric context, we consider a direct sum decomposition of the space
$\cC$ into its odd and even pieces. More precisely, let $F_0 = \{ (a_m) : a_m = 0\ \hbox{\rm if}\ m \
\hbox{\rm is even} \}$ and $F_1 = \{ (a_m): a_m = 0\ \hbox{\rm if}\ m \ \hbox{\rm is odd} \}$. Thus,
$F_0 = \Fix(C(1, 2))$, $F_1 = \Fix(C(0, 2))$, and $\cC = F_0 \oplus F_1$.  Further, for $i = 0,1$, the
$L^2$-distance of $(a_m(t))$ to $F_i$ is $d_i(t)$, and so Figures~\ref{fig:family}(a) and
\ref{fig:family}(b) demonstrate that the projections of the family of solutions onto $F_0$ are all
approximately equal, and that the time evolution of the distances to $F_1$ are all approximately
the same. In other words, the family of solutions comes very close to lying in the generalized cylinder
$\Lambda \times F_1 $, where $\Lambda\subset F_0$ is the full \relper\ of solution 16 (i.e., its
invariant three-torus), and so the bifurcations that (presumably) created these solutions respected this
cylinder.  Since the values of $d_1$ are consistently larger than those of $d_0$, the solutions are
evolving near this cylinder reasonably close to the $F_1$ subspace.

Under the Fourier transform, the decomposition $\cC = F_0 \oplus F_1$ corresponds to a decomposition
$\cS = F'_0 \oplus F'_1$, where for $\alpha\in \cS$ the projections onto $F'_0$ and $F'_1$ are given by 
$(1/2) (\alpha(x) - \alpha(x + \pi))$ and $(1/2) (\alpha(x) + \alpha(x + \pi))$, respectively.  Thus,
within the family the evolutions of $(1/2) (A(x, t) - A(x + \pi, t))$ are all very similar while those
of $(1/2) (A(x, t) + A(x + \pi, t))$ differ.

The existence in the family of a pair with $T\approx 2 T_f$ and a pair with $T\approx 6 T_f$ raises the
question of solutions with $T\approx T_f$ and $T\approx 4 T_f$.  The question of solutions with
$T \approx T_f$ asks whether the pair 14/16 is itself the result of a period doubling bifurcation.
Solution 4 has $T \approx T_f$ and it, along with the family, constitutes the five solutions with the
least total instability.  Using its drift and following it for two of its time periods, solution 4
satisfies $ A_{4}(x, 0) = \exp(5.868 \ii) A_{4}(x,0.108)$.  On the other hand, shifting the drift of
solution 16 by its stabilizer yields for one time period the very similar
\begin{equation}  \label{alt16}
    A_{16}(x, 0) = \exp(5.857 \ii) A_{16}(x,0.114).
\end{equation}
Thus, in terms of its drift, period, and total instability solution 4 fits in with the family, but it
lacks the near symmetries that characterize the family and is located nowhere near the family's cylinder.
The situation for solutions with $T\approx 4 T_f$ is somewhat similar.  After reducing mod $2 \pi$, the
drift and period of solutions 30 and 31 are approximately twice those of solution 16 as given in
\refEqn{alt16}.  Thus, while the seven solutions with total instability less than typical have drifts and
periods in (approximate) integer multiples, only four of the solutions show clear signs of being related
by well understood bifurcations.

\subsection{Weighted Averages of Observables}
\label{sec:weighted_avgs}

As mentioned in the introduction, the unstable periodic orbits embedded in a chaotic attractor can be
used to approximate statistical averages that characterize the long-term dynamics of the 
system~\cite{bk:chaosbook,cvitanovic00,zoldi98}.  In an initial effort to investigate the usefulness of
the \relsolns\ in this regard, we use a weighted averaging scheme proposed in~\cite{zoldi98} for systems
for which a symbolic dynamics is not known.

A given quantity $\Phi_{\mathrm{typ}}$ attached to a typical trajectory is approximated by
\begin{equation}  \label{eq:Phi}
    \Phi_{\mathrm{typ}}  \approx  <\Phi>_{\sss N} \ := \ \sum_{i=1}^{N} p_i \Phi_i,
\end{equation}
where 
\begin{equation*}
    p_i  =  \frac{\frac{1}{\mathrm{SL}_i}}{\sum_{k=1}^{N} \frac{1}{\mathrm{SL}_k}} \ ,
\end{equation*}
$\Phi_i$ is the value of the quantity $\Phi$ computed for the $i$-th periodic solution, and $\mathrm{SL}_i$
is the sum of the positive Lyapunov exponents of the $i$-th periodic solution.  The value $1/\mathrm{SL}_i$
approximates the fraction of time that a typical trajectory spends in a neighborhood of the $i$-th
periodic solution.

The quantities $\Phi$ we compute are: 
the largest Lyapunov exponent $\lambda_1$;
the Lyapunov dimension $\mathrm{D_L} = j + (\sum_{i=1}^{j} \lambda_i)/|\lambda_{j+1}|$,
where $\lambda_1 \ge \lambda_2 \ge \ldots \ge \lambda_m$ are the Lyapunov exponents and $j$ is the
largest integer such that $\sum_{i=1}^{j} \lambda_i \ge 0$;
the time average of the Ginzburg-Landau energy functional,
$\mathrm{E} = \frac{1}{T} \int_0^T \int_0^{2\pi} (-R|A|^2 + |\partial A/\partial x|^2 + |A|^4/2)\, \dd x \, \dd t$;
the time average of the $L^2$-norm of the vector field that defines the CGLE,
$L^2$-VF $= \frac{1}{T} \int_0^T ( \int_0^{2\pi} |\, RA
+ (1+\ii\nu)\partial^2 A/\partial x^2 - (1+\ii\mu)A|A|^2\, |^2 \, \dd x)^{1/2} \dd t$; 
and the time average of the $L^2$-norm of a solution $A(x,t)$ of the CGLE, 
$L^2$-A $= \frac{1}{T} \int_0^T (\int_0^{2\pi} |A|^2 \, \dd x)^{1/2} \dd t$.
Table~\ref{tbl:wavgs} summarizes the results obtained using the $N=77$ \relsolns\ found.  
(We remark that the rest point $\mmbf{0}$ and the plane waves (cf. Section~\ref{sec:dynatparam}) were not included in the computation of the weighted averages.)  From the
relative errors (fourth column in Table~\ref{tbl:wavgs}) one could say that, except for the value of
$\lambda_1$, the approximations obtained are quite good.
\begin{table}[ht!]
    \footnotesize
    \begin{center}
        \begin{tabular}{|c|r|r|c|}
        \hline 
        \multicolumn{1}{|c|}{$\Phi$} & \multicolumn{1}{|c|}{$\Phi_{\mathrm{typ}}$} &
	\multicolumn{1}{|c|}{$<\Phi>_{\sss 77}$} & 
	\multicolumn{1}{|c|}{$\left |\frac{\Phi_{\mathrm{typ}}- <\Phi>_{\sss 77}}{\Phi_{\mathrm{typ}}} \right |$}   \\
        \hline \hline
        $\lambda_1$    &    5.3573  &    6.9130 &  0.2940  \\ 
        $\mathrm{D_L}$ &   11.5216  &   11.5707 &  0.0043  \\
        E              & -237.4924  & -247.0737 &  0.0403  \\
        $L^2$-VF       &  280.0915  &  292.6268 &  0.0448  \\ 
        $L^2$-$A$      &    6.3152  &    6.3763 &  0.0097  \\
	\hline
        \end{tabular} 
        \vspace*{1ex} 
        \caption{Comparison between $\Phi_{{\mathrm{typ}}}$ and approximation~\refEqn{eq:Phi}.}
	\label{tbl:wavgs}
    \end{center}
\end{table}

While these initial results are promising, a detailed investigation of the predictive value of the
\relsolns\ is warranted.  For example, other averaging schemes which do not require knowledge of symbolic dynamics, such as those presented in~\cite{dettmann98,dettmann97},  should also be considered.  Another question concerns the role that unstable dimension variability played in
the computation of the trajectory averages $\Phi_{{\mathrm{typ}}}$ since, as discussed
in~\cite{udvpaper1}, convergence to erroneous values for statistical averages computed from long-time
simulations is a possibility in the presence of unstable dimension variability.  As for the \relsolns,
since they are computed using the expansions~\refEqn{eq:xFseries} and \refEqn{eq:am_ansatz} (as
opposed to integrating the system of ODEs~\refEqn{eq:odes}) we do not expect unstable dimension
variability to have had an adverse effect in the computation of the values $\Phi_i$.

\subsection{The Iterative Solver and Dynamical Properties of Solutions}
\label{sec:solverdyn}

With the rather large range of stability properties exhibited by our collection of \relsolns, it is
interesting to consider whether these properties impact the ability of the nonlinear equations solver to
converge to solutions of \Feqzero.  The 77 \relsolns\ listed in Table~\tblrelsolns\ resulted from 350
runs performed using the nonlinear least squares solver \texttt{lmder} and relative close returns (with a
measure of closeness of $\delta=0.5$) as starting values.  In total, 146 of the 350 runs resulted in 
convergence to solutions of \Feqzero\ (this count includes convergence to the 77 distinct \relsolns, runs
that converged again to these solutions, and two runs that yielded single-frequency solutions of the CGLE),
and 204 runs converged to minima of $\F^{\mathrm{T}} \F$ for which $\F \ne \mmbf{0}$.  For those runs that
converged to solutions of \Feqzero, the final value of $||\F||_2$ was on the order of $10^{-9}$ or smaller.
The total number of iterations taken by the solver \texttt{lmder} ranged between 10 and 300. 

For solutions with periods in the range $0.40 < T < 0.46$, the value $N_t = 64$ (instead of $N_t=48$;
see \S~\ref{sec:solveF}) was required to obtain well defined solutions, in the sense that the decay
in the temporal spectra of the solutions found  was around five orders of magnitude.  For a generic
compact attractor there will be finitely many \relpers\ in any finite range of periods.  Of the total
number of runs performed with the solver \texttt{lmder}, the last 55 yielded three new solutions.  This
gives around a $5.4\%$ return on the number of new solutions found from the last set of runs, which 
suggests that we were approaching the point where no new solutions with periods less than $0.46$ could
be found with our choice of solution method.  This assertion is difficult to prove, however, based only
on empirical testing.

Because we solve the equations \refEqn{eq:cgle_nleqns} using starting values derived from relative close
returns of the system~\refEqn{eq:odes} coming from random initial data, the method we use is biased
towards finding \relsolns\ on the main attractor. Nonetheless, there is no reason to expect the
iterative solver to respect the dynamics of the CGLE, so solutions disjoint from the main attractor
are certainly possible. However, there do seem to be quantities associated with the CGLE dynamics 
that are, in certain cases, connected to the solver behavior. 

Solution 14 (shown in Figure~\ref{fig:sol14_realvsimag})
was by far the most frequently found by the solver, with convergence to it 25 times out of the 146 runs
that converged to solutions.  It was also the first solution found as well as being the least unstable
solution by both measures shown in Figure~\ref{fig:TvsLyap}.  The instability of a \relsoln\ can be
connected (cf. \S~\ref{sec:jacobian}) with the eigenvalues of the Jacobian matrix of $\F$
at the root, which in turn is, at least heuristically, connected with the eigenvalues of the Hessian of
the Newton method iterator at the root.  Thus, low instability could be connected with the presence of a
large basin of attraction for the iterative solver.  On the other hand, again heuristically, under the
dynamics of the CGLE the least unstable \relsoln\ would be the one that is easiest for other orbits to 
approach, and thus low instability would tend to indicate that the solution is in a high traffic area of
the attractor, that is, in an area with a large concentration of invariant measure.  As noted in
\S~\ref{sec:family_solns}, observations of the time evolution of typical solutions do reveal their
occasional return to a sequence of patterns similar to those manifested by solution 14. This means that
the starting values we used for the solver are more likely to be in a neighborhood of solution 14, which
would favor convergence to that solution.  It is not clear which mechanism, a large basin or an
abundance of nearby starting values, is most responsible for the frequency of occurrence of solution 14
among the computed \relsolns.

While solution 14 was the most favored solution, there are other instances of multiple
convergence. However, these do not, in general, have the same  correlation with the total instability.
On one hand, the second most frequently found solution was solution 16, with convergence to it 9 times.
This solution does have low instability and is in fact paired with solution 14 in the family described
in \S~\ref{sec:family_solns}.  On the other hand, the next two most frequently found solutions,
solutions 3 and 11, had 6 convergences each, but they do not have low instability. In addition, solution
58 has total instability only slightly more than that of solution 14, but the solver found it just once.
So while we can, in certain cases, connect the solver behavior to the CGLE dynamics, it is, in general,
extremely difficult to understand the behavior of iterative solvers in high dimensions.

%%% new section %%%
\section{Discussion and Conclusions}

We have described a method of finding relative periodic solutions for differential equations with
continuous symmetries and have demonstrated its utility by finding numerous \relsolns\
for the CGLE, a nonlinear evolution equation with widespread physical and mathematical interest. The
various purposes to which periodic orbits are usually put, such as dynamical averaging and control, can
be adapted to using \relpers.  In our investigation we found an abundance of \relsolns\ and an apparent
lack of true periodic solutions. This indicates that in the presence of continuous symmetries the use of
\relpers\ is an attractive option. In addition, the computational burden of finding \relpers\ is not
significantly higher than that of finding true periodic orbits.  It would be of interest to adapt other
methods of finding periodic orbits to find \relpers\ and systematically compare the various methods.

The most computationally challenging aspect of the numerical procedure is the solution of a system of
nonlinear equations for the drift, period, and Fourier coefficients of a \relsoln.  The use of starting
values derived from relative close returns and the choice of solver were critical to our success in
identifying solutions.  The size of the system considered here, between 2,917 and 3,909 real variables,
is at the upper limit of feasible computations at workstation scale.  In particular, the solution of
linear systems with a (dense) Jacobian as coefficient matrix, which was required as part of the
procedure for solving the nonlinear system, could be done using direct methods.  For systems with higher
dimension than that considered here, one must eventually incorporate the iterative solution of linear
systems, along with preconditioning, and the use of parallel computations.
	  
As noted in the introduction, the computation of the \relsolns\ was undertaken as a first step in
understanding of the dynamics of the CGLE with the chosen parameters.  This prompts two questions, one
specific and the other more general.  First, is the computed set of \relsolns\ sufficient to capture 
the total dynamics in any reasonable sense, and second, is the computation of a collection of \relsolns\ 
a good starting point for analyzing and understanding the dynamics of a moderately high-dimensional
dynamical system with continuous symmetries? 

The general method of analyzing dynamics using a collection of periodic solutions has been investigated
in many ways, for many years.  A given single \po\ will usually not be typical with respect to any
globally supported ergodic measure, but under hypotheses that guarantee the existence of a
Bowen-Margulis measure (see \S~20.1 in \cite{katok}), the total collection of periodic orbits
yields invariant measures that are equidistributed, and thus the collection of \pos\ captures the
statistics of the system.  While these strong hypotheses can never be confirmed in practice, there is a
fair amount of empirical evidence that a sufficiently large collection of \pos\ can capture the global
dynamics, but it is usually impossible {\em a priori} to decide what is ``sufficiently large''.
 
The next issue is what part of the dynamics one can hope to capture with a collection of \pos.  A typical
moderate dimensional chaotic system will display a great deal of dynamical variety, and there is no
reason to expect a single, indecomposable attractor.  As noted in \S~\ref{sec:solverdyn}, the
method we use is biased towards finding solutions in 
the closure of the set of typical trajectories.  Thus, the appropriate question about the collection of
\relsolns\ is in what sense it predicts the behavior of a typical trajectory.

We have begun this investigation, and the preliminary results are promising.  As reported in
\S~\ref{sec:weighted_avgs}, typical values of the Lyapunov dimension and various functionals 
were accurately computed using the 77 \relsolns\ and the weighted averaging scheme proposed 
in~\cite{zoldi98}.  We have had no success in finding symbolic dynamics, and preliminary attempts to use
the sequence of patterns given by the various \relsolns\ to build up the pattern evolution of a typical
trajectory have been unsuccessful.  Thus, the question of the predictive value of our collection of 
\relsolns\ remains open.

The second question raised above we can answer with an unqualified ``yes''.  Dynamical systems with even
moderate dynamical dimension present significant challenges not adequately dealt with using the usual
low dimensional theory.  The family of \relsolns\ has allowed us to begin unlocking the intricacies of
a flow with Lyapunov dimension of 11.52 (including the flow and symmetry directions).  The detailed 
study of the dynamics of systems with comparable intrinsic dimension is still a rare occurrence in the
literature.  In this regard we remark that it is common in certain segments of the literature to
use the terms ``high-dimensional chaos'' or ``hyperchaotic'' for any system with unstable dimension
larger than one, and the unstable dimension in our system is typically 5 and varies from 3 to 8 among
the \relsolns.

We have found  the \relsolns\ invaluable as a starting point of dynamical analysis for a variety of
reasons. First, one can be quite sure of the accuracy of the numerical computations because the
solutions can be verified both by integrating the differential equations for a short time interval,
$[0,T]$, and by confirming that the drift, period, and Fourier coefficients are solutions of the system 
of nonlinear algebraic equations.  Next, each computed \relsoln\ can be studied in isolation as one
piece of the dynamics, as one part of the pattern evolution. Finally, the observations of the individual
pieces leads to hypotheses and conjectures about the whole picture. These can be tested using the
collection of \relsolns\ as a sample space and then the results compared to typical trajectories.

As an illustration of this process we briefly list some of the lines of inquiry that have been opened
by our study of the collection of \relsolns.  Is there a correlation between spatial aspects of the
patterns of a solution and its temporal dynamics, for example, between winding numbers and total
instability?  How does the unstable dimension variability of the collection of \relsolns\ contribute to
the existence of zero finite-time Lyapunov exponents for the system and other consequences of
non-hyperbolicity?  What are the global dynamics of our system, for example, how many nontrivial
indecomposable attractors are there and how do the invariant fixed symmetry planes fit in with the
attractor(s)?  What role do the coherent structures or relative equilibria play in the overall
{\em temporal} dynamics of the system; are they contained in the main attractor, or are they isolated
and thus dynamically nontypical?  Finally, how are the \relsolns\ and overall dynamics created via
bifurcations as $R$ increases from zero to $16$?

Nonlinear dynamics has made significant progress in the study of systems with low intrinsic dimension,
but the understanding of even moderate dimensional systems still presents significant challenges.  We
have demonstrated that finding a collection of \relpers\ is both computationally feasible and
conceptually valuable, and is thus a valuable tool in the important task of understanding moderate
dimensional dynamical systems with continuous symmetries.

%%% Appendix %%%
\section*{Appendix A: Properties of Solutions}

Table~\tblrelsolns\ lists the period $T$ and drift $(\varphi,S)$ for each of the 77 \relsolns\ found. 
The solutions are listed by increasing value of $T$ and have been assigned an identifying number (first
column).  Also listed is the dimension of the unstable manifold of each solution, along with the largest
Lyapunov exponent and total instability (which is given by the sum of the positive Lyapunov exponents).

While most of the \relsolns\ have trivial $\torus^2$-stabilizers, there are a few exceptions. Solutions
3, 15, and 16 have stabilizer $C(1,2)$ (cf. \S~\ref{sec:cglestabs}), and therefore satisfy
$A(x,t) = -A(x+\pi,t)$.  Solution 9 has stabilizer $C(1,3)$; it satisfies
$A(x,t) = \e^{\ii 2\pi/3}A(x+2\pi/3,t)$.  In addition, solutions 4, 10, and 15 are even about certain points
in the spatial domain and solution 15 is odd about certain points in the spatial domain.
\begin{table}[ht!]
  \footnotesize
  \begin{center}
     \begin{tabular}{|r|r|r|r|c|r|r|}
     \hline    
     \multicolumn{1}{|c|}{ } & \multicolumn{1}{|c|}{ } & 
     \multicolumn{1}{|c|}{ } & \multicolumn{1}{|c|}{ } & 
     \multicolumn{1}{|c|}{ } & \multicolumn{1}{|c|}{Largest} &
     \multicolumn{1}{|c|}{ } \\
    
     \multicolumn{1}{|c|}{ } & \multicolumn{1}{|c|}{ } & 
     \multicolumn{1}{|c|}{ } & \multicolumn{1}{|c|}{ } & 
     \multicolumn{1}{|c|}{Unstable} & \multicolumn{1}{|c|}{Lyapunov} &
     \multicolumn{1}{|c|}{Total} \\
    
     \multicolumn{1}{|c|}{Id} & \multicolumn{1}{|c|}{$T$} &
     \multicolumn{1}{|c|}{$\varphi$} & \multicolumn{1}{|c|}{$S$} & 
     \multicolumn{1}{|c|}{Dimension} & \multicolumn{1}{|c|}{Exponent} &
     \multicolumn{1}{|c|}{Instability} \\
     \hline \hline

      1 & 0.02333 & 5.36228 & 3.85441 & 4 &  8.85592 & 28.53243  \\
      2 & 0.05394 & 2.88491 & 3.09563 & 5 &  5.00321 & 21.31125  \\
      3 & 0.05394 & 0.00111 & 3.97095 & 5 & 11.00430 & 30.30192  \\
      4 & 0.05403 & 2.93438 & 3.14159 & 4 &  3.57030 &  9.56834  \\
      5 & 0.05471 & 4.60939 & 1.45371 & 5 &  6.45679 & 27.60085  \\
      6 & 0.05561 & 4.51654 & 4.70614 & 5 &  7.76507 & 25.39928  \\
      7 & 0.06080 & 0.24364 & 2.38876 & 4 & 11.31755 & 36.02733  \\
      8 & 0.08255 & 4.79599 & 3.08241 & 5 & 10.08510 & 39.81868  \\
      9 & 0.08748 & 0.28765 & 2.44316 & 6 &  9.72294 & 36.31152  \\
     10 & 0.08950 & 5.02514 & 3.14159 & 3 & 10.72851 & 27.85965  \\
     11 & 0.10458 & 2.60234 & 3.17193 & 4 &  5.83393 & 18.52259  \\
     12 & 0.10797 & 2.65754 & 3.12097 & 3 &  6.59523 & 19.55427  \\
     13 & 0.11065 & 6.05532 & 0.00321 & 6 &  5.10800 & 19.17581  \\
     
     \hline
     \end{tabular} \\
     \vspace*{1ex} 
     \normalfont\scshape{Table \tblrelsolns} \\
     \normalfont\itshape{Properties of \relsolns.} 
  \end{center}
\end{table}
\begin{table}[ht!]
  \footnotesize
  \begin{center}
     \begin{tabular}{|r|r|r|r|c|r|r|}
     \hline    
     \multicolumn{1}{|c|}{ } & \multicolumn{1}{|c|}{ } & 
     \multicolumn{1}{|c|}{ } & \multicolumn{1}{|c|}{ } & 
     \multicolumn{1}{|c|}{ } & \multicolumn{1}{|c|}{Largest} &
     \multicolumn{1}{|c|}{ } \\
    
     \multicolumn{1}{|c|}{ } & \multicolumn{1}{|c|}{ } & 
     \multicolumn{1}{|c|}{ } & \multicolumn{1}{|c|}{ } & 
     \multicolumn{1}{|c|}{Unstable} & \multicolumn{1}{|c|}{Lyapunov} &
     \multicolumn{1}{|c|}{Total} \\
    
     \multicolumn{1}{|c|}{Id} & \multicolumn{1}{|c|}{$T$} &
     \multicolumn{1}{|c|}{$\varphi$} & \multicolumn{1}{|c|}{$S$} & 
     \multicolumn{1}{|c|}{Dimension} & \multicolumn{1}{|c|}{Exponent} &
     \multicolumn{1}{|c|}{Instability} \\
     \hline \hline

     14 & 0.11282 & 2.60639 & 3.10577 & 4 &  1.88230 &  5.18967  \\
     15 & 0.11461 & 2.25004 & 3.14159 & 3 &  6.90750 & 15.73770  \\
     16 & 0.11492 & 2.71671 & 3.14159 & 5 &  3.20606 & 10.12255  \\
     17 & 0.13560 & 1.82110 & 2.04598 & 6 & 10.56790 & 42.15027  \\
     18 & 0.13730 & 0.73763 & 0.75086 & 6 & 13.54288 & 53.74900  \\     
     19 & 0.14037 & 1.87712 & 5.73712 & 7 & 10.57182 & 40.07190  \\
     20 & 0.14916 & 5.81089 & 1.26773 & 5 & 17.19835 & 50.08455  \\
     21 & 0.15160 & 1.94728 & 5.40241 & 5 &  6.75385 & 19.66934  \\
     22 & 0.16098 & 1.95738 & 3.85354 & 5 & 10.61619 & 44.38707  \\
     23 & 0.16263 & 2.56407 & 5.09938 & 6 & 11.43816 & 35.46420  \\
     24 & 0.16971 & 5.48891 & 3.93522 & 8 &  7.86379 & 30.95111  \\
     25 & 0.17500 & 3.80256 & 0.98842 & 6 & 10.36722 & 44.23884  \\
     26 & 0.18249 & 3.40328 & 6.14235 & 5 & 10.19717 & 36.97288  \\
     27 & 0.18317 & 3.61256 & 5.24560 & 6 & 12.29563 & 42.02092  \\
     28 & 0.20146 & 1.90508 & 2.41894 & 6 & 14.46845 & 59.36105  \\
     29 & 0.21038 & 0.70319 & 1.90349 & 5 & 14.85499 & 60.35458  \\
     30 & 0.21178 & 5.16441 & 0.07994 & 4 &  5.87695 & 14.88747  \\
     31 & 0.21766 & 5.61399 & 0.00002 & 5 &  3.70755 & 13.08357  \\
     32 & 0.22045 & 5.15235 & 0.06622 & 3 &  8.28842 & 16.37539  \\
     33 & 0.22482 & 4.41470 & 1.10856 & 7 & 13.46647 & 47.71035  \\
     34 & 0.24430 & 5.53858 & 3.60540 & 6 & 11.40020 & 41.36390  \\
     35 & 0.24598 & 0.88868 & 4.47924 & 5 &  8.78079 & 35.50550  \\
     36 & 0.24761 & 2.24375 & 2.32864 & 6 & 15.96544 & 59.60573  \\
     37 & 0.25234 & 4.06727 & 2.73577 & 5 &  7.27440 & 28.49855  \\
     38 & 0.25374 & 1.47983 & 2.26736 & 6 & 10.89903 & 43.88382  \\
     39 & 0.26014 & 5.12178 & 2.60985 & 4 &  5.39730 & 16.59190  \\
     40 & 0.26623 & 3.07369 & 3.27604 & 6 & 11.19883 & 50.85904  \\
     41 & 0.27286 & 3.69060 & 3.55692 & 6 &  9.88356 & 37.26567  \\
     42 & 0.27504 & 2.57465 & 3.21456 & 5 &  5.92789 & 18.84754  \\
     43 & 0.27679 & 2.90953 & 3.10889 & 4 &  6.77859 & 24.98180  \\
     44 & 0.28744 & 4.84490 & 1.76031 & 5 &  7.34615 & 23.42338  \\
     45 & 0.28803 & 2.74448 & 1.37428 & 5 &  8.98143 & 31.75536  \\
     46 & 0.29746 & 4.15104 & 4.09627 & 6 &  6.98488 & 22.53727  \\
     47 & 0.29751 & 4.32786 & 1.58049 & 5 &  6.18589 & 20.33417  \\
     48 & 0.30045 & 4.39783 & 4.62959 & 4 &  5.77264 & 18.52750  \\
     49 & 0.30392 & 3.72216 & 5.49134 & 6 &  8.10462 & 31.10158  \\
     50 & 0.30492 & 2.74872 & 3.31013 & 5 &  6.89564 & 21.04823  \\
     51 & 0.30695 & 4.31730 & 5.12453 & 4 &  9.02700 & 24.32585  \\
     52 & 0.31343 & 0.72363 & 1.98525 & 5 &  8.41168 & 25.64814  \\
     53 & 0.31517 & 1.87776 & 1.69240 & 5 &  7.79446 & 24.72635  \\
     54 & 0.31924 & 5.68308 & 2.25711 & 6 & 11.97732 & 46.26127  \\
     55 & 0.32350 & 2.21845 & 1.26058 & 5 &  8.44337 & 31.00286  \\
     56 & 0.33150 & 2.42184 & 3.37971 & 3 &  6.90072 & 15.94646  \\
     57 & 0.33217 & 4.50138 & 1.88596 & 5 &  6.86741 & 19.11005  \\
     58 & 0.34035 & 1.72898 & 3.15957 & 3 &  3.92346 &  5.99484  \\
     59 & 0.34148 & 1.74043 & 3.18504 & 3 &  2.56575 &  6.92418  \\
     60 & 0.34825 & 6.00790 & 5.07384 & 6 & 10.50662 & 47.95424  \\
     61 & 0.34892 & 2.45502 & 3.11195 & 4 &  5.98084 & 18.44338  \\
     62 & 0.35567 & 0.76100 & 0.97929 & 6 &  8.70424 & 25.31565  \\
     63 & 0.36395 & 1.11216 & 5.51746 & 5 &  4.75266 & 16.16998  \\
     64 & 0.36621 & 3.81362 & 2.39863 & 6 &  5.79290 & 27.12694  \\
     65 & 0.37306 & 5.25206 & 1.42035 & 5 &  9.96569 & 30.85402  \\
     66 & 0.37413 & 1.14509 & 1.19984 & 5 &  8.30154 & 18.70350  \\
     67 & 0.37619 & 1.78357 & 0.93027 & 5 &  6.29217 & 16.68876  \\
     68 & 0.37893 & 1.44760 & 0.83458 & 4 &  6.92225 & 20.55719  \\
 
     \hline
     \end{tabular} \\
     \vspace*{1ex} 
     \normalfont\scshape{Table \tblrelsolns} \\
     \normalfont\itshape{Properties of \relsolns, continued.}
     \vspace*{1ex} 
  \end{center}
\end{table}
\begin{table}[ht!]
  \footnotesize
  \begin{center}
     \begin{tabular}{|r|r|r|r|c|r|r|}
     \hline    
     \multicolumn{1}{|c|}{ } & \multicolumn{1}{|c|}{ } & 
     \multicolumn{1}{|c|}{ } & \multicolumn{1}{|c|}{ } & 
     \multicolumn{1}{|c|}{ } & \multicolumn{1}{|c|}{Largest} &
     \multicolumn{1}{|c|}{ } \\
    
     \multicolumn{1}{|c|}{ } & \multicolumn{1}{|c|}{ } & 
     \multicolumn{1}{|c|}{ } & \multicolumn{1}{|c|}{ } & 
     \multicolumn{1}{|c|}{Unstable} & \multicolumn{1}{|c|}{Lyapunov} &
     \multicolumn{1}{|c|}{Total} \\
    
     \multicolumn{1}{|c|}{Id} & \multicolumn{1}{|c|}{$T$} &
     \multicolumn{1}{|c|}{$\varphi$} & \multicolumn{1}{|c|}{$S$} & 
     \multicolumn{1}{|c|}{Dimension} & \multicolumn{1}{|c|}{Exponent} &
     \multicolumn{1}{|c|}{Instability} \\
     \hline \hline

     69 & 0.38444 & 1.83435 & 1.06175 & 5 &  7.33543 & 20.40029  \\
     70 & 0.38599 & 1.21873 & 1.00009 & 4 &  7.66270 & 17.44702  \\
     71 & 0.39128 & 2.38456 & 5.34383 & 4 &  5.78585 & 18.67097  \\
     72 & 0.39675 & 4.79784 & 3.35390 & 6 &  9.55893 & 30.02312  \\
     73 & 0.40557 & 3.21458 & 6.24425 & 4 &  6.80844 & 22.92050  \\
     74 & 0.42108 & 0.44223 & 4.10959 & 4 & 10.55221 & 19.44133  \\
     75 & 0.42447 & 2.27338 & 3.25431 & 5 &  7.58996 & 26.84563  \\
     76 & 0.44044 & 3.32603 & 2.31594 & 6 &  7.92846 & 26.18850  \\
     77 & 0.45840 & 5.73253 & 6.25633 & 5 &  8.01104 & 25.60890  \\
     \hline
     \end{tabular} \\
     \vspace*{1ex} 
     \normalfont\scshape{Table \tblrelsolns} \\
     \normalfont\itshape{Properties of \relsolns, continued.}
  \end{center}
\end{table}

Table~\tblsf\ lists properties of the solution $A(x,t) = 0$, which corresponds to Id=0, the plane waves
$A(x,t) = \hat{a}_{p,\pm1}\e^{\pm \ii\omega t}e^{\ii px}$, which correspond to the identifying numbers 1--4,
and the two single-frequency solutions $A(x,t) = B(x)\e^{-\ii\omega t}$ (Id=5,6) that were computed
numerically.  Solution 6 has stabilizer $C(1,2)$ and is even and odd about about certain points in the
spatial domain.  Solution 5 is even about certain points in the spatial domain.
\begin{table}[hbt!]
  \footnotesize
  \begin{center}
     \begin{tabular}{|r|c|c|c|r|r|r|}
     \hline    
     \multicolumn{1}{|c|}{ } & 
     \multicolumn{1}{|c|}{ } & \multicolumn{1}{|c|}{ } & 
     \multicolumn{1}{|c|}{ } & \multicolumn{1}{|c|}{Largest} &
     \multicolumn{1}{|c|}{ } \\
   
     \multicolumn{1}{|c|}{ } & 
     \multicolumn{1}{|c|}{ } & \multicolumn{1}{|c|}{ } & 
     \multicolumn{1}{|c|}{Unstable} & \multicolumn{1}{|c|}{Lyapunov} &
     \multicolumn{1}{|c|}{Total} \\
    
     \multicolumn{1}{|c|}{Id} & 
     \multicolumn{1}{|c|}{$T$} & \multicolumn{1}{|c|}{$\omega$} &  
     \multicolumn{1}{|c|}{Dimension} & \multicolumn{1}{|c|}{Exponent} &
     \multicolumn{1}{|c|}{Instability} \\
     \hline \hline
    
     0 &    --   &    --   &  14 & 16.0000 & 168.0000  \\
     1 & 0.07854 & 80.0000 & \ 8 & 54.7935 & 320.2662  \\
     2 & 0.09240 & 68.0000 & \ 8 & 51.7819 & 298.3522  \\
     3 & 0.19635 & 32.0000 & \ 8 & 41.2843 & 220.1846  \\
     4 & 0.22440 & 28.0000 & \ 8 & 26.0693 & 124.4494  \\
     5 & 0.12517 & 50.1968 &  10 & 15.8062 & 109.0126  \\
     6 & 0.12609 & 49.8320 & \ 7 & 15.0947 &  61.4864  \\
     \hline 
     \end{tabular} \\
  \end{center}
  \vspace*{1ex} 
  \begin{center}
      \normalfont\scshape{Table \tblsf} \\
  \end{center}
  \normalfont\itshape{Properties of solution $A(x,t) = 0$ (Id=0),
        plane waves $A(x,t) = \hat{a}_{p,\pm1}\e^{\pm \ii\omega t}e^{\ii px}$
        (Id=1,2,3,4 correspond to p=0,1,2,3, respectively), and
        single-frequency solutions $A(x,t) = B(x)\e^{-\ii\omega t}$ (Id=5,6).}
\end{table}

%%% Acknowledgements %%%
\section*{Acknowledgements}
V. L\'{o}pez thanks O. Stoyanov for helpful discussions and suggestions on preliminary versions of the
paper and the CSE program at the University of Illinois for providing excellent computing facilities.
The authors thank one of the reviewers for providing references~\cite{dettmann98,dettmann97}.

%%% Bibliography %%%
\bibliography{cgle}
\bibliographystyle{plain}

\end{document}